\documentclass[11pt]{amsart}
\newtheorem{Theorem}{Theorem}[section]

\newtheorem{Lemma}[Theorem]{Lemma}

\newtheorem{Remark}[Theorem]{Remark}
\newtheorem{Corollary}[Theorem]{Corollary}

\newtheorem{Assumption 2}[Theorem]{Assumption 2}

\numberwithin{equation}{section}

\usepackage{stmaryrd}
\usepackage{ulem}
\usepackage{yfonts}
\usepackage{amsbsy}
\usepackage{pstricks}
\usepackage{graphicx}
\usepackage{pst-xkey}
\usepackage{pst-all,pst-gr3d}
\usepackage[dvips,arrow,matrix]{xy}

\def\k#1{\kern#1em}
\def\Ib#1{{I\kern-.25em#1}}
\def\Ibb#1{{I\kern-.23em#1}}

\def\CC{{\mathbb C}}

\def\NN{{\mathbb N}}

\def\RR{{\mathbb{R}}}

\def\vci{\vrule  width.02em height1.47ex depth-.0ex}
\def\11{{\rm\k{.2}\vci\k{-.37}1}}

\def\fin{{\begin{flushright}
\it{Q.E.D.}
\end{flushright}}}

\begin{document}

\address{Universit\'e de Bordeaux, Institut de Math\'ematiques, UMR CNRS 5251, F-33405 Talence Cedex}

\email{bachelot@math.u-bordeaux1.fr}

\title{Wave Propagation and Scattering  for the {\it RS2} Brane
  Cosmology Model}

\author{Alain BACHELOT}

\begin{abstract}
We study the wave equation for the gravitational fluctuations in the
Randall-Sundrum brane cosmology model. We solve the global Cauchy
problem and we establish that the solutions are the sum of a slowly
decaying massless
wave localized near the brane, and a superposition of massive
dispersive waves. We compute the kernel of the truncated resolvent. We prove some $L^1-L^{\infty}$, $L^2-L^{\infty}$ decay
estimates and  global $L^p$ Strichartz type inequalities. We develop the complete
scattering theory : existence and asymptotic completness of the wave
operators, computation of the scattering matrix, determination of the
resonances on the logarithmic Riemann surface.
\end{abstract}
\maketitle

\pagestyle{myheadings}
\markboth{\centerline{\sc Alain Bachelot}}{\centerline{\sc Wave
    Propagation and Scattering for the {\it RS2}  Brane Cosmology Model}}



\section{Introduction}

The fundamental question in mathematical cosmology is the stability of
the models of the universe. It consists in solving the global Cauchy
problem for the Einstein equations. This is an extremely hard task
because of the deep nonlinearity of these equations. We recall the
impressive work by D. Christodoulou and S. Klainerman \cite{christodoulou}
on the global nonlinear stability of the Minkowski space-time which,
however, is the simplest model. In this paper, our aim is much more
modest. We adress the question of the linear stability of a famous
model of the brane cosmology, i.e. we investigate the global
properties of the gravitational fluctuations that are the solutions of a linear hyperbolic equation with a singular potential. We consider the  so-called {\it RS2 brane world},
proposed by L. Randall and R. Sundrum in \cite{rs}, in which our
observable universe is a 4-D Minkowski brane embedded in a 5-D Anti-de
Sitter space bulk.
This model plays a considerable role in cosmology :
it is a phenomenological realization of M theory ideas ; it also
provides a framework for exploring the holographical principle and
the AdS/CFT correspondence. The seminal papers by L. Randall and
R. Sundrum \cite{rs1} and \cite{rs} have generated a huge
litterature. Among important publications, we can cite the following
reviews, \cite{brax}, \cite{langlois}, \cite{maartens}, and the book
by P.D. Mannheim
 \cite{man}.\\

The {\it RS2 brane world},
introduced by L. Randall and R. Sundrum in \cite{rs} is
described by the 5-dimensional lorentzian manifold endowed with a
warped metric :
\begin{equation*}
{\mathcal M}=\RR_T\times\RR_{\mathbf X}^3\times\RR_y,\;\;
ds^2_{\mathcal M}=e^{-2k\mid y\mid}\left(dT^2-d{\mathbf X}^2\right)-dy^2.
  \label{RSG}
\end{equation*}
The warp coefficient $k$ is a strictly positive real number. This metric is not smooth at $y=0$ and we have to distinguish two
submanifolds : the
{\it positive tension Minkowski brane} that corresponds to our flat world
\begin{equation*}
{\mathbf M}=\RR_T\times\RR_{\mathbf X}^3\times\{y=0\},\;\;ds^2_{\mathbf M}=dT^2-d{\mathbf X}^2,
  \label{M}
\end{equation*}
and the {\it bulk} $\mathbf{B}$, associated with the extra transverse and non compact
dimension $y$, in which the brane is imbedded :
\begin{equation*}
{\mathbf B}={\mathbf B^{+}}\cup  {\mathbf B^{-}},\;\;{\mathbf B^{\pm}}=\RR_T\times\RR_{\mathbf X}^3\times\{\pm
y>0\},\;\;ds^2_{\mathbf B^{\pm}}=e^{\mp 2ky}\left(dT^2-d{\mathbf X}^2\right)-dy^2.
  \label{B}
\end{equation*}
${\mathbf B^{+}}$ and ${\mathbf B^{-}}$, that are isometric, are parts of the Anti-De Ditter universe $AdS$,
with constant negative curvature $-k^2$. To see that, we introduce new
coordinates
$$
U^0=\frac{1}{2}e^{-ky}\left( e^{2ky}+\frac{1}{k^2}+\mid
  \mathbf{X}\mid^2-T^2\right),\;\;
U^i= \frac{1}{k}e^{-ky}\mathbf{X}^i,\;\;1\leq i\leq 3,
$$
$$
U^4=\frac{1}{2}e^{-ky}\left(e^{2ky}-\frac{1}{k^2}+\mid \mathbf{X}\mid^2-T^2\right),
\;\; U^5=
\frac{T}{k}e^{-ky},
$$
and we check that $\mathbf{B}^{+}$ is included in the region $U^0>U^4$
of $AdS$ which is defined as the quadric
$$
AdS:\;\;\left(U^0\right)^2+\left(U^5\right)^2-\sum_{i=1}^4\left(U^i\right)^2=\frac{1}{k^2}
$$
embedded in the 6 dimensional flat space $\RR^6_\mathbf{U}$ with
the metric
$$
ds^2=\left(dU^0\right)^2+\left(dU^5\right)^2-\sum_{i=1}^4\left(dU^i\right)^2.
$$
To understand the causal
structure of this universe,  we can also
construct $\mathcal{M}$ by gluing together the time-like boundary of
two copies of a piece $\mathbf{B}^+$ of {\it AdS}, of which the boundary
is isometric to the Minkowski manifold. This construction is depicted
in  Figure 1 by a Penrose diagram where the radial null geodesics
travel at $\pm 45^o$ angles. The dotted line is the time-like
infinity, the dashed lines have to be identified. We see that unlike the whole Anti-De
Sitter space time, $\mathcal{M}$ is globally hyperbolic and there is
no problem of causality (see \cite{DADS} for a discussion on a
global Cauchy problem in $AdS$).\\

\begin{center}
\psset{xunit=0.5091cm, yunit=0.5091cm}
\begin{pspicture}(-9.35,-8.25)(9.35,8.25)%
\pstVerb{1 setlinejoin}%
\psset{linecolor={black},linewidth=0.2pt}
\psline(-7,-6)(-7,6)
\psline(7,-6)(7,6)
{\psset{linewidth=0.2mm, linestyle=dotted}
\psline(1,-6)(1,6)}
{\psset{linewidth=0.2pt, linestyle=dashed}
\psline(-7,6)(-1,6)}
{\psset{linewidth=0.2pt, linestyle=dashed}
\psline(-7,-6)(-1,-6)}
{\psset{linewidth=0.2pt, linestyle=dashed}
\psline(1,-6)(7,-6)}
{\psset{linewidth=0.2pt, linestyle=dashed}
\psline(1,6)(7,6)}
{\psset{linewidth=0.2mm, linestyle=dotted}
\psline(-1,-6)(-1,6)}
\psset{linestyle=solid}
\pcline(-7,-0)(-1,6)
\aput{:U}{$T=+\infty$}
\pcline(-7,-0)(-1,6)
\bput{:U}{$y=-\infty$}
\pcline(-7,-0)(-1,-6)
\bput{:U}{$T=-\infty$}
\pcline(-7,-0)(-1,-6)
\aput{:U}{$y=-\infty$}
\psline{<->}(-1.6,-0)(1.6,0)
\psset{linewidth=1pt}
\psbezier(-1,6)(-2,5)(-2,-5)(-1,-6)
\psset{linewidth=0.2pt}
\rput(0,-0.5){\tiny\it glue together}
\psset{linewidth=1pt}
\psbezier(1,6)(2,5)(2,-5)(1,-6)
\psset{linewidth=0.2pt}
\pcline(1,6)(7,0)
\aput{:U}{$T=+\infty$}
\pcline(1,6)(7,0)
\bput{:U}{$y=+\infty$}

\pcline(1,-6)(7,0)
\bput{:U}{$T=-\infty$}
\pcline(1,-6)(7,0)
\aput{:U}{$y=+\infty$}
\rput(0,0.5){\small BRANE}
\rput(-4,6.8){\it AdS}
\rput(4,6.8){\it AdS}
\rput(-4,0){$\mathbf{B}^-$}
\rput(+4,0){$\mathbf{B}^+$}
\rput{90}(2.2,0){$y=0$}
\rput{90}(-2.2,0){$y=0$}
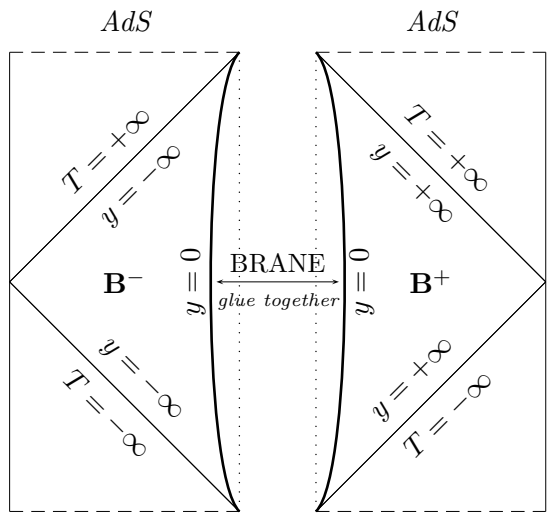
\rput(0,-8.5){{\it Figure 1.} Conformal diagram of $\mathcal{M}$.}
\end{pspicture}
\end{center}
\vspace{1cm}

We are interested in the
gravitational fluctuations around this background. Using the
linearized Einstein equations, Randall and Sundrum
established that these gravity waves obey to the master equation :
\begin{equation}
\left[e^{2k\mid y\mid}\left(\partial^2_T-\Delta_{\mathbf X}\right)-\partial_y^2-4k\delta_0(y)+4k^2\right]\Psi=0,
  \label{MEQ}
\end{equation}
where the Dirac  distribution denotes the simple layer on the brane. Our work is devoted to a complete analysis of this equation. The main
assertions that are stated heuristically in the papers by the physicists
(see e.g. \cite{man}, \cite{man-sim}, \cite{rs}, \cite{sea}) are the
following : (i) the general solution $\Psi$ of the master equation can
be expressed as
\begin{equation}
\Psi(T,\mathbf{X},y)=\Psi_{\mathbf{M}}(T,\mathbf{X},y)+\Psi_{\mathbf{B}}(T,\mathbf{X},y),
\label{repres}
\end{equation}
where $\Psi_{\mathbf{M}}$ is the  {\it massless graviton} that has the
form
\begin{equation*}
\Psi_{\mathbf M}(T,\mathbf{X},y)=e^{-2k\mid
  y\mid}\psi_{0}(T,\mathbf{X}),\;\;\partial^2_T\psi_0-\Delta_{\mathbf X}\psi_0=0,
\end{equation*}
and $\Psi_{\mathbf{B}}$ is the  {\it Kaluza-Klein tower} that is a
superposition of free massive Klein-Gordon fields propagating along the brane :
\begin{equation*}
\Psi_{\mathbf
  B}(T,\mathbf{X},y)=\sum_{\pm}\int_0^{\infty}f_m^{\pm}(y)\psi_{m}^{\pm}(T,\mathbf{X})dm,\;\;f_m^{\pm}(-y)=\pm f_m^{\pm}(y),\;\;\partial^2_T\psi_{m}^{\pm}-\Delta_{\mathbf X}\psi_{m}^{\pm}+m^2\psi_m^{\pm}=0 ;
\end{equation*}
(ii) while the massless graviton remains localized near the brane, the
energy of the Kaluza-Klein tower decays in the neighborhood of the
brane as $T$ tends to the infinity. \\

In this paper we establish these results in a rigorous way, and we
develop the complete scattering theory for the master equation. In
section 2, we construct the suitable functional framework, and we
prove that the hamiltonian is a well defined self-adjoint operator. In
section 3, we perform its spectral analysis and  the proof of the
representation (\ref{repres}). We also compute  explicitely the kernel
of the truncated resolvent, and we show the existence of its analytic continuation on the
universal covering $\widetilde{\CC^*}$ of $\CC^*$, outside a lattice of half hyperbolas. In the next section, we prove that the
Kaluza-Klein tower decays as $t^{-\frac{3}{2}}$ by establishing some
$L^1-L^{\infty}$ and $L^2-L^{\infty}$ estimates in suitable weighted
spaces. Moreover we get a Strichartz type estimate near the brane in
$L^{\infty}\left([-R,R]_y;L^4\left(\RR_T\times\RR^3_{\mathbf{X}}\right)\right)$,
and for the Kaluza-Klein tower in
$L^{\frac{10}{3}}\left(\RR_T\times\RR^3_{\mathbf{X}}\times[-R,R]_y\right)$. 
Section 5 is devoted to the scattering theory : we prove the existence
and asymptotic completeness of the wave operators describing the
scattering of the Kaluza-Klein tower by the brane. In the last part,
we calculate the scattering matrix, and we determine the set of
resonances : it is a lattice of radial half straight lines on
$\widetilde{\CC^*}$, of which the origins are the $z$-zeros of the Hankel
functions $H_{\nu}^{(j)}(z)$, $\nu,j=1,2$. All our results of asymptotic behaviours, suggest that this brane cosmology model is linearly
stable. The non-linear stability of the Minkowski brane is a huge open
problem.



\section{The Cauchy problem.}

It is convenient to use the Poincar\'e coordinates $(t,\mathbf{x},z)$ with
$$
t=kT,\;\;\mathbf{x}=k\mathbf{X},\;\;z=\frac{y}{\mid y\mid}\left(e^{k\mid y\mid}-1\right),
$$
for which the Randall-Sundrum Universe is described by the conformally
flat manifold
$$
{\mathcal M}=\RR_t\times\RR_{\mathbf x}^3\times\RR_z,\;\;
ds^2=\frac{1}{k^2}\left(\frac{1}{1+\mid z\mid}\right)^2\left(dt^2-d{\mathbf x}^2-dz^2\right).
$$
We change the unknown field by puting
$$
\Phi(t,{\mathbf x},z)=e^{\frac{k}{2}\mid y\mid}\Psi(T,{\mathbf X},y).
 $$
Since the simple layer $\delta_0$ is a homogeneous distribution of
degree $-1$ and $\delta_0(y)=k\delta_0(z)$, we see that $\Psi$ is solution of (\ref{MEQ}) iff $\Phi$ is solution of :
\begin{equation}
\left(\square_{t, {\mathbf x}, z}+\frac{15}{4}\left(\frac{1}{1+\mid
      z\mid}\right)^2-3\delta_0(z)\right)\Phi=0,\;\;(t,{\mathbf x},z)\in\RR\times\RR^3\times\RR.
  \label{EQB}
\end{equation}
Therefore Equation (\ref{MEQ}) is equivalent to the D'Alembertian on the five-dimensional Minkowski space-time,
$\square_{t, {\mathbf x}, z}:=\partial^2_t-\Delta_{\mathbf
  x}-\partial^2_z$, perturbed by a singular potential, $\frac{15}{4}\left(\frac{1}{1+\mid
      z\mid}\right)^2-3\delta_0(z)$, the so-called
{\it volcano potential}. In this section we investigate the Cauchy
problem for the equation (\ref{EQB}), with the initial data
\begin{equation}
\Phi(0,\mathbf{x},z)=\Phi_0(\mathbf{x},z),\;\;\partial_t\Phi(0,\mathbf{x},z)=\Phi_1(\mathbf{x},z),\;\;({\mathbf x},z)\in\RR^3\times\RR.
  \label{ci}
\end{equation}

We introduce the differential operator
\begin{equation}
  P(\partial):=-\Delta_{\mathbf
    x}-\partial^2_z+\frac{15}{4}\left(\frac{1}{1+\mid
      z\mid}\right)^2-3\delta_0(z),
  \label{H}
\end{equation}
considered as a densely defined operator ${\mathbf H}_0$ on $L^2(\RR^4)$, endowed with
the domain :
\begin{equation}
{\mathfrak D}({\mathbf H_0}):=\left\{u\in
    H^1\left(\RR^3_{\mathbf x}\times\RR_z\right);\;\;P(\partial)u\in
    L^2\left(\RR^3_{\mathbf x}\times\RR_z\right)\right\},
  \label{DH}
\end{equation}
which makes sense since $H^1(\RR^4)\subset
C^0\left(\RR_z;H^{\frac{1}{2}}(\RR^3_{\mathbf x})\right)$ and so
$u({\mathbf x},0)\delta_0(z)$ is a well defined distribution in
$H^{-1}(\RR^4)$.

To analyse this operator and to be able to construct the suitable
functional framework, we recall some definitions of spaces of
distributions. Given an Hilbert space $X$,  the
Beppo Levi space $BL^1(\RR^n;X)$ is defined as the completion of the space of the
$X$-valued test functions on $\RR^n$, $C^{\infty}_0(\RR^n;X)$, for the Dirichlet norm $\parallel \nabla v\parallel_{L^2(\RR^n;X)}$ (see \cite{denylions}).
On the other hand, we introduce the weighted
Sobolev space $W^1_0(\RR;X)$  that is the subspace of $\mathcal{D}'(\RR;X)$ satisfying :
\begin{equation}
\parallel f\parallel_{W^1_0(\RR;X)}^2:=\int_{-\infty}^{\infty}\mid
f'(z)\mid^2_X+\frac{15}{4}\left(\frac{1}{1+\mid z\mid}\right)^2\mid
f(z)\mid^2_Xdz<\infty.
  \label{wh}
\end{equation}
We have $W^1_0(\RR;X)\subset C^0(\RR;X)$, and thanks to the Hardy
inequality, 
\begin{equation}
\forall f\in
L^2(\RR_z;X),\;\;\int_{-\infty}^{\infty}\frac{1}{z^2}\left\vert\int_0^zf(\xi)d\xi\right\vert_X^2dz\leq
4\int_{-\infty}^{\infty}\mid f(z)\mid^2_Xdz,
  \label{HAR}
\end{equation}
there exists $C>0$ such that
\begin{equation}
\forall f\in W^1_0(\RR;X),\;\;f(0)=0\Rightarrow
\parallel f\parallel_{W^1_0(\RR;X)}\leq C\parallel f'\parallel_{L^2(\RR;X)}.
  \label{ha}
\end{equation}


\begin{Lemma}
The operator  $({\mathbf H}_0, {\mathfrak D}({\mathbf H}_0))$ is a positive
self-adjoint operator on $L^2(\RR^4)$ and $0$ is not an eigenvalue. The domain of ${\mathbf
  H}^{\frac{1}{2}}_0$ is $H^1(\RR^4)$ and for any $u\in H^1(\RR^4)$ the
three following quantities are equal to $\parallel{\mathbf
  H}^{\frac{1}{2}}_0u \parallel^2_{L^2(\RR^4)}$ :
\begin{equation}
\parallel \nabla_{\mathbf x}u\parallel^2_{
  L^2\left(\RR^3_{\mathbf{x}}\times\RR_z\right)}+\parallel
\partial_zu+\frac{3}{2}\frac{z}{\mid z\mid}\left(\frac{1}{1+\mid z \mid}\right)u\parallel^2_{
  L^2\left(\RR^3_{\mathbf{x}}\times\RR_z\right)},
\label{qnorm}
\end{equation}

\begin{equation}
\parallel u\parallel^2_{
  L^2\left(\RR_z;BL^1\left(\RR^3_{\mathbf{x}}\right)\right)}+\parallel
u\parallel^2_{W^1_0\left(\RR_z;L^2\left(\RR^3_{\mathbf
        x}\right)\right)}-3\parallel u(\mathbf{x},0)\parallel^2_{L^2(\RR^3_{\mathbf{x}})},
  \label{qqqq}
\end{equation}

\begin{equation}
\parallel u\parallel^2_{
  L^2\left(\RR_z;BL^1\left(\RR^3_{\mathbf{x}}\right)\right)}+\parallel
u-u(\mathbf{x},0)(1+\mid z\mid)^{-\frac{3}{2}}\parallel^2_{W^1_0\left(\RR_z;L^2\left(\RR^3_{\mathbf
        x}\right)\right)}.
  \label{qqg}
\end{equation}
Moreover for any $m>0$, we have the elliptic estimate :
\begin{equation}
  \min\left(\frac{m^2}{3+m^2},\frac{m^2}{4}\right)\parallel u \parallel^2_{H^1(\RR^4)}\leq
\parallel{\mathbf
  H}^{\frac{1}{2}}_0u \parallel^2_{L^2(\RR^4)}+m^2\parallel u \parallel^2_{L^2(\RR^4)}.
  \label{clo}
\end{equation}
  \label{SA}
\end{Lemma}

{\it Proof of Lemma \ref{SA}.}
We know that when
$u\in H^1\left(\RR^3_{\mathbf x}\times]0,\infty[_z\right)$ satisfies
 $\Delta_{{\mathbf x},z}u\in L^2\left(\RR^3_{\mathbf
      x}\times]0,\infty[_z\right)$, where $\Delta_{{\mathbf x},z}$ is
  the euclidean Laplace operator on $\RR^3_{\mathbf
      x}\times\RR_z$, then $\partial_zu\in
C^0\left([0,\infty[_z;H^{-\frac{1}{2}}\left(\RR^3_{\mathbf
      x}\right)\right)$ (see \cite{lions}, Theorem 7.3, p. 201).
We deduce that $u\in {\mathfrak D}({\mathbf H}_0)$ iff $u\in
    H^1\left(\RR^3_{{\mathbf x}}\times\RR_z\right)$ and
\begin{equation}
\Delta_{{\mathbf
        x},z}u_{| \RR^3_{\mathbf x}\times\RR^*_z}\in
    L^2\left(\RR^3_{\mathbf
        x}\times\RR_z^*\right),
  \label{DHcarl}
\end{equation}
\begin{equation}
\partial_zu({\mathbf
      x},0^+)-\partial_zu({\mathbf x},0^-)
+3u({\mathbf x},0)=0.
  \label{DHcar}
\end{equation}
Now for $u,v\in {\mathfrak D}({\mathbf H}_0)$, using the jumps formula and (\ref{DHcar}), we get by the Green formula :
\begin{equation*}
\left<{\mathbf H}_0u,v\right>_{L^2}=Q(u,v),
\end{equation*}
where we have introduced the quadratic form $Q$ on $H^1(\RR^4)$ given by 
\begin{equation}
Q(u,v):=\int_{\RR^4}\nabla_{\mathbf
  x}u\overline{\nabla_{\mathbf
    x}v}+\partial_zu\overline{\partial_zv}+\frac{15}{4}\left(\frac{1}{1+\mid z\mid}\right)^2u\overline{v}d{\mathbf x}dz-3\int_{\RR^3}u({\mathbf x},0)\overline{v({\mathbf x},0)}d{\mathbf x},
  \label{QQ}
\end{equation}
and we see that ${\mathbf H}_0$ is symmetric. To pursue the proof, it is
useful to present some results on the quadratic forms.

Given a Hilbert space $X$, we consider the quadratic form
\begin{equation}
f\in W^1_0(\RR;X),\;\;q(f,f):=\int_{-\infty}^{\infty}\mid
f'(z)\mid^2_X+\frac{15}{4}\left(\frac{1}{1+\mid z\mid}\right)^2\mid
f(z)\mid^2_Xdz-3\mid f(0)\mid^2_X.
  \label{qk}
\end{equation}
We immediately check with an integration by part that :
\begin{equation}
q(f,f)=\int_{-\infty}^{\infty}\left\vert
  f'(z)+\frac{3}{2}\frac{z}{\mid z\mid}\left(\frac{1}{1+\mid z
      \mid}\right)f(z)\right\vert_X^2dz\geq 0.
  \label{qpopo}
\end{equation}
We simply use the inequalities
$$
\mid f'(z)\mid_X\leq  \left\vert f'(z)+\frac{3}{2}\frac{z}{\mid z\mid}\left(\frac{1}{1+\mid z
      \mid}\right)f(z)\right\vert_X+\frac{3}{2}\mid f(z)\mid_X,$$ $$\mid f'(z)\mid_X\mid f(z)\mid_X\leq\frac{1}{2\alpha}\mid
f'(z)\mid_X^2+\frac{\alpha}{2}\mid f(z)\mid_X^2$$ for any $\alpha>0$,
to get an elliptic estimate : given $m>0$, we write 
$$
\int_{-\infty}^{\infty}\mid f'(z)\mid_X^2\left(1-\frac{3}{2\alpha}\right)
+\mid f(z)\mid_X^2\left(\frac{9}{4}-\frac{3\alpha}{2}+m^2\right)dz\leq q(f,f)+m^2\int_{-\infty}^{\infty}\mid f(z)\mid_X^2dz,
$$
and we choose $\alpha=\frac{3+m^2}{2}$ to obtain :
\begin{equation}
  \min\left(\frac{m^2}{3+m^2},\frac{m^2}{4}\right)\int_{-\infty}^{\infty}\mid f'(z)\mid^2_X+\mid
f(z)\mid_X^2dz\leq q(f,f)+m^2\int_{-\infty}^{\infty}\mid f(z)\mid_X^2dz.
  \label{coeq}
\end{equation}
(\ref{qpopo}) leads also to
the following crucial result. If we introduce
\begin{equation}
f_0(z):=(1+\mid
z\mid)^{-\frac{3}{2}},
  \label{fo}
\end{equation}
then we have for any $x\in X$, $f\in W^1_0(\RR;X)$ :
\begin{equation}
q(f_0x,f)=0,
  \label{fo}
\end{equation}
and thus in particular
\begin{equation}
q(f,f)=\int_{-\infty}^{\infty}\mid
f'(z)-f'_0(z)f(0)\mid^2_X+\frac{15}{4}\left(\frac{1}{1+\mid z\mid}\right)^2\mid
f(z)-f_0(z)f(0)\mid^2_Xdz.
  \label{qqkk}
\end{equation}
In the sequel we take $X=L^2(\RR^3_{\mathbf x})$. Then the equivalence
between (\ref{qnorm}), (\ref{qqqq}) and (\ref{qqg})
follows from (\ref{qpopo}) and (\ref{qqkk}), hence  $\mathbf H_0$ is positive
and $0$ is not an eigenvalue. Moreover (\ref{coeq}) gives
\begin{equation}
 \min\left(\frac{m^2}{3+m^2},\frac{m^2}{4}\right) \int_{\RR^4}\mid \nabla_{\mathbf{x},z} u({\mathbf x},z)\mid^2+\mid u({\mathbf x},z)\mid^2d{\mathbf x}dz\leq Q(u,u)+m^2\int_{\RR^4}\mid u({\mathbf x},z)\mid^2d{\mathbf x}dz.
  \label{coer}
\end{equation}

 Now to prove the
self-adjointness, we consider $g\in L^2(\RR^4)$, $\varepsilon=\pm 1$,
and we have to solve
\begin{equation}
u\in {\mathfrak D}(\mathbf{H}_0),\;\;P(\partial)u+\varepsilon iu=g.
  \label{pb}
\end{equation}
This problem is equivalent to find $u\in {\mathfrak D}(\mathbf{H}_0)$ such that
\begin{equation}
A(u,\varphi):=Q(u,\varphi)+\varepsilon
i\int_{\RR^4}u(\mathbf{x},z)\overline{\varphi(\mathbf{x},z)}d\mathbf{x}dz=
\int_{\RR^4}g(\mathbf{x},z)\overline{\varphi(\mathbf{x},z)}d\mathbf{x}dz,\;\;\forall
\varphi\in {\mathcal D}(\RR^4).
  \label{pol}
\end{equation}
Since the sesquilinear form $A$ is continuous on $H^1(\RR^4)$,
(\ref{pol}) is equivalent to 
\begin{equation}
A(u,v):=Q(u,v)+\varepsilon
i\int_{\RR^4}u(\mathbf{x},z)\overline{v(\mathbf{x},z)}d\mathbf{x}dz=
\int_{\RR^4}g(\mathbf{x},z)\overline{v(\mathbf{x},z)}d\mathbf{x}dz,\;\;\forall
v\in H^1(\RR^4).
  \label{pole}
\end{equation}
Since (\ref{coer})
assures that $A$ is coercive on $H^1(\RR^4)$, the Lax-Milgram theorem
implies that there exists $u\in H^1(\RR^4)$ solution of
(\ref{pole}). Taking $v$ an arbitrar test function in
$\mathcal{D}(\RR^3_{\mathbf x}\times\RR^*_z)$ we deduce that
$$
-\Delta_{\mathbf
    x}u-\partial^2_zu+\frac{15}{4}\left(\frac{1}{1+\mid
      z\mid}\right)^2u+\varepsilon iu=g\;\;in\;\;{\mathcal D}'(\RR^3_{\mathbf
    x}\times\RR^*_z).
$$
Thus (\ref{DHcarl}) is satisfied and an integration by parts in
(\ref{pole}) gives
$$
\left<\partial_zu({\mathbf
      x},0^+)-\partial_zu({\mathbf x},0^-)
+3u({\mathbf x},0),\overline{v(\mathbf{x},0)}\right>_{H^{-\frac{1}{2}}(\RR^3),H^{\frac{1}{2}}(\RR^3)}=0,\;\;\forall
v\in H^1(\RR^4).
$$
We conclude that $u$ belongs to the domain of $\mathbf{H}_0$, which is self-adjoint.

Finally (\ref{coer})
assures that the quadratic form $Q$ on $L^2(\RR^4)$ with its natural form domain
${\mathfrak D}(Q)=H^1(\RR^4)$ is closed. The standard results of the spectral theory
(see e.g. \cite{pear}, section 2.7) states that the domain of
$\mathbf{H}_0^\frac{1}{2}$ is ${\mathfrak D}(Q)$ and (\ref{clo}) follows from (\ref{coer}).

\fin


This result allows to easily solve the Cauchy problem (\ref{EQB}), (\ref{ci}),
for $\Phi_0\in H^1(\RR^4)$,  $\Phi_1\in L^2(\RR^4)$ by putting
\begin{equation*}
\Phi(t)=\cos\left(t\mathbf{H}_0^\frac{1}{2}\right)\Phi_0+\mathbf{H}_0^{-\frac{1}{2}}\sin\left(t\mathbf{H}_0^{\frac{1}{2}}\right)\Phi_1.
\end{equation*}
It is obvious that $\Phi\in C^0(\RR_t;H^1(\RR^4))\cap
C^1(\RR_t;L^2(\RR^4))$ and satisfies the conservation of the energy :
\begin{equation*}
Q(\Phi(t),\Phi(t))+\parallel
\partial_t\Phi(t)\parallel_{L^2(\RR^4)}^2=
Q(\Phi_0,\Phi_0)+\parallel
\Phi_1\parallel_{L^2(\RR^4)}^2.
\end{equation*}

Since this conserved quantity is positive, it is natural to consider the Cauchy
problem in the larger functional space, associated with this energy. (\ref{qnorm}) shows that $\sqrt{Q(u,u)}$
is a norm on $H^1(\RR^4)$, therefore we introduce the Hilbert space ${\mathfrak W}^1(\RR^4)$ that is the closure of
$H^1(\RR^4)$ for the norm
\begin{equation}
\parallel u\parallel^2_{{\mathfrak W}^1}:=\parallel \nabla_{\mathbf x}u\parallel^2_{
  L^2\left(\RR^3_{\mathbf{x}}\times\RR_z\right)}+\parallel
\partial_zu+\frac{3}{2}\frac{z}{\mid z\mid}\left(\frac{1}{1+\mid z \mid}\right)u\parallel^2_{
  L^2\left(\RR^3_{\mathbf{x}}\times\RR_z\right)}.
  \label{wun}
\end{equation}
${\mathfrak W}^1(\RR^4)$ is a space of
distributions on $\RR^4$, and 
\begin{equation*}
{\mathfrak W}^1\left(\RR^4\right)=\left\{ u\in
  L^2\left(\RR_z;BL^1\left(\RR^3_{\mathbf{x}}\right)\right);\;\;\partial_zu+\frac{3}{2}\frac{z}{\mid z\mid}\left(\frac{1}{1+\mid z \mid}\right)u\in L^2(\RR^4)\right\}.
  \end{equation*}
We make some warnings on this space :
By the classical embedding of the Beppo
Levi space :
\begin{equation*}
BL^1\left(\RR^3_{\mathbf{x}}\right)\subset L^2\left(\RR^3_{\mathbf x},\frac{1}{1+\mid\mathbf{x}\mid^2}d\mathbf{x}\right),
\end{equation*}
we can see that the functions  $u\in {\mathfrak W}^1(\RR^4)$ satisfy
\begin{equation*}
u,\;\partial_zu\in  L^2\left(\RR_z;L^2\left(\RR^3_{\mathbf x},\frac{1}{1+\mid\mathbf{x}\mid^2}d\mathbf{x}\right)\right),
\end{equation*}
hence
\begin{equation*}
u\in C^0\left(\RR_z; L^2\left(\RR^3_{\mathbf x},\frac{1}{1+\mid\mathbf{x}\mid^2}d\mathbf{x}\right)\right),
\end{equation*}
and thus $u(\mathbf{x},0)\delta_0(z)$ is well defined in $H^{-1}_{loc}(\RR^4)$.
But we have to be carefull by using expressions (\ref{qqqq}) or (\ref{qqg}) to evaluate
$Q(u,u)$ when $u\in {\mathfrak W}^1(\RR^4)\setminus H^1(\RR^4)$, because there
exists $u\in {\mathfrak W}^1(\RR^4)$ such that
$$
\partial_zu\notin L^2(\RR^4),\;\;u\notin
L^2\left(\RR^3_{\mathbf{x}}\times\RR_z,\frac{1}{1+z^2}d{\mathbf x}dz\right),\;\;u(\mathbf{x},0)\notin
L^2(\RR^3_{\mathbf{x}}).
$$
To see that, we take $\chi\in{\mathcal D}(\RR^3)$ such that $\parallel
\chi\parallel_{L^2}=\parallel\nabla \chi\parallel_{L^2}=1$. For
any $n\in\NN$, we put
$u_n(\mathbf{x},z)=n^{-\frac{1}{2}}\chi(\frac{\mathbf x}{n})(1+\mid
z\mid)^{-\frac{3}{2}}$. We compute $Q(u_n,u_n)=1$,
$\parallel\partial_zu_n\parallel^2_{L^2(\RR^4)}=\frac{9}{8}n^2$, 
$\parallel\frac{1}{1+\mid
  z\mid}u_n\parallel^2_{L^2(\RR^4)}=\frac{1}{2}n^2$,
$\parallel u_n(.,0)\parallel^2_{L^2(\RR^3_{\mathbf x})}=n^2$.\\


We return to the Cauchy problem (\ref{EQB}), (\ref{ci}) for
$\Phi_0\in {\mathfrak W}^1(\RR^4)$, $\Phi_1\in L^2(\RR^4)$. On
${\mathfrak W}^1(\RR^4)\times L^2(\RR^4)$ we introduce the
operator
\begin{equation*}
\mathbf{A}_0:=\frac{1}{i}
\left(
\begin{array}{cc}
0&1\\
-P(\partial)&0
\end{array}
\right),\;\;
{\mathfrak D}(\mathbf{A}_0)={\mathfrak D}(\mathbf{H}_0)\times H^1(\RR^4).
\end{equation*}
From the properties of $\mathbf{H}_0$ we deduce that $\mathbf{A}_0$ is
a densely defined, symmetric operator on
${\mathfrak W}^1(\RR^4)\times L^2(\RR^4)$ and $H^1(\RR^4)\times
L^2(\RR^4)\subset {\rm Ran}(\mathbf{A}_0\pm i)$ where ${\rm Ran}(f)$
denotes the range of any map $f$. Therefore $\mathbf{A}_0$ is
essentially self-adjoint and its self-adjoint closure $\mathbf{A}$
generates a unitary group $e^{it\mathbf{A}}$. Then
\begin{equation*}
\left(
\begin{array}{c}
\Phi(t)\\
\partial_t\Phi(t)
\end{array}
\right)
=
e^{it\mathbf{A}}
\left(
\begin{array}{c}
\Phi_0\\
\Phi_1
\end{array}
\right)
\end{equation*}
is a solution of the Cauchy problem that satisfies
\begin{equation}
\Phi\in \Xi:=\left\{\Phi\in C^0\left(\RR_t;{\mathfrak W}^1(\RR^4)\right),\;\;
\partial_t\Phi\in C^0\left(\RR_t;L^2(\RR^4)\right)\right\},
  \label{regf}
\end{equation}
and the conservation of the energy :
\begin{equation}
\forall t\in\RR,\;\;\;\parallel \Phi(t)\parallel_{{\mathfrak W}^1(\RR^4)}^2+\parallel
\partial_t\Phi(t)\parallel_{L^2(\RR^4)}^2=\parallel \Phi_0\parallel_{{\mathfrak W}^1(\RR^4)}^2+\parallel
\Phi_1\parallel_{L^2(\RR^4)}^2.
\label{ener}
\end{equation}
We prove the uniqueness by the usual way. If $\Phi$ is a solution of (\ref{EQB}), (\ref{ci}),
(\ref{regf}), we take a test function $\theta\in C^{\infty}_0(\RR)$,
such that
$\int\theta(t)dt=1$ and for all integer $n>0$, we put
$\Phi_n(t)=\int\Phi(s)\theta(n(t-s))ds$. Since $\Phi_n\in
C^{\infty}\left(\RR_t;{\mathfrak W}^1(\RR^4)\right)$,  $\partial_t\Phi_n\in
C^{\infty}\left(\RR_t;L^2(\RR^4)\right)$ we can multiply equation (\ref{EQB})
by $\partial_t\Phi_n$ and an integration in $(\mathbf{x},z)$ leads to
the conservation 
$$
\forall t\in\RR,\;\;\;\parallel \Phi_n(t)\parallel_{{\mathfrak W}^1(\RR^4)}^2+\parallel
\partial_t\Phi_n(t)\parallel_{L^2(\RR^4)}^2=\parallel \Phi_n(0)\parallel_{{\mathfrak W}^1(\RR^4)}^2+\parallel
\partial_t\Phi_n(0)\parallel_{L^2(\RR^4)}^2.
$$
Since $\Phi_n$ and $\partial_t\Phi_n$ tend to $\Phi$ and
$\partial_t\Phi$ as $n\rightarrow\infty$, respectively in
$C^{0}\left(\RR_t;{\mathfrak W}^1(\RR^4)\right)$
and $C^{0}\left(\RR_t;L^2(\RR^4)\right)$, we conclude that
(\ref{ener}) holds and $\Phi=0$ when $\Phi_0=\Phi_1=0$.\\

 We summarize our results :

\begin{Theorem}
Given $\Phi_0\in {\mathfrak W}^1(\RR^4)$, $\Phi_1\in L^2(\RR^4)$, there exists a
unique solution $\Phi$ of the Cauchy problem (\ref{EQB}), (\ref{ci}), (\ref{regf}).
Moreover this solution satisfies (\ref{ener}).
  \label{THEOCAU}
\end{Theorem}

The solutions given by this theorem are called {\it finite energy
  solutions.}

\begin{Remark}
We could interpret this Cauchy problem for a wave equation with the
singular potential $\frac{15}{4}\left(\frac{1}{1+\mid
      z\mid}\right)^2-3\delta_0(z)$, as two mixed problems with smooth
coefficients on the half-space, and boundary condition on the brane $z=0$. If we introduce
\begin{equation}
2\Phi_{\pm}(t,{\mathbf x},z):=\Phi(t,{\mathbf x},z)\pm \Phi(t,{\mathbf x},-z),
  \label{papa}
\end{equation}
$\Phi$ is solution of (\ref{EQB}) iff $\Phi_{\pm}$ are solutions of
the boundary problems in $\RR_t\times\RR^3_{\mathbf x}\times]0,\infty[_z$ :
\begin{equation}
\square_{t, {\mathbf x}, z}\Phi_{\pm}+\frac{15}{4}\left(\frac{1}{1+
      z}\right)^2\Phi_{\pm}=0,\;\;(t,{\mathbf x},z)\in\RR\times\RR^3\times]0,\infty[,
  \label{upm}
\end{equation}
\begin{equation}
\partial_z\Phi_+(t,{\mathbf x},0)+\frac{3}{2}\Phi_+(t,{\mathbf x},0)=0,
  \label{cup}
\end{equation}
\begin{equation}
\Phi_-(t,{\mathbf x},0)=0.
  \label{cum}
\end{equation}
The homogeneous Dirichlet problem for $\Phi_-$ is trivial, but the
Robin problem (with the ``bad'' sign) for $\Phi_+$, that is the part of the wave that is
physically pertinent, needs a carefull analysis, similar to the
previous one, of the quadratic form :

\begin{equation*}
\int_{\RR^3_{\mathbf x}}\int_0^{\infty}\lvert\nabla_{\mathbf
  x}\Phi_+(\mathbf{x},z)\rvert^2+
\lvert\partial_z\Phi_+(\mathbf{x},z)\rvert^2
+\frac{15}{4}\left(\frac{1}{1+\mid
    z\mid}\right)^2\lvert\Phi_+(\mathbf{x},z)\rvert^2dzd{\mathbf x}-\frac{3}{2}\int_{\RR^3}\lvert\Phi_+(\mathbf{x},0)\rvert^2d{\mathbf x}.
\end{equation*}
This approach can be useful for the numerical
experiments (e.g. \cite{sea}). 
  \label{rem}
\end{Remark}

We now present  some classes of solutions. First of all we note that the equation
is invariant with respect to the transform $z\rightarrow -z$,
therefore if $\Phi$ is solution, then $\Phi_+$ and $\Phi_-$ defined by
(\ref{papa}) are also solutions, respectively called {\it z-even wave} and
{\it z-odd wave}. In particular we have
$\Phi_-(t,\mathbf{x},0)=0$, hence $\Phi_-$ is a wave, equal to zero on the
brane, that is solution of
\begin{equation}
\square_{t, {\mathbf x}, z}\Phi_-+\frac{15}{4}\left(\frac{1}{1+\mid
      z\mid}\right)^2\Phi_-=0,\;\;(t,\mathbf{x},z)\in \RR\times\RR^3\times\RR.
  \label{eqmoin}
\end{equation}
This equation is a smooth perturbation of the D'Alembertian with a
non-decreasing potential with cartesian anisotropy (see \cite{richard}). Thanks to (\ref{qqqq}) and the Hardy
inequality (\ref{HAR}), the ${\mathfrak W}^1(\RR^4)$-norm is equivalent to the
$BL^1(\RR^4)$-norm on the subspace of the $z$-odd functions, and
Theorem \ref{THEOCAU} assures that the Cauchy problem for
(\ref{eqmoin}) is well posed in $BL^1(\RR^4)\times L^2(\RR^4)$. We shall see that its solutions are
asymptotically free.

Much more interesting is the even part $\Phi_+$
of $\Phi$.
We can very simply perform a large family of waves that are confined
in the vicinity of  the brane. For any $\varphi_0\in BL^1(\RR^3_{\mathbf x})$, 
$\varphi_1\in L^2(\RR^3_{\mathbf x})$, we put
\begin{equation*}
\Phi(t,\mathbf{x},z)=\phi_0(t,\mathbf{x})f_0(z),
\end{equation*}
with $f_0(z):=(1+\mid
z\mid)^{-\frac{3}{2}}$, and $\phi_0\in C^0(\RR_t; BL^1(\RR^3_{\mathbf
  x}))$,  $\partial_t\phi_0\in C^0(\RR_t; L^2(\RR^3_{\mathbf
  x}))$ is solution of the wave equation on the Minkowski brane :
\begin{equation*}
\partial^2_t\phi_0-\Delta_{\mathbf{x}}\phi_0=0,\;\;\phi_0(0,\mathbf{x})=\varphi_0(\mathbf{x}),\;\;\partial_t\phi_0(0,\mathbf{x})=\varphi_1(\mathbf{x}).
\end{equation*}
We call {\it massless graviton wave} any solution of (\ref{EQB}),
(\ref{ci}) with initial data
\begin{equation*}
\Phi_0\in BL^1(\RR^3_{\mathbf x})\otimes \CC f_0(z),\;\;\Phi_1\in
L^2(\RR^3_{\mathbf x})\otimes \CC f_0(z).
\end{equation*}
Its energy is
obviously localized near the brane on which it is propagating : for
all $a<0$, $b>0$, there exists $C(a,b)>0$ such that
\begin{equation*}
\forall t\in\RR,\;\;
\parallel \nabla_{\mathbf x}\Phi(t)\parallel_{L^2(\RR^3_{\mathbf x}\times[a,b]_z)}^2+\parallel
\partial_t\Phi(t)\parallel_{L^2(\RR^3_{\mathbf x}\times[a,b]_z)}^2=C(a,b)\left(\parallel \phi_0\parallel_{BL^1(\RR^3)}^2+\parallel
\phi_1\parallel_{L^2(\RR^3)}^2\right).
\end{equation*}
In opposite, a {\it Kaluza-Klein wave} is a solution for which there
exists $a<0$, $b>0$ such that
\begin{equation}
\lim_{\mid t\mid \rightarrow\infty}
\parallel \nabla_{\mathbf x}\Phi(t)\parallel_{L^2(\RR^3_{\mathbf x}\times[a,b]_z)}+\parallel
\partial_t\Phi(t)\parallel_{L^2(\RR^3_{\mathbf x}\times[a,b]_z)}=0.
  \label{kk}
\end{equation}
We shall see that in this case, (\ref{kk}) is true for any $a,b$ and the Kaluza-Klein waves are asymptotic to a
free wave of the Minkowski space time $\RR^{1+4}$. The main
results of the next part state that any finite energy solution, is the sum
of  a massless graviton and a Kaluza-Klein wave. Since the unitary
group $e^{it\mathbf{A}}$ leaves invariant $\left(BL^1\otimes\CC
  f_0\right)\times \left(L^2\otimes\CC
  f_0\right)$, we have to study its action on the orthogonal of this
subspace. We introduce :
\begin{equation}
\mathfrak{K}^1(\RR^4):=\left(BL^1\left(\RR^3_{\mathbf x}\right)\otimes
\CC f_0(z)\right)^{\perp_{{\mathfrak W}^1}},\;\;\mathfrak{K}^0(\RR^4):=\left(L^2\left(\RR^3_{\mathbf x}\right)\otimes
\CC f_0(z)\right)^{\perp_{L^2}}.
  \label{kaka}
\end{equation}
 $f\in {\mathfrak W}^1(\RR^4)$ belongs to
$\mathfrak{K}^1(\RR^4)$ iff
\begin{equation*}
\forall \varphi\in C^{\infty}_0(\RR^3_{\mathbf{x}}),\;\;<\varphi(\mathbf{x})f_0(z),f>_{{\mathfrak W}^1}=0.
\end{equation*}
Since we have
$$
<\varphi(\mathbf{x})f_0(z),f>_{{\mathfrak W}^1}=-<\Delta_{\mathbf x}\int_{-\infty}^{\infty}f(.,z)f_0(z)dz,\varphi>_{\mathcal{D}'(\RR^3_{\mathbf{x}}),\mathcal{D}(\RR^3_{\mathbf{x}})},
$$
and $\int_{-\infty}^{\infty}f(.,z)f_0(z)dz$ is an integral absolutely
converging in $BL^1(\RR^3_{\mathbf x})$, we conclude that
\begin{equation}
\forall f\in {\mathfrak W}^1(\RR^4),\;\;f\in\mathfrak{K}^1(\RR^4)\Leftrightarrow
\int_{-\infty}^{\infty}f(.,z)f_0(z)dz=0.
  \label{kun}
\end{equation}
It is obvious that the $z$-odd functions belongs to
$\mathfrak{K}^{0}$ but we are mainly interested by the $z$-even
functions that are associated with the Kaluza-Klein waves.

Finally we can solve the inhomogeneous Cauchy problem describing the
propagation of the gravitational perturbations due to a source
$S(t,\mathbf{x})$,
localized on the brane.

\begin{Theorem}
Given $\Phi_0\in {\mathfrak W}^1(\RR^4)$, $\Phi_1\in L^2(\RR^4)$,
$S\in C^0\left(\RR_t;H^1(\RR^3_{\mathbf x})\right)\cap C^1\left(\RR_t;L^2(\RR^3_{\mathbf x})\right)$ such that
  $(\partial^2_t-\Delta_{\mathbf x})S \in
  C^0\left(\RR_t;L^2(\RR^3_{\mathbf x})\right)$ there exists a
unique solution $\Phi\in \Xi$ of the
Cauchy problem
\begin{equation}
\left(\square_{t, {\mathbf x}, z}+\frac{15}{4}\left(\frac{1}{1+\mid
      z\mid}\right)^2-3\delta_0(z)\right)\Phi=S(t,\mathbf{x})\otimes \delta_0(z),\;\;(t,{\mathbf x},z)\in\RR\times\RR^3\times\RR,
  \label{EQBS}
\end{equation}
\begin{equation}
\Phi(0,\mathbf{x},z)=\Phi_0(\mathbf{x},z),\;\;\partial_t\Phi(0,\mathbf{x},z)=\Phi_1(\mathbf{x},z),\;\;({\mathbf x},z)\in\RR^3\times\RR.
  \label{cis}
\end{equation}

  \label{ssf}
\end{Theorem}

{\it Proof of Theorem \ref{ssf}.} We take $\theta\in C^2_0(\RR_z)$ such that
$\theta(0)=-\frac{1}{3}$. Then $\Phi$ is solution iff
$\Psi(t,\mathbf{x},z):=\Phi(t,\mathbf{x},z)-S(t,\mathbf{x})\otimes\theta(z)$
is solution of
$$
\left(\square_{t, {\mathbf x}, z}+\frac{15}{4}\left(\frac{1}{1+\mid
      z\mid}\right)^2-3\delta_0(z)\right)\Psi=-(\partial^2_t-\Delta_{\mathbf x})S\otimes\theta-S\otimes\left[-\theta''+\frac{15}{4}\left(\frac{1}{1+\mid
      z\mid}\right)^2\theta\right]:=F,
$$
$$
\Psi(0,\mathbf{x},z)=\Phi_0(\mathbf{x},z)-S(0,\mathbf{x})\theta(z):=\Psi_0(\mathbf{x},z),\;\;
\partial_t\Psi(0,\mathbf{x},z)=\Phi_1(\mathbf{x},z)-\partial_tS(0,\mathbf{x})\theta(z):=\Psi_1(\mathbf{x},z).
$$
Since $\Psi_0\in{\mathfrak W}^1(\RR^4)$, $ \Psi_1\in L^2(\RR^4)$ and
$F\in C^0\left(\RR_t; L^2(\RR^4)\right)$, we get the solution
of this  Cauchy problem easily  with the Duhamel formula.

\fin



\section{Spectral expansion.}

In this part we prove that any finite energy solution is the sum of a
massless graviton and a Kaluza-Klein tower. The tool is the spectral
decomposition of $\mathbf{H}_0$. We also compute the kernel of the
truncated (in brane energy) resolvent.

\begin{Theorem} There exists $f^{\pm}_m(z)\in
  C^0\left([0,\infty[_m\times\RR_z\right)$ with $f^{\pm}_0(z)=0$,  such that for any
  $\Phi_0\in{\mathfrak W}^1(\RR^4)$, $\Phi_1\in L^2(\RR^4)$, there
  exists $\phi_0\in C^0\left(\RR_t;BL^1\left(\RR^3_{\mathbf x}\right)\right)$
  with $\partial_t\phi_0\in
  C^0\left(\RR_t;L^2\left(\RR^3_{\mathbf x}\right)\right)$, and for almost all $m>0$,
  $\phi_m\in C^0\left(\RR_t;H^1\left(\RR^3_{\mathbf x}\right)\right)\cap
  C^1\left(\RR_t;L^2\left(\RR^3_{\mathbf x}\right)\right)$ solutions of
\begin{equation*}
\partial^2_t\phi_0-\Delta_{\mathbf x}\phi_0=0,\;\;\partial^2_t\phi_{m}^{\pm}-\Delta_{\mathbf x}\phi_{m}^{\pm}+m^2\phi_m^{\pm}=0,
\end{equation*}
such that 
\begin{equation}
\int_0^{\infty}\parallel\nabla_{t,\mathbf{x}}\phi_m^{\pm}(t)\parallel^2_{L^2(\RR^3)}
+m^2\parallel\phi_m^{\pm}(t)\parallel^2_{L^2(\RR^3)}dm<\infty,
  \label{Ldem}
\end{equation}
and the solution $\Phi\in\Xi$ of the Cauchy problem (\ref{EQB}), (\ref{ci}), can be written as
\begin{equation}
\Phi(t,{\mathbf x},z)=\phi_0(t,{\mathbf
  x})f_0(z)+\sum_{\pm}\lim_{M\rightarrow\infty}\int_0^{M}\phi_m^{\pm}(t,{\mathbf x})f_m^{\pm}(z)dm,
    \label{decompo}
  \end{equation}
where the limit holds in $\Xi$. Moreover we have
\begin{equation}
\parallel \Phi_0\parallel_{\mathfrak{W}^1(\RR^4)}^2+
\parallel \Phi_1\parallel_{L^2(\RR^4)}^2=
\parallel\nabla_{t,\mathbf{x}}\phi_0(t)\parallel^2_{L^2(\RR^3)}+2\sum_{\pm}
\int_0^{\infty}\parallel\nabla_{t,\mathbf{x}}\phi_m^{\pm}(t)\parallel^2_{L^2(\RR^3)}
+m^2\parallel\phi_m^{\pm}(t)\parallel^2_{L^2(\RR^3)}dm.
\label{deconer}
\end{equation}

\label{theosp} 
\end{Theorem}

$\phi_0(t,{\mathbf
  x})f_0(z)$ and $\left(\phi_m^{\pm}(t,{\mathbf x})f_m^{\pm}(z)\right)_{0<m}$ are
solutions of equation (\ref{EQB}). The first one is  the massless
graviton, its energy is finite,  and the second one is called the
tower of massive Kaluza-Klein modes. Since $f_m^{\pm}\notin
L^2(\RR_z)$, the energy of these modes is infinite.\\

{\it Proof of Theorem \ref{theosp}.}
First we develop the spectral analysis of the
one-dimensional operator
\begin{equation}
{\mathbf h}:=-\frac{d^2}{dz^2}+\frac{15}{4}\left(\frac{1}{1+\mid
      z\mid}\right)^2-3\delta_0(z),\;\;{\mathfrak D}({\mathbf h}):=\left\{u\in
  H^1(\RR);\;{\mathbf h}u\in L^2(\RR)\right\},
  \label{h!}
\end{equation}
in particular, we determine its pure point spectrum
$\sigma_{pp}({\mathbf h})$ and its absolutely continuous spectrum
$\sigma_{ac}({\mathbf h})$.


\begin{Lemma}
$\left({\mathbf h},{\mathfrak D}({\mathbf h})\right)$ is a self-adjoint positive
operator on $L^2(\RR)$, and we have :
\begin{equation*}
{\mathfrak D}\left({\mathbf h}^{\frac{1}{2}}\right)=H^1(\RR),
\end{equation*}
\begin{equation}
\sigma_{pp}({\mathbf h})=\{0\},\;\;Ker({\mathbf h})=\CC f_0,
  \label{spp}
\end{equation}
\begin{equation}
\sigma_{ac}({\mathbf h})=[0,\infty[.
  \label{spp}
\end{equation}
  \label{lemah}
\end{Lemma}

{\it Proof of Lemma \ref{lemah}.} We use the quadratic form
(\ref{qk}) with $X=\CC$. For $f,g\in {\mathfrak D}(\mathbf{h})$, an
integration by part and (\ref{qpopo}) give
\begin{equation*}
<\mathbf{h}f,g>_{L^2}=q(f,g),\;\;<\mathbf{h}f,f>_{L^2}\geq 0,
\end{equation*}
thus $\mathbf h$ is symmetric and positive. Now given $g\in L^2(\RR)$,
$f_{\pm}\in {\mathfrak D}(\mathbf{h})$ is solution of
$\mathbf{h}f_{\pm}\pm if_{\pm}=g$ iff
\begin{equation*}
\forall v\in H^1(\RR),\;\;a_{\pm}(f_{\pm},v):=q(f_{\pm},v)\pm i<f_{\pm},v>_{L^2}=<g,v>_{L^2}.
\end{equation*}
(\ref{coeq}) implies that $a_{\pm}$ is continuous and coercive on
$H^1(\RR)$, hence this problem
has a unique solution and $\mathbf{h}$ is self-adjoint. Moreover, since
$q$ is a closed form, and its form domain is $H^1(\RR)$, a classical
result assures that 
${\mathfrak D}\left({\mathbf h}^{\frac{1}{2}}\right)=H^1(\RR)$.
To investigate the spectrum, we solve for $m\geq 0$ the equation
\begin{equation*}
\mathbf{h}u_m=m^2u_m,\;\;u_m\in{\mathcal D}'(\RR).
\end{equation*}
When $m=0$, an explicit calculation with the jump formula gives
\begin{equation*}
u_0(z)=\lambda(1+\mid z\mid)^{-\frac{3}{2}}+\mu \frac{z}{\mid z\mid}\left[(1+\mid
  z\mid)^{\frac{5}{2}}
-(1+\mid z\mid)^{-\frac{3}{2}}\right],\;\;\lambda,\mu\in\CC.
\end{equation*}
For
$0<m$, we use the Bessel equation satisfied by the Hankel functions
$H^{(1)}_2$, $H^{(2)}_2$ :
\begin{equation*}
x^2u''+xu'+(x^2-4)u=0,\;\;x\in\RR,
\end{equation*}
to check that
\begin{equation*}
\pm z>0\Rightarrow
u_m(z)=\lambda_{\pm}^{(1)}\sqrt{1+\mid z\mid}H^{(1)}_2\left(m(1+\mid
  z\mid)\right)+
\lambda_{\pm}^{(2)}\sqrt{1+\mid z\mid}H^{(2)}_2\left(m(1+\mid
  z\mid)\right),\;\;\lambda_{\pm}^{(j)}\in\CC.
\end{equation*}
Since we have the asymptotics (\cite{watson}, 7.2)
\begin{equation}
H^{(1)}_2(x)\sim -\sqrt{\frac{2}{\pi x}}e^{i(x-\frac{\pi}{4})},\;\;\;
H^{(2)}_2(x)\sim -\sqrt{\frac{2}{\pi x}}e^{-i(x-\frac{\pi}{4})},\;\;\;x\rightarrow\infty,
  \label{asym}
\end{equation}
we conclude that $0$ is the only eigenvalue and $u_0\in
Ker(\mathbf{h})$ iff $\mu=0$.\\

 To investigate the continuous part of the spectrum of $\mathbf h$, it is convenient to split
 the functions into even and odd parts like in Remark \ref{rem} by putting
 \begin{equation*}
\sqrt{2}u_{\pm}(z)=u(z)\pm u(-z),\;\;\;\mathbf{P}u:=(u_+,u_-)
\end{equation*}
and the map $\mathbf{P}:u\mapsto\mathbf{P}u$ is an
isometry from $L^2(\RR)$ onto $L^2(\RR^+)\times
L^2(\RR^+)$.
Since $u\in {\mathfrak D}(\mathbf{h})$ iff $u\in H^1(\RR)$, $u''_{\vert \RR^*}\in
L^2(\RR^*)$, $u'(0^+)-u'(0^-)+3u(0)=0$, we introduce the self-adjoint
operators $\left(\mathbf{h}_{\pm}, {\mathfrak D}(\mathbf{h}_{\pm})\right)$, on $L^2(\RR^+)$ defined by :
\begin{equation*}
{\mathbf h_{\pm}}:=-\frac{d^2}{dz^2}+\frac{15}{4}\left(\frac{1}{1+
      z}\right)^2,
\end{equation*}
\begin{equation*}
{\mathfrak D}({\mathbf h_+}):=\left\{u_+\in
  H^2(\RR^+);\;u'_+(0)+\frac{3}{2}u_+(0)=0\right\},\;\;{\mathfrak D}({\mathbf h_-}):=\left\{u_-\in
  H^2(\RR^+);\;u_-(0)=0\right\}.
\end{equation*}
We have
\begin{equation}
\mathbf{Ph}u=\left(\mathbf{h_+}u_+,\mathbf{h}_-u_-\right),
  \label{interp}
\end{equation}
therefore  the operator $\left(\mathbf{h},
{\mathfrak D}(\mathbf{h})\right)$ is unitarily equivalent with the selfadjoint operator
$\left(\mathbf{h_+},\mathbf{h_-}\right)$ on $L^2(\RR^+)\times
L^2(\RR^+)$. As previous, we can easily show that
\begin{equation*}
\sigma_{pp}(\mathbf{h_+})=\{0\},\;\;\sigma_{pp}(\mathbf{h_-})=\emptyset,
\end{equation*}
\begin{equation*}
{\mathfrak D}\left(({\mathbf
    h_+})^{\frac{1}{2}}\right)=H^1(\RR^+),\;\;{\mathfrak D}\left(({\mathbf h_-})^{\frac{1}{2}}\right)=H^1_0(\RR^+).
\end{equation*}
Since $\sigma_{ac}(\mathbf{h})=\sigma_{ac}(\mathbf{h}_+)\cap
\sigma_{ac}(\mathbf{h}_-)$, the proof of (\ref{spp}) is reduced to
investigating the spectrum of $\mathbf{h}_{\pm}$. To establish that
$\sigma_{ac}(\mathbf{h}_{\pm})=[0,\infty[$, we use a nice result on
the Schr\"odinger operator (Theorem 7.3 of \cite{pear}) : it sufficient to prove that
for any $m>0$ and all  solutions $u_1,\,u_2\neq 0$ of 
\begin{equation}
-u''+\frac{15}{4}\left(\frac{1}{1+
      z}\right)^2u=m^2u,\;\;z>0,
  \label{sch}
\end{equation}
we have 
\begin{equation*}
0<\liminf_{b\rightarrow\infty}\frac{\int_0^b\mid u_1(z)\mid^2dz}{\int_0^b\mid u_2(z)\mid^2dz}.
\end{equation*}
Since the solutions $u\neq 0$ of (\ref{sch}) have the form
\begin{equation}
u(z)=\lambda^{(1)}(u)\sqrt{1+z}H^{(1)}_2\left(m(1+z)\right)+
\lambda^{(2)}(u)\sqrt{1+z}H^{(2)}_2\left(m(1+
  z)\right),\;\;\lambda^{(j)}\in\CC,
  \label{solh}
\end{equation}
with $\mid\lambda^{(1)}\mid+\mid\lambda^{(2)}\mid\neq 0$,
we deduce from the asymptotics (\ref{asym}) that
\begin{equation*}
\int_0^b\mid u(z)\mid^2dz\sim \frac{2k}{\pi m}\left(\mid\lambda^{(1)}(u)\mid^2+\mid\lambda^{(2)}(u)\mid^2\right)b,\;\;b\rightarrow\infty,
\end{equation*}
therefore
\begin{equation*}
\liminf_{b\rightarrow\infty}\frac{\int_0^b\mid
  u_1(z)\mid^2dz}{\int_0^b\mid u_2(z)\mid^2dz}=
\frac{\mid\lambda^{(1)}(u_1)\mid^2+\mid\lambda^{(2)}(u_1)\mid^2}{\mid\lambda^{(1)}(u_2)\mid^2+\mid\lambda^{(2)}(u_2)\mid^2}>0.
\end{equation*}
\fin

We now construct the explicit spectral representations of $\mathbf
h_{\pm}$. We denote $L^2_{ac}$ the absolutely continuous sub-space of
$\mathbf h_+$,
\begin{equation*}
L^2_{ac}:=\left\{f\in L^2(\RR^+_z),\;\;\int_0^{\infty}f(z)f_0(z)dz=0\right\}.
\end{equation*}
We introduce the functions $u_{\pm}(z,m)$ defined for $0<m$, $0<z$ by :

\begin{equation}
u_+(z,m):=\frac{1}{2}\sqrt{m(1+z)}\frac{
H_1^{(2)}\left(m\right)
H_2^{(1)}\left(m(1+z)\right)
-H_1^{(1)}\left(m\right)H_2^{(2)}\left(m(1+z)\right)}{H_1^{(1)}\left(m\right)}e^{i\theta_1(m)},
 \label{uplus}
\end{equation}

\begin{equation}
u_-(z,m):=\frac{1}{2}\sqrt{m(1+z)}\frac{
H_2^{(2)}\left(m\right)
H_2^{(1)}\left(m(1+z)\right)
-H_2^{(1)}\left(m\right)H_2^{(2)}\left(m(1+z)\right)}{H_2^{(1)}\left(m\right)}e^{i\theta_2(m)},
 \label{umoins}
\end{equation}
\begin{equation}
\theta_j(m):=\arg\left( H_j^{(1)}\left(m\right)\right)-\frac{\pi}{2}.
  \label{tetaj}
\end{equation}

It will be very important to know the asymtotic behaviours of these functions.

\begin{Lemma}
$u_+(z,m)-\sqrt{m}(1+z)^{-\frac{3}{2}}$ and $u_{-}$ belong
to $C^2\left([0,\infty[_z\times[0,\infty[_m\right)$ and satisfy the
asymptotic properties :
\begin{equation}
\forall R\geq
0,\;\;\exists C_R>0,\;\;\forall m\in]0,1],\;\;\sup_{0\leq z\leq
  R}\left\vert u_+(z,m)+\sqrt{m}(1+z)^{-\frac{3}{2}}\right\vert\leq C_Rm^{\frac{5}{2}},
  \label{mup}
\end{equation}

\begin{equation}
\forall R\geq
0,\;\;\exists C_R>0,\;\;\forall m\in]0,1],\;\;\sup_{0\leq z\leq
  R}\left\vert u_-(z,m)-\frac{1}{8}m^{\frac{5}{2}}\left((1+z)^{-\frac{3}{2}}-(1+z)^{\frac{5}{2}}\right)\right\vert\leq C_Rm^{\frac{9}{2}},
  \label{mun}
\end{equation}
moreover there exists $C>0$ such that for all $z\geq 0$, $m\geq M>0$
\begin{equation}
\sup_{0\leq z<\infty}\left\vert
u_+(z,m)+\sqrt{\frac{2}{\pi}}\cos(mz)\right\vert
+
\left\vert
u_-(z,m)-\sqrt{\frac{2}{\pi}}\sin(mz)\right\vert\leq
\frac{C}{m},
  \label{zup}
\end{equation}

\begin{equation}
\sup_{M\leq m<\infty}\left\vert
u_{+[-]}(z,m)-\sqrt{\frac{2}{\pi}}\sin\left(
mz+m-\frac{5\pi}{4}-\arg H_{1[2]}^{(1)}\left(m\right)
\right)
\right\vert\leq
\frac{C}{Mz},
  \label{estmz}
\end{equation}

  \label{repspec}
\end{Lemma}


{\it Proof of Lemma \ref{repspec}.} Since the Hankel functions have no
real zero, $u_{\pm}$ are well defined in
$C^{\infty}(]0,\infty[_m\times [0,\infty[_z)$. To get the asymptotic
behaviours at low energy,
it is convenient to express $u_{\pm}$ in terms of Bessel and Neuman
functions. We have :

\begin{equation*}
u_+(z,m)=\sqrt{m(1+z)}\frac{J_1\left(m\right)N_2\left(m(1+z)\right)-N_1\left(m\right)J_2\left(m(1+z)\right)}{N_1\left(m\right)-iJ_1\left(m\right)}e^{i\theta_1(m)},
\end{equation*}
\begin{equation*}
u_-(z,m)=\sqrt{m(1+z)}\frac{J_2\left(m\right)N_2\left(m(1+z)\right)-N_2\left(m\right)J_2\left(m(1+z)\right)}{N_2\left(m\right)-iJ_2\left(m\right)}e^{i\theta_2(m)}.
\end{equation*}
We recall that the Bessel functions are entiere functions
(\cite{watson}, 3.1) :
\begin{equation*}
J_{\nu}(x)=\sum_{n=0}^{\infty}\frac{(-1)^n}{n!(n+\nu)!}\left(\frac{x}{2}\right)^{\nu+2n},
\end{equation*}
and we get from the Neumann expansion of the Bessel functions of
second kind (\cite{olver} p. 283, \cite{watson} 3.1), that the Neuman functions are analytic functions defined on the Riemann
surface of the logarithm with the following asymptotics near zero :
\begin{equation*}
N_1(x)+\frac{2}{\pi x}-\frac{x}{\pi}\log(x)=O(x)\in C^2([0,\infty[_x),\;\;x\rightarrow 0^+,
\end{equation*}
\begin{equation*}
N_2(x)+\frac{4}{\pi x^2}+\frac{1}{\pi}-\frac{x^2}{4
\pi}\log(x)=O(x^2)\in C^2([0,\infty[_x),\;\;x\rightarrow 0^+.
\end{equation*}
Elementary calculations lead to :
\begin{equation}
u_+(z,m)+\sqrt{m}(1+z)^{-\frac{3}{2}}=O\left(m^{\frac{5}{2}}(1+z)^{\frac{5}{2}}\right)\in
C^2([0,\infty[_m\times[0,\infty[_z),\;\;m\rightarrow
0,
  \label{ump}
\end{equation}

\begin{equation}
u_-(z,m)-\frac{1}{8}m^{\frac{5}{2}}\left((1+z)^{-\frac{3}{2}}-(1+z)^{\frac{5}{2}}\right)
=O\left(m^{\frac{9}{2}}(1+z)^{\frac{5}{2}}\right)\in C^2([0,\infty[_m\times[0,\infty[_z),\;\;m\rightarrow
0.
  \label{umn}
\end{equation}
To get the asymptotics at high energy or large $z$, we use the
estimate (\cite{olver} p. 238) :
\begin{equation*}
H_{\nu}^{(1)}(x)=\sqrt{\frac{2}{\pi x}}e^{ix}e^{-i\left(\nu+\frac{1}{2}\right)\frac{\pi}{2}}+O\left(x^{-\frac{3}{2}}\right),\;\;x\rightarrow\infty.
\end{equation*}
Replacing the Hankel functions by these asymptotics in (\ref{uplus})
and (\ref{umoins}), we obtain
\begin{equation*}
u_+(z,m)=-\sqrt{\frac{2}{\pi}}\cos(mz)+O\left(\frac{1}{m}\right)+O\left(\frac{1}{m(1+z)}\right),\;\;m\rightarrow\infty,
\end{equation*}

\begin{equation*}
u_-(z,m)=\sqrt{\frac{2}{\pi}}\sin(mz)+O\left(\frac{1}{m}\right)+O\left(\frac{1}{m(1+z)}\right),\;\;m\rightarrow\infty,
\end{equation*}

\begin{equation}
\left\{
\begin{array}{c}
u_{+}(z,m)=\sqrt{\frac{2}{\pi}}\sin\left(mz+m-\frac{5\pi}{4}-\arg
  H_{1}^{(1)}\left(m\right)\right)+O\left(\frac{1}{m(1+z)}\right),\;\;z\rightarrow\infty,\\
u_{-}(z,m)=\sqrt{\frac{2}{\pi}}\sin\left(mz+m-\frac{5\pi}{4}-\arg
  H_{2}^{(1)}\left(m\right)\right)+O\left(\frac{1}{m(1+z)}\right),\;\;z\rightarrow\infty.
\end{array}
\right.
  \label{uzp}
\end{equation}

\fin

We are now ready to introduce the distorded Fourier transforms
associated with $\mathbf{h}_{\pm}$. For $f\in C^0_0\left([0,\infty[_z\right)$, $\tilde{f}\in
  C^0_0\left([0,\infty[_m\right)$, we put
\begin{equation*}
0<m,\;\;F_{\pm}(f)(m):=\int_0^{\infty}f(z)u_{\pm}(z,m)dz,
 \end{equation*}

 \begin{equation*}
0<z,\;\;\check{F}_{\pm}(\tilde{f})(z):=\int_0^{\infty}\tilde{f}(m)u_{\pm}(z,m)dm.
\end{equation*}


\begin{Lemma}
$F_+$ (respectively $F_-$) can be extended into an isometry from
$L^2_{ac}$ (respectively from $L^2(\RR^+_z)$) onto $L^2(\RR^+_m)$ and
$\check{F}_{\pm}=F_{\pm}^{-1}$. $\mathbf{h}_{\pm}$ is implemented by
the operator of multiplication by $m^2$ :
\begin{equation}
f\in {\mathfrak D}(\mathbf{h}_{\pm}),\;\;F_{\pm}\left(\mathbf{h}_{\pm}f\right)(m)=m^2F_{\pm}(f)(m).
  \label{mulspec}
\end{equation}

  \label{repspec}
\end{Lemma}


{\it Proof of Lemma \ref{repspec}.}
  We construct the spectral representation by the usual way (see
  e.g. \cite{pear} Theorem 7.4). First, we have to find a normalized $\varphi(z,\lambda)$
  upper solution of the Schr\"odinger equation
$$
-\partial^2_z\varphi+\frac{15}{4}\left(\frac{1}{1+z}\right)^2\varphi=\lambda\varphi,\;\;z>0,
$$
that is a solution satisfying
$$
for\;\;almost\;\;\lambda>0,\;\;\varphi(z,\lambda)=\lim_{\varepsilon\rightarrow
  0^+}\varphi(z,\lambda+i\varepsilon)\;\;in\;\;L^2_{loc}(\RR^+_z),
$$
with
$$
\forall\varepsilon>0,\;\;\varphi(z,\lambda+i\varepsilon)\in
L^2(\RR^+_z),
$$
and
$$
\varphi\overline{\partial_z\varphi}-\overline\varphi\partial_z\varphi=-i.
$$
From (\ref{asym}) and (\ref{solh}), we obtain
$$
0<\lambda,\;\;0<z,\;\;\varphi(z,\lambda)=\frac{1}{2}\sqrt{\pi}\sqrt{1+z}H_2^{(1)}\left(\sqrt{\lambda}(1+z)\right).
$$
The next step consists in finding a real solution $v_{\pm}(z,\lambda)$  of the Schr\"odinger
equation, satisfying the appropriate boundary condition at $z=0$,
i.e. $\partial_zv_+(0,\lambda)+\frac{3}{2}v_+(0,\lambda)=0$, $v_-(0,\lambda)=0$, and
normalized to have spectral amplitude $A:=2\mid
v_{\pm}\partial_z\varphi-\varphi\partial_z v_{\pm}\mid=1$. Up to the
sign, this solution is unique, and a tedious
computation based on (\ref{solh}) leads to
$$
v_-(z,\lambda):=\frac{1}{4}\sqrt{\pi(1+z)}\left[\frac{H_2^{(2)}\left({\sqrt\lambda}\right)}{H_2^{(1)}\left({\sqrt\lambda}\right)}H_2^{(1)}\left({\sqrt\lambda}(1+z)\right)
-H_2^{(2)}\left({\sqrt\lambda}(1+z)\right)\right]e^{i\theta_2(\sqrt\lambda)},
$$
\begin{equation*}
\begin{split}
v_+(z,\lambda):=&\\
\frac{1}{4}\sqrt{\pi(1+z)}&\left[\frac{2H_2^{(2)}\left({\sqrt\lambda}\right)+{\sqrt\lambda}\left(H_2^{(2)}\right)'\left({\sqrt\lambda}\right)}{2H_2^{(1)}\left({\sqrt\lambda}\right)+{\sqrt\lambda}\left(H_2^{(1)}\right)'\left({\sqrt\lambda}\right)}H_2^{(1)}\left({\sqrt\lambda}(1+z)\right)
-H_2^{(2)}\left({\sqrt\lambda}(1+z)\right)\right]e^{i\theta_1(\sqrt\lambda)},
\end{split}
\end{equation*}
where $\theta_j$ is given by (\ref{tetaj}).
Finally we put
$u_{\pm}(z,m):=2\sqrt{\frac{m}{\pi}}v_{\pm}(z,m^2)$, and we get
(\ref{umoins}), and also (\ref{uplus}) by using the recurrence relation
$$2H_2^{(j)}(x)+x\frac{d}{dx}H_2^{(j)}(x)=xH_1^{(j)}(x).$$ Then $F_{\pm}$
are unitary representations that satisfy (\ref{mulspec}).

\fin

Now we are ready to prove the theorem. We define
\begin{equation*}
\varphi_j^{\mathbf M}(\mathbf{x}):=\int_{-\infty}^{\infty}\Phi_j(\mathbf{x},\zeta)f_0(\zeta)d\zeta,\;\;
\Phi_j^{\mathbf M}(\mathbf{x},z):=\varphi_j^{\mathbf M}(\mathbf{x})f_0(z),\;\;\Phi_j^{\mathbf B}:=\Phi_j-\Phi_j^{\mathbf M}.
\end{equation*}
The solution $\Phi$ of the Cauchy problem (\ref{EQB}), (\ref{ci}),
(\ref{regf}) can be written as
\begin{equation*}
\Phi=\Phi^{\mathbf M}+\Phi^{\mathbf B},\;\;\Phi^{\mathbf M}(t,\mathbf{x},z)=\phi_0(t,\mathbf{x})f_0(z),
\end{equation*}
where $\phi_0\in C^0\left(\RR_t;BL^1\left(\RR^3_{\mathbf x}\right)\right)$
  with $\partial_t\phi_0\in
  C^0\left(\RR_t;L^2\left(\RR^3_{\mathbf x}\right)\right)$, is solution of
\begin{equation}
\partial^2_t\phi_0-\Delta_{\mathbf
  x}\phi_0=0,\;\;\phi_0(0,\mathbf{x})=\varphi_0^{\mathbf
  M}(\mathbf{x}),\;\;\partial_t\phi_0(0,\mathbf{x})=\varphi_1^{\mathbf
  M}(\mathbf{x}),
\end{equation}
and $\Phi^{\mathbf B}\in C^0\left(\RR_t;{\mathfrak K}^1\left(\RR^4\right)\right)$
  with $\partial_t\Phi^{\mathbf B}\in
  C^0\left(\RR_t;{\mathfrak K}^0\left(\RR^4\right)\right)$, is
  solution of  (\ref{EQB}) with initial data
  \begin{equation*}
\Phi^{\mathbf B}(0,\mathbf{x},z)=\Phi^{\mathbf B}_0(\mathbf{x},z),\;\;
\partial_t\Phi^{\mathbf B}(0,\mathbf{x},z)=\Phi^{\mathbf B}_1(\mathbf{x},z).
 \end{equation*}
If we put
\begin{equation*}
2\Phi_{\pm}(t,{\mathbf x},z):=\Phi^{\mathbf B}(t,{\mathbf x},z)\pm
\Phi^{\mathbf B}(t,{\mathbf x},-z),
\end{equation*}
we have
\begin{equation*}
\Phi^{\mathbf
  B}(t,\mathbf{x},z)=\Phi_+(t,\mathbf{x},\mid
z\mid)+\frac{z}{\mid z\mid}\Phi_-(t,\mathbf{x},\mid z\mid).
\end{equation*}
We note that $\Phi_+(t,.)$ and $\Phi_-(t,.)$ are orthogonal in
$\mathfrak{W}^1\left(\RR^4\right)$ and $L^2\left(\RR^4\right)$, and
\begin{equation*}
\parallel \Phi_0^{\mathbf B}\parallel_{\mathfrak{W}^1(\RR^4)}^2+
\parallel \Phi_1^{\mathbf B}\parallel_{L^2(\RR^4)}^2=
2\sum_{\pm}
\int_0^{\infty}\int_{\RR^3_{\mathbf x}}\mid\nabla_{t,\mathbf{x}}\Phi_{\pm}\mid^2
+\left\vert\partial_z\Phi_{\pm}+\frac{3}{2}\frac{z}{\mid z\mid}\left(\frac{1}{1+z}\right)\Phi_{\pm}\right\vert^2 dzd\mathbf{x}.
\end{equation*}
Given $t\in\RR$, for almost $\mathbf{x}\in\RR^3$, the maps
$z\mapsto\nabla_{\mathbf{x}}\Phi_-(t,\mathbf{x},z)$ and
 $z\mapsto\nabla_{\mathbf{x}}\Phi_+(t,\mathbf{x},z)$ respectively  belong to
$L^2(\RR^+_z)$ and to $L^2_{ac}$. Thus we can apply Lemma
\ref{repspec} and  for almost all $m>0$ we introduce
\begin{equation*}
\phi^{\pm}_m(t,\mathbf{x}):=\lim_{M\rightarrow\infty}\int_0^{M}\Phi_{\pm}(t,\mathbf{x},z)u_{\pm}(z,m)dz,
\end{equation*}
that is converging in $L^2\left(\RR^+_m;BL^1(\RR^3_{\mathbf x})\right)$. Moreover, 
since $\Phi_{\pm}\in C^0\left(\RR_t;{\mathfrak K}^1\left(\RR^4\right)\right)$
  with $\partial_t\Phi_{\pm}\in
  C^0\left(\RR_t;{\mathfrak K}^0\left(\RR^4\right)\right)$, then
  \begin{equation}
\phi^{\pm}_m\in
 C^0\left(\RR_t;L^2\left(\RR^+_m;BL^1\left(\RR^3_{\mathbf x}\right)\right)\right),
 \;\; \partial_t\phi^{\pm}_m\in
  C^0\left(\RR_t;L^2\left(\RR^+_m;L^2\left(\RR^3_{\mathbf
        x}\right)\right)\right),
    \label{rgfipl}
  \end{equation}
 and
  \begin{equation}
\phi^{\pm}_m(0,\mathbf{x})=\frac{1}{2}\lim_{M\rightarrow\infty}\int_0^{M}\left(\Phi_0^{\mathbf
  B}(\mathbf{x},z)\pm\Phi_0^{\mathbf
  B}(\mathbf{x},-z)\right) u_{\pm}(z,m)dz\in L^2\left(\RR^+_m;BL^1\left(\RR^3_{\mathbf x}\right)\right),
    \label{cifi}
  \end{equation}
 \begin{equation}
\partial_t\phi^{\pm}_m(0,\mathbf{x})=\frac{1}{2}\lim_{M\rightarrow\infty}\int_0^{M}\left(\Phi_1^{\mathbf
  B}(\mathbf{x},z)\pm\Phi_1^{\mathbf
  B}(\mathbf{x},-z)\right)u_{\pm}(z,m)dz\in L^2\left(\RR^+_m;L^2\left(\RR^3_{\mathbf x}\right)\right).
    \label{cifit}
  \end{equation}
Therefore, if we introduce
\begin{equation}
f^+_m(z):=u_+(\mid z\mid,m),\;\;f^-_m(z):=\frac{z}{\mid
  z\mid}u_-(\mid z\mid,m),
  \label{fufu}
\end{equation}
Lemma \ref{mulspec} assures that
\begin{equation*}
\nabla_{t,\mathbf{x}}\Phi_{\pm}(t,\mathbf{x},z)=\lim_{M\rightarrow\infty}\int_0^M\nabla_{t,\mathbf{x}}\phi^{\pm}_m(t,\mathbf{x})f_m^{\pm}(z)dm\;\;in\;\;C^0\left(\RR_t;L^2(\RR^4_{\mathbf{x},z})\right),
\end{equation*}
\begin{equation*}
\int_0^{\infty}\int_{\RR^3_{\mathbf
    x}}\mid\nabla_{t,\mathbf{x}}\Phi_{\pm}(t,\mathbf{x},z)\mid^2
dzd\mathbf{x}=\int_0^{\infty}\int_{\RR^3_{\mathbf
    x}}\mid\nabla_{t,\mathbf{x}}\phi^{\pm}_m(t,\mathbf{x},m)\mid^2
dmd\mathbf{x}.
\end{equation*}
We remark that for $u\in{\mathfrak D}(\mathbf{h}_{\pm})$ we have 
$$
\int_0^{\infty}\left\vert f'(z)+\frac{3}{2}\frac{1}{1+z}f(z)\right\vert^2dz=
<\mathbf{h}_{\pm}f,f>_{L^2(\RR^+)}=\int_0^{\infty}\left\vert F_{\pm}f(m)\right\vert^2m^2dm.
$$
This equality can be extended by density into an isometry from  the
closure of ${\mathfrak D}(\mathbf{h}_{\pm})$ for the norm associated
with the first integral, onto $L^2(\RR^+_m, m^2dm)$. We deduce that
$m\phi^{\pm}_m\in
  C^0\left(\RR_t;L^2(\RR^+_m;L^2\left(\RR^3_{\mathbf
        x}\right)\right)$ hence with (\ref{rgfipl}) 
  \begin{equation}
\forall a>0,\;\;\phi^{\pm}_m\in
 C^0\left(\RR_t;L^2([a,\infty[_m;H^1\left(\RR^3_{\mathbf x}\right)\right),
    \label{regfih}
  \end{equation}
and 
 \begin{equation*}
\int_0^{\infty}\int_{\RR^3_{\mathbf x}}\left\vert
    \partial_z\Phi_{\pm}(t,\mathbf{x},z)+\frac{3}{2}\frac{1}{1+z}\Phi_{\pm}(t,\mathbf{x},z)\right\vert^2dzd\mathbf{x}=
\int_0^{\infty}\int_{\RR^3_{\mathbf
    x}}m^2\mid\phi^{\pm}_m(t,\mathbf{x},m)\mid^2
dmd\mathbf{x},
  \end{equation*}
therefore (\ref{Ldem}), (\ref{decompo}), (\ref{deconer}) are
established.\\

It remains to prove that $\phi_m^{\pm}$ is a finite energy solution of the
Klein-Gordon equation for almost all $m>0$. Thanks to (\ref{deconer}),
we see that the map $(\Phi_0,\Phi_1)\mapsto (\phi_m^{\pm},\partial_t\phi_m^{\pm})$ is
continuous from ${\mathfrak W}^1(\RR^4)\times L^2(\RR^4)$ to
$C^0\left(\RR_t;L^2\left(\RR^+_m;BL^1(\RR^3_{\mathbf x})\right)\right)\times C^0\left(\RR_t;L^2\left(\RR^+_m;L^2(\RR^3_{\mathbf x})\right)\right)$, hence it is sufficient to prove
that  
\begin{equation}
(\partial_t^2-\Delta_{\mathbf x}+m^2)\phi_m^{\pm}=0\;\;in\;\;{\mathcal D}'(\RR_t\times\RR^3_{\mathbf x}\times]0,\infty[_m).
  \label{kgm}
\end{equation}
 for a
dense set of initial data. We choose $\Phi_1\in
C^{\infty}_0(\RR^4)$, and $\Phi_0\in {\mathfrak D}(\mathbf
{H}_0)$, compactly supported in $\{(\mathbf{x},z);\;\mid\mathbf
x\mid,\;\mid z\mid\leq R\}$, and such that $\Phi_0\in
C^0_0\left(\RR_z;C^{\infty}_0(\RR^3_{\mathbf x}\right)$. It is easy to
see that the set of all such data is dense in
$\mathfrak{W}^1(\RR^4)$ : given  $\Phi_0\in {\mathfrak D}(\mathbf
{H}_0)$, we take $\chi\in C^{\infty}_0(\RR^4)$ equal to $1$ on a
neighborhood of $0$ and $\theta\in C^{\infty}_0(\RR^3_{\mathbf x})$,
$0\leq \theta$, $\int\theta(\mathbf{x})d\mathbf{x}=1$. Then $\Phi_{n,p}(\mathbf{x},z):=p^3\int\theta(p(\mathbf{x-x'}))\chi(\frac{\mathbf  x'}{n},\frac{z}{n})\Phi_0(\mathbf{x'},z)d{\mathbf x'}$ belongs to ${\mathfrak D}(\mathbf
{H}_0)\cap{\mathcal E}'(\RR^4)\cap C^0_0\left(\RR_z;C^{\infty}_0(\RR^3_{\mathbf x}\right)
$, and $\Phi_{n,p}$ tends to $\Phi_0$ in
$H^1(\RR^4)$ as $n,p\rightarrow\infty$. The solution of the Cauchy
problem satisfies
\begin{equation*}
\Phi\in C^2\left(\RR_t;L^2(\RR^4)\right)\cap C^1\left(\RR_t;H^1(\RR^4)\right),
\end{equation*}
and since $\Delta_{\mathbf x}$ and $P(\partial)$ defined by (\ref{H})
are commuting,
$\Delta_{\mathbf x}\Phi$ is also a finite energy solution, compactly
supported in space at each time, hence
\begin{equation*}
\Delta_{\mathbf x}\Phi\in C^0\left(\RR_t;L^2(\RR^4)\right)\cap C^1\left(\RR_t;H^1(\RR^4)\right).
\end{equation*}
This implies that
\begin{equation*}
\Phi,\,\partial_t\Phi\in
C^0\left(\RR_t\times\RR_z;H^{\frac{5}{2}}(\RR^3_{\mathbf
    x}\right)\subset C^0(\RR^5).
\end{equation*}
We deduce that for all $t\in\RR$, $\mathbf{x}\in\RR^3$, the map
$z\mapsto\Phi_{\pm}(t,\mathbf{x},z)$ belongs to
$\mathfrak{D}(\mathbf{h}_{\pm})$, and for $\mid t\mid\leq T$
\begin{equation*}
\phi_m^{\pm}(t,\mathbf{x})=\frac{1}{2}\int_0^{R+T}\left(\Phi(t,\mathbf{x},z)\pm\Phi(t,\mathbf{x},-z)\right)u_{\pm}(z,m)dm\in
C^1\left(]-T,T[;H^1(\RR^3_{\mathbf x})\right),
\end{equation*}
\begin{equation*}
(\partial_t^2-\Delta_{\mathbf
  x}+m^2)\phi_m^{\pm}=\int_0^{R+T}\left(\partial_t^2+P(\partial)\right)
\left(\Phi(t,\mathbf{x},z)\pm\Phi(t,\mathbf{x},-z)\right)u_{\pm}(z,m)dm=0.
\end{equation*}

It remains to prove that in the general case where
$(\Phi_0,\Phi_1)\in{\mathfrak W}^1(\RR^4)\times L^2(\RR^4)$, $\phi_m^{\pm}$ belongs to $C^0\left(\RR_t;H^1\left(\RR^3_{\mathbf x}\right)\right)\cap
  C^1\left(\RR_t;L^2\left(\RR^3_{\mathbf x}\right)\right)$ for almost
  all $m>0$. We have established that for $0<a$, $\phi_m^{\pm}\in C^0\left(\RR_t;L^2\left([a,\infty[_m;H^1\left(\RR^3_{\mathbf x}\right)\right)\right)\cap
  C^1\left(\RR_t;L^2\left([a,\infty[_m;L^2\left(\RR^3_{\mathbf
          x}\right)\right)\right)$ is solution of (\ref{cifi}),
  (\ref{cifit}) and (\ref{kgm}). We remark that this Cauchy problem is
  well posed in this functional framework because of the conservation
  of the energy :
  \begin{equation*}
\int_a^{\infty}\int_{\RR^3_{\mathbf x}}\mid\nabla_{\mathbf{x},t}\phi_m^{\pm}(t,\mathbf{x})\mid^2+m^2\mid\phi_m^{\pm}(t,\mathbf{x})\mid^2dmd\mathbf{x}=Cst.
 \end{equation*}
This equality can be easily proved by the usual way from the
Klein-Gordon equation when $\phi_m^{\pm}\in C^1\left(\RR_t;L^2\left([a,\infty[_m;H^1\left(\RR^3_{\mathbf x}\right)\right)\right)\cap
  C^2\left(\RR_t;L^2\left([a,\infty[_m;L^2\left(\RR^3_{\mathbf
          x}\right)\right)\right)$. In the general case, we use
  $\phi_{\varepsilon}(t,\mathbf{x}):=\varepsilon^{-1}\int_t^{t+\varepsilon}\phi_m^{\pm}(s,\mathbf{x})ds$ that belongs to this space and tends to $\phi_m^{\pm}$ in  $C^0\left(\RR_t;L^2\left([a,\infty[_m;H^1\left(\RR^3_{\mathbf x}\right)\right)\right)\cap
  C^1\left(\RR_t;L^2\left([a,\infty[_m;L^2\left(\RR^3_{\mathbf
          x}\right)\right)\right)$ as $\varepsilon\rightarrow 0$.
Now (\ref{cifit}) and (\ref{regfih}) assure that for almost $m>0$,
$\phi_m^{\pm}(0,\mathbf{x})\in H^1(\RR^3_{\mathbf x})$, $\partial_t\phi_m^{\pm}(0,\mathbf{x})\in L^2(\RR^3_{\mathbf x})$. For such an $m$ we consider $\psi_m^{\pm}\in
C^0\left(\RR_t;H^1(\RR^3_{\mathbf x})\right)\cap C^1\left(\RR_t;L^2(\RR^3_{\mathbf x})\right)$
solution of the Klein-Gordon equation with initial data
$\psi_m^{\pm}(0,\mathbf{x})=\phi_m^{\pm}(0,\mathbf{x})$,
$\partial_t\psi_m^{\pm}(0,\mathbf{x})=\partial_t\phi_m^{\pm}(0,\mathbf{x})$.
The energy estimate implies that $$\psi_m^{\pm}\in
L^2\left([a,\infty[_m;C^0\left(\RR_t;H^1(\RR^3_{\mathbf x})\right)\cap
  C^1\left(\RR_t;L^2(\RR^3_{\mathbf x})\right)\right).$$ Since this space is
included in $C^0\left(\RR_t;L^2\left([a,\infty[_m;H^1\left(\RR^3_{\mathbf x}\right)\right)\right)\cap
  C^1\left(\RR_t;L^2\left([a,\infty[_m;L^2\left(\RR^3_{\mathbf
          x}\right)\right)\right)$, the uniqueness implies
  $\psi_m^{\pm}=\phi_m^{\pm}$, therefore  we have proved that 
 $\phi_m^{\pm}$ belongs to $C^0\left(\RR_t;H^1\left(\RR^3_{\mathbf x}\right)\right)\cap
  C^1\left(\RR_t;L^2\left(\RR^3_{\mathbf x}\right)\right)$ for almost
  all $m>0$.

\fin
We achieve this part devoted to the
spectral analysis of $\mathbf{H}_0$, by the computation of the kernel of
its resolvent. Since  $\mathbf{H}_0$ is a positive self-adjoint
operator, $\left(\mathbf{H}_0-\lambda^2\right)^{-1}$ is well defined
in $\mathcal{L}\left(L^2(\RR^4)\right)$ for all $\lambda\in\CC^*$ with
$0<\arg\lambda<\pi$. If we add a cut-off in energy relatively to the
brane,
i.e. we consider
$\left(\mathbf{H}_0-\lambda^2\right)^{-1}\mathbf{1}_{[0,R]}\left(\left\vert\nabla_{\mathbf{x}}\right\vert\right)$,
we can express the kernel explicitely.

We recall that the Hankel functions are holomorphic on
the whole Riemann surface of the logarithm $\widetilde{\CC^*}$. We introduce
\begin{equation}
\begin{split}
\mathbf{K}(m;z,z'):=\frac{\pi}{4i}\sqrt{(1+z)(1+z')}&
\left[H_2^{(2)}\left(m\right)\right.H_2^{(1)}\left(m(1+z\wedge
    z')\right)\\
&-\left.H_2^{(1)}\left(m\right)H_2^{(2)}\left(m(1+z\wedge z')\right)\right]H_2^{(1)}\left(m(1+z\vee z')\right)
\end{split}
  \label{KKKKKK}
\end{equation}
where  $z\wedge z':=\min(z,z')$, $z\vee
z':=\max(z,z')$.
For $x\in\CC^*$, we denote $\sqrt{x}$ the  branch of the
square root defined by $0\leq \arg\sqrt{x}<\pi$ when $0\leq
\arg x<2\pi$.\\

\begin{Theorem}
For any $R>0$, $\lambda\in\CC^*$, $0<\arg\lambda<\pi$, $F\in L^1\cap
L^2(\RR^4)$, we have
\begin{equation*}
\left(\mathbf{H}_0-\lambda^2\right)^{-1}\mathbf{1}_{[0,R]}\left(\left\vert\nabla_{\mathbf{x}}\right\vert\right)F(\mathbf{x},z)=\int_{\RR^4}\mathbf{K}_R(\lambda;\mathbf{x},z;\mathbf{x'},z')F(\mathbf{x'},z')d\mathbf{x'}dz',
\end{equation*}
where this integral converges absolutely and the kernel of the truncated resolvent
is given by 
\begin{equation*}
\begin{split}
\mathbf{K}_R(\lambda;\mathbf{x},z;\mathbf{x'},z'):=\\
\int_0^R\frac{\sin\left(r\mid\mathbf{x-x'}\mid\right)}{4\pi^2\mid
  \mathbf{x-x'}\mid}&\left[\frac{zz'}{\mid
    zz'\mid}\frac{1}{H_2^{(1)}\left(\sqrt{\lambda^2-r^2}\right)}-
\frac{1}{H_1^{(1)}\left(\sqrt{\lambda^2-r^2}\right)}\right]\mathbf{K}\left(\sqrt{\lambda^2-r^2};\mid
z\mid, \mid z'\mid\right)rdr.
\end{split}
\end{equation*}
\label{kernel}
\end{Theorem}

{\it Proof of Theorem \ref{kernel}.}
Given $m\in\CC$, $0<\arg m<\pi$, we compute the kernel
$\mathbf{K}_{\pm}(m;z,z')$ of the
resolvents $\left(\mathbf{h}_{\pm}-m^2\right)^{-1}$ by the usual way
(see \cite{pear}, p. 262). First we determine  $f_j(m;z)$ the
solutions of equation (\ref{sch}) with $f_1(m;0)=0$,
$\partial_zf_1(m;0)=1$, $f_2(m;0)=1$, $\partial_zf_2(m;0)=0$.  Using
(\ref{solh}), the wronskian relation
$H_{\nu}^{(1)}(x)\frac{d}{dx}H_{\nu}^{(2)}(x)-H_{\nu}^{(2)}(x)\frac{d}{dx}H_{\nu}^{(1)}(x)=-\frac{4i}{\pi
  x}$, and the identity
$x\frac{d}{dx}H_2^{(j)}(x)+2H_2^{(j)}(x)=xH_1^{(j)}(x)$, we find after
tedious calculations :
$$
f_1(m;z)=\frac{\pi}{4i}\sqrt{1+z}\left[H_2^{(2)}\left(m\right)H_2^{(1)}\left(m(1+z)\right)-H_2^{(1)}\left(m\right)H_2^{(2)}\left(m(1+z)\right)\right],
$$
\begin{equation*}
\begin{split}
f_2(m;z)=-\frac{\pi}{4i}\sqrt{1+z}&\left[\left(mH_1^{(2)}(m)-\frac{3}{2}H_2^{(2)}(m)\right)H_2^{(1)}(m(1+z))\right.\\
&\left.-\left(mH_1^{(1)}\left(m\right)-\frac{3}{2}H_2^{(1)}\left(m\right)\right)H_2^{(2)}\left(m(1+z)\right)\right].
\end{split}
\end{equation*}
Secondly we have to determine $C(m)\in\CC$ such that
$f_2(m;z)+C(m)f_1(m;z)\in L^2(\RR^+_z)$. 
The bounds for the Hankel functions (\cite{olver}, P. 267) assure that
for any $\delta>0$ there exists $C_{\delta}>0$ such that for all
$x\in\CC^*$, with $0<\arg x<\pi$ we have
\begin{equation}
\forall x\in\CC^*,\;\;0<\arg x<\pi,\;\;\left\vert
  H_{\nu}^{(1)}(x)\right\vert\leq C_{\delta}\left\vert e^{ix}\right\vert,\;\;\left\vert
  H_{\nu}^{(2)}(x)\right\vert\leq C_{\delta}\left\vert e^{-ix}\right\vert.
  \label{hankbo}
\end{equation}
Since  $0<\arg m<\pi$, the
function $H^{(1)}_2\left(m(1+z)\right)$ is exponentially
decreasing as $z\rightarrow\infty$, and  $H^{(2)}_2\left(m(1+z)\right)$ is exponentially
increasing as $z\rightarrow\infty$. Thus we get
$$
C(m)=\frac{mH^{(1)}_1\left(m\right)-\frac{3}{2}H^{(1)}_2\left(m\right)}{H_2^{(1)}\left(m\right)},
$$
$$
f_2(m;z)+C(m)f_1(m;z)=\sqrt{1+z}\frac{H_2^{(1)}\left(m(1+z)\right)}{H_2^{(1)}\left(m\right)}.
$$
Finally we know that
$$
\mathbf{K}_-(m;z,z')=f_1(m,z\wedge z')\left[f_2(m;z\vee
  z')+C(m)f_1(m;z\vee z')\right],
$$
$$
\mathbf{K}_+(m;z,z')=-\frac{\left[f_2(m;z\wedge z')-\frac{3}{2}f_1(m,z\wedge z')\right]\left[f_2(m;z\vee
  z')+C(m)f_1(m;z\vee z')\right]}{\frac{3}{2}+C(m)},
$$
and so we conclude that
\begin{equation}
\mathbf{K}_{+}(m;z,z')=-\frac{\mathbf{K}(m;z,z')}{H_1^{(1)}\left(m\right)},\;\;
\mathbf{K}_{-}(m;z,z')=\frac{\mathbf{K}(m;z,z')}{H_2^{(1)}\left(m\right)}.
  \label{kpm}
\end{equation}

Now given $\lambda\in\CC$, $0<\arg\lambda<\pi$, and $F\in
L^2(\RR^3_{\mathbf{x}}\times\RR_z)$ we introduce $F_{\pm}(\mathbf{x},z):=\frac{1}{2}(F(\mathbf{x},z)\pm
  F(\mathbf{x},z)$, and we solve
$
\left(\mathbf{H}_0-\lambda^2\right)\Phi_{\pm}=\mathbf{1}_{[0,R]}\left(\left\vert\nabla_{\mathbf{x}}\right\vert\right)F_{\pm}.
$ Then the solution $\Phi$ of $
\left(\mathbf{H}_0-\lambda^2\right)\Phi=\mathbf{1}_{[0,R]}\left(\left\vert\nabla_{\mathbf{x}}\right\vert\right)F
$
is given by $\Phi(\mathbf{x},z)=\Phi_+(\mathbf{x},\mid
z\mid)+\frac{z}{\mid z\mid}\Phi_-(\mathbf{x},\mid z\mid).$
We use the partial Fourier transform  $\mathcal{F}_{\mathbf{x}}f(\pmb{\xi},z)=\hat{f}(\pmb{\xi},z)$ of
$f(.,z)$ with respect to $\mathbf{x}$ given for $f\in
C^{\infty}_0(\RR^4)$ by
\begin{equation}
\mathcal{F}_{\mathbf{x}}f(\pmb{\xi},z)=
\hat{f}(\pmb{\xi},z):=\frac{1}{(2\pi)^{\frac{3}{2}}}\int_{\RR^3}e^{-i{\mathbf
    x}{\bf .}{\pmb\xi}}f({\mathbf
    x},z)d{\mathbf x}.
  \label{fourier}
\end{equation}
Then for almost all $\pmb{\xi}\in\RR^3$, the function
$r\in\RR^+\mapsto\hat{\Phi}_{\pm}(\pmb{\xi},.)$ belongs to  $\mathfrak{D}(\mathbf{h}_{\pm})$, and 
$$
(\mathbf{h}_{\pm}+\mid \pmb{\xi}\mid^2-\lambda^2)\hat{\Phi}_{\pm}(\pmb{\xi},z)=\mathbf{1}_{[0,R]}\left(\left\vert\pmb\xi\right\vert\right)\hat{F}_{\pm}(\pmb{\xi},z),\;\;0<z.
$$
In the sequel, we assume that $F\in
L^1\cap L^2(\RR^4)$, hence all the following
integrals make sense.
Taking advantage of the Hankel transform, we have for $0<z$ :
\begin{equation*}
\begin{split}
\Phi_{\pm}(\mathbf{x},z)&=
\left(\frac{1}{2\pi}\right)^{\frac{3}{2}}\int_{\mid\pmb{\xi}\mid\leq R}e^{i\mathbf{x}.\pmb{\xi}}\left(\int_0^{\infty}\mathbf{K}_{\pm}\left(\sqrt{\lambda^2-\mid\pmb{\xi}\mid^2};z,z'\right)\hat{F}_{\pm}(\pmb{\xi},z')dz'\right)d\pmb{\xi}\\
&=\left(\frac{1}{2\pi}\right)^{3}\int_{\RR^3}\int_0^{\infty}\left[\int_{\mid
    \pmb{\xi}\mid\leq
  R}e^{i(\mathbf{x-x'}).\pmb{\xi}}\mathbf{K}_{\pm}\left(\sqrt{\lambda^2-\mid\pmb{\xi}\mid^2};z,z'\right)d\pmb{\xi}\right]F_{\pm}(\mathbf{x'},z')dz'd\mathbf{x'}\\
&=\left(\frac{1}{2\pi}\right)^{\frac{3}{2}}\int_{\RR^3}\int_0^{\infty}\frac{1}{\mid\mathbf{x-x'}\mid^{\frac{1}{2}}}\left[\int_0^{R}J_{\frac{1}{2}}(\mid\mathbf{x-x'}\mid
r)\mathbf{K}_{\pm}\left(\sqrt{\lambda^2-r^2};z,z'\right)r^{\frac{3}{2}}dr\right]F_{\pm}(\mathbf{x'},z')dz'd\mathbf{x'}.
\end{split}
\end{equation*}
We conclude that for all $\mathbf{x}\in\RR^3$, $z\in\RR$, we have :
\begin{equation*}
\Phi(\mathbf{x},z)=\int_{\RR^4}\mathbf{K}_R(\lambda;\mathbf{x},z;\mathbf{x'},z')F(\mathbf{x'},z')d\mathbf{x'}dz',
\end{equation*}
with
\begin{equation*}
\mathbf{K}_R(\lambda;\mathbf{x},z;\mathbf{x'},z'):=\int_0^R\frac{\sin\left(r\mid\mathbf{x-x'}\mid\right)}{4\pi^2\mid
  \mathbf{x-x'}\mid}\left[\mathbf{K}_+\left(\sqrt{\lambda^2-r^2};\mid z\mid, \mid z'\mid\right)+\frac{zz'}{\mid zz'\mid}\mathbf{K}_-\left(\sqrt{\lambda^2-r^2};\mid z\mid, \mid z'\mid\right)\right]rdr.
\end{equation*}
Now the result follows from (\ref{kpm}) and (\ref{hankbo}) that assure
that for any compact $A\subset\{\lambda\in\CC^*,\;\;0<\arg
\lambda<\pi\}$, and all $R,R_1\in [0,\infty[$, there exist $C,\gamma>0$
such that
\begin{equation}
\sup_{\mathbf{x},\mathbf{x'}\in\RR^4}\sup_{\mid z\mid\leq
  R_1}\sup_{\lambda\in A}\left\vert
  \mathbf{K}_R(\lambda;\mathbf{x},z;\mathbf{x'},z') \right\vert\leq C
e^{-\gamma \mid z'\mid}.
  \label{kbo}
\end{equation}
\fin

Finally we investigate the domain of analyticity of the continuation
of the truncated resolvents (see Figure 3, section VI).
\begin{Corollary}
For any $R_j>0$,  the truncated resolvent
$$\mathbf{1}_{[0,R_1]}\left(\mid\mathbf{x}\mid+\mid z\mid\right)\left(\mathbf{H}_0-\lambda^2\right)^{-1}\mathbf{1}_{[0,R_0]}\left(\left\vert\nabla_{\mathbf{x}}\right\vert\right)\mathbf{1}_{[0,R_2]}\left(\mid\mathbf{x}\mid+\mid z\mid\right)$$
considered as a  $\mathcal{L}\left(L^2\left(\RR^4_{\mathbf{x},z}\right)\right)$-valued function of $\lambda$,
has an analytic continuation on
$$\mathcal{O}_{R_0}:=\left\{\lambda\in\widetilde{\CC^*};\;\;
  H_{\nu}^{(1)}\left(\sqrt{\lambda^2-r^2}\right)\neq 0
    \;\;\forall r\in[0,R_0],\;\nu=1,2\right\}.$$
  \label{comres}
\end{Corollary}

{\it Proof of Corollary \ref{comres}.} By the Dunford theorem
(\cite{yoshida}, p. 128), it is sufficient to show that given $F\in
L^2(\RR^4)$, the map
$$
\lambda\longmapsto\mathbf{1}_{[0,R_1]}\left(\mid\mathbf{x}\mid+\mid
  z\mid\right)\int_{\RR^4}
\mathbf{K}_{R_0}(\lambda;\mathbf{x},z;\mathbf{x'},z')\mathbf{1}_{[0,R_2]}\left(\mid\mathbf{x'}\mid+\mid z'\mid\right)F(\mathbf{x'},z')d\mathbf{x'}dz'
$$
is a $L^2(\RR^4_{\mathbf{x},z})$-valued, holomorphic function on
$\mathcal{O}_{R_0}$. This property is easily proved with (\ref{kbo}) by remarking that
the map
$$
\lambda\longmapsto\mathbf{1}_{[0,R_1]}\left(\mid\mathbf{x}\mid+\mid
  z\mid\right)
\mathbf{K}_{R_0}(\lambda;\mathbf{x},z;\mathbf{x'},z')\mathbf{1}_{[0,R_2]}\left(\mid\mathbf{x'}\mid+\mid z'\mid\right)
$$
is a $L^{\infty}(\RR^4_{\mathbf{x},z}\times\RR^4_{\mathbf{x'},z'})$-valued, holomorphic function on
$\mathcal{O}_{R_0}$.
\fin


\section{Decay near the Brane and Strichartz estimates}

It is clear that  the massless graviton decays uniformly as
$t^{-1}$ for regular data, and belongs to $L^4(\RR^5)$ according to \cite{strichartz}. In this
part we investigate the decay of the Kaluza-Klein tower. We know that for initial data in $C^{\infty}_0(\RR^4)$, the solutions
of the D'Alembertian on the Minkowski space-time $\RR^{1+4}$ decay
uniformly in space as $t^{-\frac{3}{2}}$ (Von Wahl estimate
\cite{vonwahl}). In this section we first prove that the same
estimate holds for the Kaluza-Klein tower near the brane. These
results can appear to be  surprising since, although these waves are superpositions of
Klein-Gordon field on $\RR^{1+3}$ for which this decay holds, there is
a continuum of mass on $]0,\infty[$, without gap to separate the zero
mass. The key of the phenomenon is the behaviour at zero of the
spectral kernels $f_m^{\pm}(z)=O(\sqrt{m})$. Since the Von Wahl
estimates involve the norms of the initial data in some Sobolev spaces
based on $L^1$, we have to construct the functional framework adapted
to our problem. We introduce for $N=0,1$ and $\varepsilon>0$ :

\begin{equation}
\begin{split}
\mathcal{X}^N_{\varepsilon}:=\left\{\Phi\in{\mathfrak
    K}^0\cap\mathfrak{D}\left(\mathbf{H}_0^{N+1}\right)\right. &;\;\;\mid\alpha\mid\leq
2+N,\;1\leq l\leq 3+N-\mid\alpha\mid\Longrightarrow \\
&\left.(1+\mid z\mid)^{\varepsilon}\partial^{\alpha}_{\mathbf
  x}\Phi\in L^1(\RR^4),\;\partial_z^l\partial^{\alpha}_{\mathbf
  x}\Phi\in
L^1(\RR^3_{\mathbf{x}}\times\RR^*_z)
\right\},
\end{split}
  \label{xm}
\end{equation}
where the derivatives with respect to $z$ are taken in the sense of
the distributions in
$\RR^3_{\mathbf{x}}\times\RR^*_z$. This space is endowed with its
natural norm $\|\Phi\|_{L^2(\RR^4)}+\|\mathbf{H}_0^{N+1}\Phi\|_{L^2(\RR^4)}+\sum\|(1+\mid z\mid)^{\varepsilon}\partial^{\alpha}_{\mathbf
  x}\Phi\|_{L^1(\RR^4)}+\sum\|\partial_z^l\partial^{\alpha}_{\mathbf
  x}\Phi\|_{L^1(\RR^3\times\RR^*)}$.
 We are now ready to state the
 decay estimate of  $L^1-L^{\infty}$ type :


\begin{Theorem}
For any $R>0$ there exists $C_R>0$ such that for any
$\varepsilon\in]0,\frac{1}{2}]$, and for all
$\Phi_0\in{\mathcal X}^1_{\varepsilon}$, $\Phi_1\in{\mathcal X}^0_{\varepsilon}$,  the solution
$\Phi$ of the Cauchy problem (\ref{EQB}), (\ref{ci}), (\ref{regf})
satisfies :
\begin{equation*}
\begin{split}
\parallel \Phi(t,.)\parallel_{L^{\infty}(\RR^3_{\mathbf{x}}\times[-R,R]_z)}&\leq\\
C_R\mid t\mid^{-\frac{3}{2}}  & \left[  
\frac{1}{\varepsilon}\sum_{\mid\alpha\mid+j\leq 3}\parallel
(1+\mid z\mid)^{\varepsilon}\partial^{\alpha}_{\mathbf
  x}\Phi_j\parallel_{L^1(\RR^4)}
+\sum_{\mid\alpha\mid+j\leq 3}\sum_{l=1}^{4-\mid\alpha\mid-j}\parallel
\partial_z^l\partial^{\alpha}_{\mathbf
  x}\Phi_j\parallel_{L^1(\RR^3_{\mathbf{x}}\times\RR^*_z)}   \right].
\end{split}
\end{equation*}
\label{infbrane}
\end{Theorem}

The first term of the right-hand side of this estimate, $\frac{1}{\varepsilon}\sum_{\mid\alpha\mid+j\leq 3}\parallel
(1+\mid z\mid)^{\varepsilon}\partial^{\alpha}_{\mathbf
  x}\Phi_j\parallel_{L^1(\RR^4)}$, controls the contribution of the
light Kaluza-Klein modes with the mass $m\in]0,1]$, while the second one, $\sum_{\mid\alpha\mid+j\leq 3}\sum_{l=1}^{4-\mid\alpha\mid-j}\parallel
\partial_z^l\partial^{\alpha}_{\mathbf
  x}\Phi_j\parallel_{L^1(\RR^3_{\mathbf{x}}\times\RR^*_z)}$, is due to the heavy Kaluza-Klein modes with the mass $m\geq 1$.\\

{\it Proof of Theorem \ref{infbrane}.} According to Theorem \ref{theosp},
we write the solution as
\begin{equation}
\Phi(t,{\mathbf
  x},z)=\lim_{M\rightarrow\infty}\Phi_M(t,{\mathbf
  x},z),\;\;\;\Phi_M(t,{\mathbf
  x},z):=\sum_{\pm}\int_0^{M}\phi_m^{\pm}(t,{\mathbf x})f_m^{\pm}(z)dm,
\label{express}
 \end{equation}
where for $m>0$,  $\phi_m^{\pm}$ is solution of the
Klein-Gordon equation :
\begin{equation}
\partial_t^2\phi_m^{\pm}-\Delta_{\mathbf
  x}\phi_m^{\pm}+m^2\phi_m^{\pm}=0,\;\;
\phi_m^{\pm}(0,\mathbf{x})=\phi_m^{0,\pm}(\mathbf{x}),\;\;
\partial_t\phi_m^{\pm}(0,\mathbf{x})=\phi_m^{1,\pm}(\mathbf{x}).
  \label{exkg}
\end{equation}
Then
$v(t,\mathbf{x}):=\phi_m^{\pm}(t/m,\mathbf{x}/m)$ is solution of
$\partial_t^2v-\Delta_{\mathbf
  x}v+v=0$,
$v(0,\mathbf{x})=v_0(\mathbf{x}):=\phi_m^{0,\pm}(\mathbf{x}/m)$,
$\partial_tv(0,\mathbf{x})=v_1(\mathbf{x}):=\frac{1}{m}\phi_m^{1,\pm}(\mathbf{x}/m)$.
The $L^1-L^{\infty}$ estimate due to Von Wahl (see \cite{vonwahl}) assures
that
\begin{equation*}
\parallel v(t,\mathbf{x})\parallel_{L^{\infty}(\RR^3_{\mathbf{x}})}\leq
C(1+\mid t\mid)^{-\frac{3}{2}}\sum_{\mid \alpha\mid+j\leq 3}\parallel \partial^{\alpha}_{\mathbf{x}}v_j(\mathbf{x})\parallel_{L^1(\RR^3_{\mathbf{x}})}.
\end{equation*}
We deduce that
\begin{equation}
\parallel \phi_m^{\pm}(t,\mathbf{x})\parallel_{L^{\infty}(\RR^3_{\mathbf{x}})}\leq
C(1+m\mid t\mid)^{-\frac{3}{2}}\sum_{\mid \alpha\mid+j\leq 3}m^{3-\mid\alpha\mid-j}\parallel \partial^{\alpha}_{\mathbf{x}}\phi_m^{j,\pm}(\mathbf{x})\parallel_{L^1(\RR^3_{\mathbf{x}})}.
  \label{vonwal}
\end{equation}
(\ref{mup}) and (\ref{mun}) imply that for any $R>0$ there exists
$C_R>0$ such that
\begin{equation}
 0<m\leq 1,\;\; \sup_{\mid z\mid\leq R} \mid
 f_m^{\pm}(z)\mid\leq
C_R\sqrt{m},
\label{estf}
\end{equation}
and we also have :
\begin{equation}
\sup_{1\leq m}\sup_{z\in\RR}\mid f_m^{\pm}(z)\mid<\infty.
  \label{estff}
\end{equation}
We deduce that
\begin{equation}
\begin{split}
\parallel\Phi_M(t,\mathbf{x},z)\parallel_{L^{\infty}(\RR^3_{\mathbf
    x}\times [-R,R]_z)}\leq\\
&C_R\sum_{\pm}\sum_{\mid\alpha\mid+j\leq 3}\int_0^1(1+m\mid
t\mid)^{-\frac{3}{2}}\sqrt{m}\parallel \partial_{\mathbf
  x}^{\alpha}\phi_m^{j,\pm}\parallel_{L^1(\RR^3_{\mathbf x})}dm\\
&+\int_1^M(1+m\mid
t\mid)^{-\frac{3}{2}}m^{3-\mid\alpha\mid-j}\parallel \partial_{\mathbf
  x}^{\alpha}\phi_m^{j,\pm}\parallel_{L^1(\RR^3_{\mathbf x})}dm.
\end{split}
  \label{dekol}
\end{equation}
To estimate the integral for $m\in]0,1]$, we write
\begin{equation*}
\begin{split}
 \partial_{\mathbf
  x}^{\alpha}\phi_m^{j,\pm}(\mathbf{x})&=
\frac{1}{2}\int_0^{\frac{1}{k}\left(\frac{1}{m}-1\right)}
 \left(\partial_{\mathbf
  x}^{\alpha}\Phi_j(\mathbf{x},z)\pm \partial_{\mathbf
  x}^{\alpha}\Phi_j(\mathbf{x},-z)\right)u_{\pm}(z,m)dz\\
&+
\frac{1}{2}\int_{\frac{1}{k}\left(\frac{1}{m}-1\right)}^{M}
 \left(\partial_{\mathbf x}^{\alpha}\Phi_j(\mathbf{x},z)\pm
   \partial_{\mathbf
     x}^{\alpha}\Phi_j(\mathbf{x},-z)\right)u_{\pm}(z,m)dz.
\end{split}
\end{equation*}
(\ref{ump}) and (\ref{umn}) assure that for any
$\alpha\in[0,\frac{1}{2}]$, we have
$$
m(1+z)\leq 1\Rightarrow \mid u_{+}(z,m)\mid\leq
\sqrt m(1+z)^{-\frac{3}{2}}+Cm^{\frac{5}{2}}(1+\mid z\mid)^{\frac{5}{2}},
$$
$$
m(1+z)\leq 1\Rightarrow \mid u_{-}(z,m)\mid\leq
\frac{1}{8}\left(m\right)^{\frac{5}{2}}(1+z)^{\frac{5}{2}}+Cm^2,
$$
hence for any
$\varepsilon\in[0,\frac{1}{2}]$, we have
\begin{equation}
m(1+z)\leq 1\Rightarrow \mid u_{\pm}(z,m)\mid\leq Cm^{\varepsilon}(1+k\mid
z\mid)^{\varepsilon}.
  \label{lege}
\end{equation}
On the other hand (\ref{zup})
implies that 
\begin{equation}
\sup_{m(1+z)\geq 1}\mid u_{\pm}(z,m)\mid\leq
C<\infty.
  \label{ukk}
\end{equation}
Therefore we get for all $0<m\leq 1$ and $0<\varepsilon\leq\frac{1}{2}$ :
\begin{equation*}
\parallel \partial_{\mathbf
  x}^{\alpha}\phi_m^{j,\pm}\parallel_{L^1(\RR^3_{\mathbf x})}\leq
Cm^{\varepsilon}
\parallel(1+\mid z\mid)^{\varepsilon}\partial_{\mathbf{x}}^{\alpha}\Phi_j\parallel_{L^1(\RR^4_{\mathbf{x},z})},
\end{equation*}
hence
\begin{equation}
\int_0^1(1+m\mid
t\mid)^{-\frac{3}{2}}\sqrt{m}\parallel \partial_{\mathbf
  x}^{\alpha}\phi_m^{j,\pm}\parallel_{L^1(\RR^3_{\mathbf x})}dm\leq
\frac{C}{\varepsilon}\mid t\mid^{-\frac{3}{2}}\parallel(1+\mid z\mid)^{\varepsilon}\partial_{\mathbf{x}}^{\alpha}\Phi_j\parallel_{L^1(\RR^4_{\mathbf{x},z})}.
  \label{mimi}
\end{equation}

We now estimate the integral for large $m$. We need a regularization
with respect to $\mathbf{x}$. Let $\left(\theta_n\right)_n\subset
C^{\infty}_0\left(\RR^3_{\mathbf{x}}\right)$ such that $0\leq
\theta_n$, $\mathrm{Supp}\theta_n\subset\{\mid\mathbf{x}\mid\leq1/n\}$,
$\int\theta_n(\mathbf{x})d\mathbf{x}=1$. We introduce
$\Phi_j^n(\mathbf{x},z):=\int\Phi_j(\mathbf{x-y},z)\theta_n(\mathbf{y})d\mathbf{y}$.
We can easily check that $\Phi_j^n$ tends to $\Phi_j$ in
$\mathcal{X}^{1-j}$ as $n$ tends to infinity. Moreover, for all $N$,
$\Delta^N_{\mathbf{x}}\partial_{\mathbf
  x}^{\alpha}\Phi_j^n$ belongs to $\mathfrak{D}(\mathbf{H}_0^{2-j})\cap
\mathfrak{K}^0$. We conclude with a classical
argument of density, that it is sufficient to establish the result when the
data satisfy $\Delta^N_{\mathbf{x}}\partial_{\mathbf
  x}^{\alpha}\Phi_j\in\mathfrak{D}(\mathbf{H}_0^{2-j})\cap
\mathfrak{K}^0$ for all $N$.
In this case, for almost all
$\mathbf{x}\in\RR^3$, the map $z\longmapsto\partial_{\mathbf
  x}^{\alpha}\Phi_j(\mathbf{x},z)\pm \partial_{\mathbf
  x}^{\alpha}\Phi_j(\mathbf{x},-z)$ belongs to $\mathfrak{D}(\mathbf{h}_{\pm}^{2-j})$. Thanks to Lemma
\ref{repspec} we have 
\begin{equation}
 \partial_{\mathbf
  x}^{\alpha}\phi_m^{j,\pm}(\mathbf{x})=m^{-2N}F_{\pm}\left(\mathbf{h}_{\pm}^{2-j}\left[
\partial_{\mathbf
  x}^{\alpha}\Phi_j(\mathbf{x},z)\pm \partial_{\mathbf
  x}^{\alpha}\Phi_j(\mathbf{x},-z)\right]\right)(m),
  \label{flop}
\end{equation}
hence with (\ref{ukk}) we get for $m\geq 1$
\begin{equation*}
\|\partial_{\mathbf
  x}^{\alpha}\phi_m^{j,\pm}\|_{L^1(\RR^3_{\mathbf{x}})}\leq m^{-2N}
\|\mathbf{h}_{\pm}^{2-j}\left[
\partial_{\mathbf
  x}^{\alpha}\Phi_j(\mathbf{x},z)\pm \partial_{\mathbf
  x}^{\alpha}\Phi_j(\mathbf{x},-z)\right]\|_{L^1(\RR^3_{\mathbf{x}}\times ]0,\infty[_z)}.
\end{equation*}
Since
\begin{equation*}
\|\mathbf{h}_{\pm}^{N}\left[
\partial_{\mathbf
  x}^{\alpha}\Phi_j(\mathbf{x},z)\pm \partial_{\mathbf
  x}^{\alpha}\Phi_j(\mathbf{x},-z)\right]\|_{L^1(\RR^3_{\mathbf{x}}\times ]0,\infty[_z)}
\leq C \sum_{l=0}^{4-2j}\|\partial^l_z\partial^{\alpha}_{\mathbf{x}}\Phi_j\|_{L^1(\RR^3_{\mathbf{x}}\times\RR^*_z)},
\end{equation*}
where the $z$-derivatives are taken in the sense of the distributions
in $\RR^4\setminus\{z=0\}$, 
we conclude that

\begin{equation*}
\begin{split}
\int_1^M(1+m\mid
t\mid)^{-\frac{3}{2}}m^{3-\mid\alpha\mid-j}\parallel \partial_{\mathbf
  x}^{\alpha}\phi_m^{j,\pm}\parallel_{L^1(\RR^3_{\mathbf x})}&dm
\leq\\
 C \sum_{l=0}^{4-2j}&\|\partial^l_z\partial^{\alpha}_{\mathbf{x}}\Phi_j\|_{L^1(\RR^3_{\mathbf{x}}\times\RR^*_z)}\int_1^M(1+m\mid
t\mid)^{-\frac{3}{2}}m^{j-1-\mid\alpha\mid}dm.
\end{split}
\end{equation*}
Since
$$
\int_1^{\infty}(1+m\mid
t\mid)^{-\frac{3}{2}}m^{j-1-\mid\alpha\mid}dm\leq C \mid
t\mid^{-\frac{3}{2}},\;\;\mid\alpha\mid+j\leq 3,\;\;j=0,1,
$$
we obtain
\begin{equation}
\sum_{\mid\alpha\mid+j\leq 3}\int_1^{\infty}(1+m\mid
t\mid)^{-\frac{3}{2}}m^{3-\mid\alpha\mid-j}\parallel \partial_{\mathbf
  x}^{\alpha}\phi_m^{j,\pm}\parallel_{L^1(\RR^3_{\mathbf x})}dm
\leq
 C\mid t\mid^{-\frac{3}{2}} \sum_{l+\mid \alpha\mid+j\leq 4}\|\partial^l_z\partial^{\alpha}_{\mathbf{x}}\Phi_j\|_{L^1(\RR^3_{\mathbf{x}}\times\RR^*_z)}.
  \label{momo}
\end{equation}
Now the theorem follows from (\ref{dekol}), (\ref{mimi}) and (\ref{momo}).

\fin

The $L^2-L^{\infty}$ estimates are very useful to the study of
non-linear problems. In the case of the massive Klein-Gordon equation
in the Minkowski space-time, they were initially proved by S. Klainerman
\cite{klainerman} for compactly supported initial data,  refined in
\cite{KGD} and \cite{hormander} and extended to the non compact data
by Georgiev \cite{georgiev}. We establish such estimates for the
Kaluza-Klein solutions. We introduce suitable weighted spaces :
\begin{equation}
\begin{split}
\mathcal{Y}^N_{\varepsilon}:=\left\{\Phi\in{\mathfrak
    K}^0\cap\mathfrak{D}\left(\mathbf{H}_0^{N+1}\right)\right. &;\;\;
\sum_{\mid\alpha\mid\leq N+1}\sum_{0\leq
    p,q}\|\chi_p(\mid\mathbf{y}\mid)\chi_q(\mid z\mid)(1+\mid
  \mathbf{y}\mid)^{\frac{3}{2}}(1+\mid
  z\mid)^{\frac{1}{2}}\partial_{\mathbf{y}}^{\alpha}\Phi_j\|_{L^2(\RR^3_{\mathbf{y}}\times\RR_z)}\\
&+\sum_{\mid\alpha\mid=N+2}\sum_{0\leq
    p,q}\|\chi_p(\mid\mathbf{y}\mid)\chi_q(\mid z\mid)(1+\mid
  \mathbf{y}\mid)^{\frac{3}{2}}(1+\mid
  z\mid)^{\frac{1}{2}+\varepsilon}\partial_{\mathbf{y}}^{\alpha}\Phi_j\|_{L^2(\RR^3_{\mathbf{y}}\times\RR_z)}\\
&\left.+\sum_{l=1}^{N+3}\sum_{\mid\alpha\mid\leq N+3-l}\sum_{0\leq p}\|\chi_p(\mid\mathbf{y}\mid)(1+\mid
  \mathbf{y}\mid)^{\frac{3}{2}}\partial_z^l\partial_{\mathbf{y}}^{\alpha}\Phi_j\|_{L^2(\RR^3_{\mathbf{y}}\times\RR^*_z)}<\infty
\right\}.
\end{split}
  \label{ym}
\end{equation}
 Here $\chi_p$ is
the dyadic partition of the unity on $[0,\infty[$, defined by
\begin{equation*}
\chi_0={\mathbf 1}_{[0,1[},\;\;1\leq
p\Rightarrow \chi_p={\mathbf 1}_{[2^{p-1},2^p[}.
\end{equation*}

\begin{Theorem}
For any $R>0$ there exists $C_R>0$ such that for any
$\varepsilon\in]0,\frac{1}{2}]$, and for all
$\Phi_0\in{\mathcal Y}^1_{\varepsilon}$, $\Phi_1\in{\mathcal Y}^0_{\varepsilon}$,  the solution
$\Phi$ of the Cauchy problem (\ref{EQB}), (\ref{ci}), (\ref{regf})
satisfies :
\begin{equation*}
\begin{split}
\|\Phi(t,\mathbf{x},.)\|_{L^{\infty}([-R,R]_z)}&\leq\\
&
C_R(\mid\mathbf{x}\mid+\mid
t\mid)^{-\frac{3}{2}}\left[\sum_{\mid\alpha\mid+j\leq 2}\sum_{0\leq
    p,q}\|\chi_p(\mid\mathbf{y}\mid)\chi_q(\mid z\mid)(1+\mid
  \mathbf{y}\mid)^{\frac{3}{2}}(1+\mid
  z\mid)^{\frac{1}{2}}\partial_{\mathbf{y}}^{\alpha}\Phi_j\|_{L^2(\RR^3_{\mathbf{y}}\times\RR_z)}\right.\\
&+\frac{1}{\varepsilon}\sum_{\mid\alpha\mid+j=3}\sum_{0\leq
    p,q}\|\chi_p(\mid\mathbf{y}\mid)\chi_q(\mid z\mid)(1+\mid
  \mathbf{y}\mid)^{\frac{3}{2}}(1+\mid
  z\mid)^{\frac{1}{2}+\varepsilon}\partial_{\mathbf{y}}^{\alpha}\Phi_j\|_{L^2(\RR^3_{\mathbf{y}}\times\RR_z)}\\
&\left.+\sum_{l=1}^4\sum_{\mid\alpha\mid+j\leq 4-l}\sum_{0\leq p}\|\chi_p(\mid\mathbf{y}\mid)(1+\mid
  \mathbf{y}\mid)^{\frac{3}{2}}\partial_z^l\partial_{\mathbf{y}}^{\alpha}\Phi_j\|_{L^2(\RR^3_{\mathbf{y}}\times\RR^*_z)}\right].
\end{split}
\end{equation*}
  \label{theollde}
\end{Theorem}
Like in the previous theorem, the term with the weight $(1+\mid
z\mid)^{\frac{1}{2}+\varepsilon}$ is used to control the contribution
of the light Kaluza-Klein modes, while the $z$-derivatives are useful
to estimate the heavy modes.\\

{\it Proof of Theorem \ref{theollde}.} We start with expression
(\ref{express}). We have to control carefully  the dependence of the decay estimates
with respect to the mass, so we need the following :

\begin{Lemma} There exists $C>0$ such that for any $m>0$, the solution
  $\phi_m$ of the Klein-Gordon equation  $\partial_t^2\phi_m-\Delta_{\mathbf
  x}\phi_m+m^2\phi_m=0$,
$\phi_m(0,\mathbf{x})=\phi_m^{0}(\mathbf{x})$,
$\partial_t\phi_m(0,\mathbf{x})=\phi_m^{1}(\mathbf{x})$, satisfies for
all $t\in\RR$, $\mathbf{x}\in\RR^3$ :
\begin{equation*}
\begin{split}
\mid\phi_m(t,\mathbf{x})\mid
&\leq
C\left(1+\frac{1}{m}\right)(1+\mid
t\mid+\mid\mathbf{x}\mid)^{-\frac{3}{2}}\sum_{\mid\alpha\mid+j\leq
  3}\parallel(1+\mid\mathbf{y}\mid)^{\frac{3}{2}}
\partial_{\mathbf{y}}^{\alpha}\phi_m^j(\mathbf{y})\parallel_{L^2(\RR^3_{\mathbf{y}})}
\\
&+C(1+m\mid
t\mid+m\mid\mathbf{x}\mid)^{-\frac{3}{2}}\sum_{\mid\alpha\mid+j\leq
  3}\sum_{p=0}^{\infty}m^{3-\mid\alpha\mid-j}\parallel
\chi_p(\mid\mathbf{y}\mid)(1+\mid\mathbf{y}\mid)^{\frac{3}{2}}\partial_{\mathbf{y}}^{\alpha}\phi_m^j(\mathbf{y})\parallel_{L^2(\RR^3_{\mathbf{y}})},
\end{split}
\end{equation*}
provided the norms of the right-hand side are finite.
  \label{llde}
\end{Lemma}

{\it Proof of Lemma \ref{llde}.}
We follow closely the method employed by Georgiev \cite{georgiev}. By the Sobolev inequality on $\RR^3$, there exists $C>0$ such that for
any $t\in\RR$, $\mathbf{x}\in\RR^3$ we have :
\begin{equation}
\mid\phi_m(t,\mathbf{x})\mid\leq C\sum_{\mid\alpha\mid\leq 2}\|\partial^{\alpha}_{\mathbf{y}}\phi_m(t,\mathbf{y})\|_{L^2(\RR^3_{\mathbf{y}})}.
  \label{sobo}
\end{equation}
We can control the $L^2$-norm of the right-hand side by using the
energy equality
\begin{equation*}
\begin{split}
\int_{\RR^3}\mid\partial_t\partial_{\mathbf{y}}^{\alpha}\phi_m(t,\mathbf{y})\mid^2+\mid\nabla_{\mathbf{y}}\partial_{\mathbf{y}}^{\alpha}\phi_m(t,\mathbf{y})\mid^2&+m^2\mid\partial_{\mathbf{y}}^{\alpha}\phi_m(t,\mathbf{y})\mid^2d\mathbf{y}\\
=&\int_{\RR^3}\mid\partial_{\mathbf{y}}^{\alpha}\phi_m^1(\mathbf{y})\mid^2+\mid\nabla_{\mathbf{y}}\partial_{\mathbf{y}}^{\alpha}\phi_m^0(\mathbf{y})\mid^2+m^2\mid\partial_{\mathbf{y}}^{\alpha}\phi_m^0(\mathbf{y})\mid^2d\mathbf{y},
\end{split}
\end{equation*}
to get
\begin{equation*}
\mid\phi_m(t,\mathbf{x})\mid\leq C\left(1+\frac{1}{m}\right)\sum_{\mid\alpha\mid+j\leq 3}\|\partial^{\alpha}_{\mathbf{y}}\phi_m^j(\mathbf{y})\|_{L^2(\RR^3_{\mathbf{y}})}.
\end{equation*}
Now we choose some function $\chi\in C^{\infty}_0(\RR^3)$ such that
$\chi(\mathbf{y})=0$ when $\mid\mathbf{y}\mid\geq 2$ and
$\chi(\mathbf{y})=1$ when $\mid\mathbf{y}\mid\leq 2$. Given $t\in\RR$
and $\mathbf{x}\in \RR^3$ fixed, we consider the
solution $\psi(s,\mathbf{y})$ of
$\partial_s^2\psi-\Delta_{\mathbf{y}}\psi+m^2\psi=0$,
$\psi(0,\mathbf{y})=\chi\left(\frac{\mathbf{y}-\mathbf{x}}{1+\mid t\mid}\right)\phi_m^0(\mathbf{y})$,
$\partial_s\psi(0,\mathbf{y})=\chi\left(\frac{\mathbf{y}-\mathbf{x}}{1+\mid
    t\mid}\right)\phi_m^1(\mathbf{y})$. The finite dependence domain
argument implies that $\phi_m(t,\mathbf{x})=\psi(t,\mathbf{x})$, thus
by applying the previous estimate to $\psi$ we obtain :
\begin{equation*}
\mid\phi_m(t,\mathbf{x})\mid\leq
C\left(1+\frac{1}{m}\right)\sum_{\mid\alpha\mid+j\leq
  3}\|\partial^{\alpha}_{\mathbf{y}}\phi_m^j(\mathbf{y})\|_{L^2(\mid\mathbf{y}-\mathbf{x}\mid\leq 2+2\mid t\mid)}.
\end{equation*}
If $\mid\mathbf{x}\mid\geq 4\mid t\mid+4$, and
$\mid\mathbf{y}-\mathbf{x}\mid\leq 2+2\mid t\mid$, we have
$\mid\mathbf{y}\mid\geq\frac{1}{2}\mid\mathbf{x}\mid$, hence
\begin{equation*}
\mid\mathbf{x}\mid\geq 4\mid t\mid+4\Rightarrow
(1+\mid\mathbf{x}\mid+\mid
t\mid)^{\frac{3}{2}}\mid\phi_m(t,\mathbf{x})\mid\leq
C\left(1+\frac{1}{m}\right)\sum_{\mid\alpha\mid+j\leq 3}\|(1+\mid\mathbf{y}\mid)^{\frac{3}{2}}\partial^{\alpha}_{\mathbf{y}}\phi_m^j(\mathbf{y})\|_{L^2(\RR^3_{\mathbf{y}})}.
\end{equation*}
When $\mid\mathbf{x}\mid\leq 4\mid t\mid+4$ and $\mid t\mid \leq 1$,
the Sobolev inequality (\ref{sobo}) gives the result. Finally, when
$\mid\mathbf{x}\mid\leq 4\mid t\mid+4$ and $\mid t\mid \geq 1$, we use
the Von Wahl estimate (\ref{vonwal}) and the inequality
\begin{equation*}
\|\phi \|_{L^1(\RR^3)}\leq\sqrt{4\pi\log(2)}\sum_{k=0}^{\infty}\|\chi_k(\mid\mathbf{y}\mid)(1+\mid\mathbf{y}\mid)^{\frac{3}{2}}\phi(\mathbf{y})\|_{L^2(\RR^3)}.
\end{equation*}

\fin

Now we use Lemma \ref{llde} and the estimates
(\ref{estf}) and (\ref{estff}), to get from (\ref{express}) :
\begin{equation*}
\begin{split}
\parallel\Phi(t,\mathbf{x},.)\parallel_{L^{\infty}([-R,R]_z)}\leq C_R&\sum_{\pm}\sum_{\mid\alpha\mid+j\leq 3}(1+\mid\mathbf{x}\mid+\mid
t\mid)^{-\frac{3}{2}}
\int_0^1\frac{1}{\sqrt m}\|(1+\mid\mathbf{y}\mid)^{\frac{3}{2}}\partial_{\mathbf{y}}^{\alpha}\phi^{j,\pm}_m\|_{L^2(\RR^3_{\mathbf{y}})}dm\\
+&\sum_{p=0}^{\infty}
\int_0^1(1+m\mid\mathbf{x}\mid+m\mid
t\mid)^{-\frac{3}{2}}m^{\frac{7}{2}-\mid\alpha\mid-j}\|\chi_p(\mid\mathbf{y}\mid)(1+\mid\mathbf{y}\mid)^{\frac{3}{2}}\partial_{\mathbf{y}}^{\alpha}\phi^{j,\pm}_m\|_{L^2(\RR^3_{\mathbf{y}})}dm\\
+&(1+\mid\mathbf{x}\mid+\mid
t\mid)^{-\frac{3}{2}}
\int_1^{\infty}\|(1+\mid\mathbf{y}\mid)^{\frac{3}{2}}\partial_{\mathbf{y}}^{\alpha}\phi^{j,\pm}_m\|_{L^2(\RR^3_{\mathbf{y}})}dm\\
+&\sum_{p=0}^{\infty}
\int_1^{\infty}(1+m\mid\mathbf{x}\mid+m\mid
t\mid)^{-\frac{3}{2}}m^{3-\mid\alpha\mid-j}\|\chi_p(\mid\mathbf{y}\mid)(1+\mid\mathbf{y}\mid)^{\frac{3}{2}}\partial_{\mathbf{y}}^{\alpha}\phi^{j,\pm}_m\|_{L^2(\RR^3_{\mathbf{y}})}dm.
\end{split}
\end{equation*}
To estimate the integral on the light mass, we use (\ref{lege}) and
(\ref{ukk}). If $P$ is a partial differential operator on
$\RR^3_{\mathbf x}$, we obtain
 for any $0\leq\epsilon$ and
$0<m\leq 1$ 
\begin{equation*}
\| P\phi^{j,\pm}_m\|_{L^2(\RR^3_{\mathbf{y}})}\leq
Cm^{\epsilon}\|(1+k\mid
z\mid)^{\epsilon}P\Phi_j\|_{L^1(\RR_z;L^2(\RR^3_{\mathbf{y}}))}\leq
Cm^{\epsilon}\sum_{p=0}^{\infty}\|\chi_p(\mid z\mid)(1+\mid
z\mid)^{\frac{1}{2}+\epsilon}P\Phi_j\|_{L^2(\RR^3_{\mathbf{y}}\times\RR_z)}.
\end{equation*}
For $P=(1+\mid\mathbf{y}\mid)^{\frac{3}{2}}\partial_{\mathbf{x}}^{\alpha}$, we take $\epsilon=0$ and we
get :
\begin{equation}
\int_0^1\frac{1}{\sqrt
  m}\|\partial_{\mathbf{x}}^{\alpha}\phi^{j,\pm}_m\|_{L^2(\RR^3_{\mathbf{y}})}dm\leq C \sum_{p=0}^{\infty}\|\chi_p(\mid z\mid)(1+\mid\mathbf{y}\mid)^{\frac{3}{2}}(1+\mid
z\mid)^{\frac{1}{2}}\partial^{\alpha}_{\mathbf{x}}\Phi_j\|_{L^2(\RR^3_{\mathbf{y}}\times\RR_z)}.
  \label{estun}
\end{equation}
For
$P=\chi_p(\mathbf{y})(1+\mid\mathbf{y}\mid)^{\frac{3}{2}}\partial_{\mathbf{y}}^{\alpha}$
with $\mid\alpha\mid+j\leq 2$,
we take $\epsilon=0$ again and we obtain :
\begin{equation*}
\begin{split}
\int_0^1(1+m\mid\mathbf{x}\mid+m\mid
t\mid)^{-\frac{3}{2}}m^{\frac{7}{2}-\mid\alpha\mid-j}\|\chi_p(\mid\mathbf{y}\mid)(1+\mid\mathbf{y}\mid)^{\frac{3}{2}}\partial_{\mathbf{y}}^{\alpha}\phi^{j,\pm}_m\|_{L^2(\RR^3_{\mathbf{y}})}dm\leq\\
C(\mid\mathbf{x}\mid+\mid
t\mid)^{-\frac{3}{2}}\sum_{q=0}^{\infty}\|\chi_q(\mid
z\mid)\chi_p(\mid\mathbf{y}\mid)(1+\mid
z\mid)^{\frac{1}{2}}(1+\mid\mathbf{y}\mid)^{\frac{3}{2}}\partial_{\mathbf{y}}^{\alpha}\Phi_j\|_{L^2(\RR^3_{\mathbf{y}}\times\RR_z)}.
\end{split}
\end{equation*}
Finally for $P=\chi_p(\mathbf{y})(1+\mid\mathbf{y}\mid)^{\frac{3}{2}}\partial_{\mathbf{y}}^{\alpha}$
with $\mid\alpha\mid+j=3$,
we take $\epsilon>0$  and we obtain :
\begin{equation*}
\begin{split}
\int_0^1(1+m\mid\mathbf{x}\mid+m\mid
t\mid)^{-\frac{3}{2}}m^{\frac{7}{2}-\mid\alpha\mid-j}\|\chi_p(\mid\mathbf{y}\mid)(1+\mid\mathbf{y}\mid)^{\frac{3}{2}}\partial_{\mathbf{y}}^{\alpha}\phi^{j,\pm}_m\|_{L^2(\RR^3_{\mathbf{y}})}dm\leq\\
\frac{C}{\epsilon}(\mid\mathbf{x}\mid+\mid
t\mid)^{-\frac{3}{2}}\sum_{q=0}^{\infty}\|\chi_q(\mid
z\mid)\chi_p(\mid\mathbf{y}\mid)(1+\mid
z\mid)^{\frac{1}{2}+\epsilon}(1+\mid\mathbf{y}\mid)^{\frac{3}{2}}\partial_{\mathbf{y}}^{\alpha}\Phi_j\|_{L^2(\RR^3_{\mathbf{y}}\times\RR_z)}.
\end{split}
\end{equation*}

To estimate the integrals on the large mass, $m\geq 1$, we use the
procedure of regularization employed in the proof of the previous
theorem. Thanks to (\ref{flop}) and Lemma \ref{repspec}, we have
\begin{equation*}
\begin{split}
\int_1^{\infty}\|(1+\mid\mathbf{y}\mid)^{\frac{3}{2}}\partial_{\mathbf{x}}^{\alpha}\phi^{j,\pm}_m\|_{L^2(\RR^3_{\mathbf{y}})}dm
&\leq \int_1^{\infty}\frac{1}{m^2}\|(1+\mid\mathbf{y}\mid)^{\frac{3}{2}}F_{\pm}\left(\mathbf{h}_{\pm}\left[
\partial_{\mathbf
  y}^{\alpha}\Phi_j(\mathbf{y},z)\pm \partial_{\mathbf
  y}^{\alpha}\Phi_j(\mathbf{y},-z)\right]\right)(m)\|_{L^2(\RR^3_{\mathbf{y}})}dm\\
&\leq\|(1+\mid\mathbf{y}\mid)^{\frac{3}{2}}F_{\pm}\left(\mathbf{h}_{\pm}\left[
\partial_{\mathbf
  y}^{\alpha}\Phi_j(\mathbf{y},z)\pm \partial_{\mathbf
  y}^{\alpha}\Phi_j(\mathbf{y},-z)\right]\right)(m)\|_{L^2(\RR^3_{\mathbf{y}}\times\RR^+_m)}\\
&\leq \|(1+\mid\mathbf{y}\mid)^{\frac{3}{2}}\mathbf{h}_{\pm}\left[
\partial_{\mathbf
  y}^{\alpha}\Phi_j(\mathbf{y},z)\pm \partial_{\mathbf
  y}^{\alpha}\Phi_j(\mathbf{y},-z)\right]\|_{L^2(\RR^3_{\mathbf{y}}\times \RR^+_z)}\\
&\leq C\sum_{l=0}^2\|(1+\mid\mathbf{y}\mid)^{\frac{3}{2}}\partial_z^l
\partial_{\mathbf
  y}^{\alpha}\Phi_j(\mathbf{y},z)\|_{L^2(\RR^3_{\mathbf{y}}\times \RR_z^*)}.
\end{split}
 \end{equation*}
At last we evaluate for $j\leq 1+\mid\alpha\mid$, $j=0,1$,
\begin{equation*}
\begin{split}
\int_1^{\infty}&(1+m\mid\mathbf{x}\mid+m\mid
t\mid)^{-\frac{3}{2}}m^{3-\mid\alpha\mid-j}\|\chi_p(\mid\mathbf{y}\mid)(1+\mid\mathbf{y}\mid)^{\frac{3}{2}}\partial_{\mathbf{y}}^{\alpha}\phi^{j,\pm}_m\|_{L^2(\RR^3_{\mathbf{y}})}dm\leq\\
&\int_1^{\infty}\frac{m^{j-1-\mid\alpha\mid} }{(1+m\mid\mathbf{x}\mid+m\mid
t\mid)^{\frac{3}{2}} }\|F_{\pm}\left(\chi_p(\mid\mathbf{y}\mid)(1+\mid\mathbf{y}\mid)^{\frac{3}{2}}\mathbf{h}_{\pm}^{2-j}\left[
\partial_{\mathbf
  y}^{\alpha}\Phi_j(\mathbf{y},z)\pm \partial_{\mathbf
  y}^{\alpha}\Phi_j(\mathbf{y},-z)\right]\right)(m)
\|_{L^2(\RR^3_{\mathbf{y}})}dm\\
&\leq C(\mid\mathbf{x}\mid+\mid t\mid)^{-\frac{3}{2}}\|F_{\pm}\left(\chi_p(\mid\mathbf{y}\mid)(1+\mid\mathbf{y}\mid)^{\frac{3}{2}}\mathbf{h}_{\pm}^{2-j}\left[
\partial_{\mathbf
  y}^{\alpha}\Phi_j(\mathbf{y},z)\pm \partial_{\mathbf
  y}^{\alpha}\Phi_j(\mathbf{y},-z)\right]\right)(m)
\|_{L^2(\RR^3_{\mathbf{y}}\times\RR^+_m)}\\
&\leq C(\mid\mathbf{x}\mid+\mid t\mid)^{-\frac{3}{2}}\sum_{l=0}^{4-2j}\|\chi_p(\mid\mathbf{y}\mid)(1+\mid\mathbf{y}\mid)^{\frac{3}{2}}\partial^l_z
\partial_{\mathbf
  y}^{\alpha}\Phi_j(\mathbf{y},z)
\|_{L^2(\RR^3_{\mathbf{y}}\times\RR^*_z)}.
\end{split}
\end{equation*}
\fin

We know that Strichartz has proved in \cite{strichartz} that for
suitable initial data, the solutions of the D'Alembertian in the
Minkowski space-time $\RR^{1+3}$ belong to $L^4(\RR^4)$, and the
solutions of the massive Klein-Gordon equation belong to $L^q(\RR^4)$,
$\frac{10}{3}\leq q\leq 4$. Moreover the massless fields in
the Minkowski space-time $\RR^{1+4}$ belong to
$L^{\frac{10}{3}}\left(\RR^{1+4}\right)$. Therefore, we expect that
near the brane, the solution of the master equation are in
$L^{\infty}\left([-R,R]_z;L^{4}\left(\RR_t\times\RR^3_{\mathbf{x}}\right)\right)$, and
the Kaluza-Klein tower that is more dispersive because of the mass,  is in 
$L^{\frac{10}{3}}\left(\RR_t\times\RR^3_{\mathbf{x}}\times[-R,R]_z\right)$.
This is indeed the case.\\

\begin{Theorem}
There exists $C>0$ and for any $R>0$, some $C_R>0$, such that the solution
$\Phi$ of the Cauchy problem (\ref{EQB}), (\ref{ci}), (\ref{regf})
satisfies the following inequalities. When $\Phi_0\in{\mathfrak
  K}^1\cap\mathfrak{D}(\mathbf{H}_0^{\frac{1}{4}})$,
$\Phi_1\in{\mathfrak
  K}^0\cap\mathfrak{D}(\mathbf{H}_0^{-\frac{1}{4}})$ we have :

\begin{equation}
\parallel\Phi\parallel_{L^{\frac{10}{3}}(\RR_t\times\RR^3_{\mathbf{x}}\times[-R,R]_z)}\leq
C_R\sum_{j=0,1}\parallel{\mathbf
  H}_0^{\frac{1}{2}(\frac{1}{2}-j)}\Phi_j  \parallel_{L^2(\RR^4)}.
\label{striun}
\end{equation}
When
$\Phi_j\in\mathfrak{K}^0\cap\mathfrak{D}(\mathbf{H}_0^{\frac{1}{2}(1-j)+\varepsilon})$
for some $\varepsilon>0$,
 we have :
\begin{equation}
\parallel\Phi\parallel_{L^{\infty}([-R,R]_z;L^4(\RR_t\times\RR^3_{\mathbf{x}}))}\leq
\frac{C_R}{\sqrt{\varepsilon}}\sum_{j=0,1}\parallel{\mathbf
  H}_0^{\frac{1}{2}(1-j)+\varepsilon}\Phi_j  \parallel_{L^2(\RR^4)}.
\label{stride}
\end{equation}
When $\Phi$ is a massless graviton, we have
\begin{equation}
\parallel\Phi\parallel_{L^{\infty}\left(\RR_z;L^4\left(\RR_t\times\RR^3_{\mathbf{x}}\right)\right)}\leq
C\sum_{j=0,1}\parallel{\mathbf
  H}_0^{\frac{1}{2}(\frac{1}{2}-j)}\Phi_j  \parallel_{L^2(\RR^4)}.
\label{strig}
\end{equation}
  \label{stribrane}
\end{Theorem}

{\it Proof of Theorem \ref{stribrane}.}
The global $L^q(\RR_t\times\RR^3_{\mathbf{x}})$ estimate due to Strichartz \cite{strichartz} assures
that the solution $v$ of the Klein-Gordon equation with the mass equal
to 1 in $\RR_t\times\RR^3_{\mathbf{x}}$, solution of $\partial_t^2v-\Delta_{\mathbf
  x}v+v=0$,
$v(0,\mathbf{x})=v_0(\mathbf{x})$,
$\partial_tv(0,\mathbf{x})=v_1(\mathbf{x})$, satisfies
\begin{equation*}
\frac{10}{3}\leq q\leq 4,\;\;\parallel v(t,\mathbf{x})\parallel_{L^q(\RR_t\times\RR^3_{\mathbf{x}})}\leq
C\sum_{j=0,1}\parallel \left(1-\Delta_{\mathbf{x}}\right)^{\frac{1}{2}(\frac{1}{2}-j)}v_j(\mathbf{x})\parallel_{L^2(\RR^3_{\mathbf{x}})}.
\end{equation*}
Therefore we deduce that the solution  $\phi_m$  of the
Klein-Gordon equation with mass $m>0$,  $\partial_t^2\phi_m-\Delta_{\mathbf
  x}\phi_m+m^2\phi_m=0$,
$\phi_m(0,\mathbf{x})=\phi_m^{0}(\mathbf{x})$,
$\partial_t\phi_m(0,\mathbf{x})=\phi_m^{1}(\mathbf{x})$, satisfies :
\begin{equation}
\frac{10}{3}\leq q\leq 4,\;\;\parallel \phi_m(t,\mathbf{x})\parallel_{L^q(\RR_t\times\RR^3_{\mathbf{x}})}\leq
Cm^{1-\frac{4}{q}}\sum_{j=0,1}\parallel \left(m^2-\Delta_{\mathbf{x}}\right)^{\frac{1}{2}(\frac{1}{2}-j)}\phi_m^j(\mathbf{x})\parallel_{L^2(\RR^3_{\mathbf{x}})}.
  \label{strim}
\end{equation}
Concerning the massless free wave $\phi_0$, solution of $\partial_t^2\phi_0-\Delta_{\mathbf{x}}\phi_0=0$, we also have :
\begin{equation}
\parallel \phi_0(t,\mathbf{x})\parallel_{L^4(\RR_t\times\RR^3_{\mathbf{x}})}\leq
C\sum_{j=0,1}\parallel \left(-\Delta_{\mathbf{x}}\right)^{\frac{1}{2}(\frac{1}{2}-j)}\phi_0^j(\mathbf{x})\parallel_{L^2(\RR^3_{\mathbf{x}})}.
  \label{strimo}
\end{equation}
 We use the representation (\ref{express}), (\ref{exkg}). We deduce from
(\ref{zup}), (\ref{fufu}) and 
(\ref{estf})  that for any $z\in[-R,R]$ we have for any $M>1$ :
\begin{equation*}
\begin{split}
\|\Phi_M(.,.,z)\|_{L^q(\RR_t\times\RR^3_{\mathbf{x}})}\leq
C_R\left[\sum_{\pm}\right.&\int_0^1
\sqrt{m}\|\phi_m^{\pm}\|_{L^q(\RR_t\times\RR^3_{\mathbf{x}})}dm
+\left\|\int_1^{M}\phi_m^{\pm}g^{\pm}(z,m)dm\right\|_{L^q(\RR_t\times\RR^3_{\mathbf{x}})}\\
&\left.+\int_1^{M}\frac{1}{m}\|\phi_m^{\pm}\|_{L^q(\RR_t\times\RR^3_{\mathbf{x}})}dm\right],
\end{split}
\end{equation*}
where $g^+(z,m)=\sqrt{\frac{2}{\pi}}\cos(mz)$, 
$g^-(z,m)=\sqrt{\frac{2}{\pi}}\sin(mz)$.
We use (\ref{strim}) with $q=\frac{10}{3}$ to get
\begin{equation*}
\begin{split}
\int_0^1
\sqrt{m}\|\phi_m^{\pm}\|_{L^{\frac{10}{3}}(\RR_t\times\RR^3_{\mathbf{x}})}dm
+
\int_1^{M}\frac{1}{m}&\|\phi_m^{\pm}\|_{L^{\frac{10}{3}}(\RR_t\times\RR^3_{\mathbf{x}})}dm\leq\\
&C\sum_{j=0,1}\left(\int_0^{M}\parallel
  \left(m^2-\Delta_{\mathbf{x}}\right)^{\frac{1}{2}(\frac{1}{2}-j)}\phi_m^{j,\pm}(\mathbf{x})\parallel_{L^2(\RR^3_{\mathbf{x}})}^2dm\right)^{\frac{1}{2}}.
\end{split}
\end{equation*}
Since Lemma \ref{mulspec} assures that the distorded Fourier
transforms are isometries, we have :
\begin{equation*}
\begin{split}
\sum_{\pm}\int_0^{\infty}\parallel
\left(m^2-\Delta_{\mathbf{x}}\right)^{\frac{1}{2}(\frac{1}{2}-j)}&\phi_m^{j,\pm}(\mathbf{x})\parallel_{L^2(\RR^3_{\mathbf{x}})}^2dm\\
&=\frac{1}{4}
\sum_{\pm}\int_0^{\infty}\parallel
\left(\mathbf{h}_{\pm}-\Delta_{\mathbf{x}}\right)^{\frac{1}{2}(\frac{1}{2}-j)}\left[\Phi_j(\mathbf{x},z)\pm\Phi_j(\mathbf{x},-z)\right]\parallel_{L^2(\RR^3_{\mathbf{x}})}^2dz\\
&=
\int_{-\infty}^{\infty}\parallel
\left(\mathbf{h}-\Delta_{\mathbf{x}}\right)^{\frac{1}{2}(\frac{1}{2}-j)}\Phi_j(\mathbf{x},z)\parallel_{L^2(\RR^3_{\mathbf{x}})}^2dz
=\frac{1}{2}\left\|
  \mathbf{H}_0^{\frac{1}{2}\left(\frac{1}{2}-j\right)}\Phi_j\right\|_{L^2(\RR^4_{\mathbf{x},z})}^2.
\end{split}
\end{equation*}
We conclude that
\begin{equation*}
\sum_{\pm}\int_0^1
\sqrt{m}\|\phi_m^{\pm}\|_{L^{\frac{10}{3}}(\RR_t\times\RR^3_{\mathbf{x}})}dm
+
\int_1^{\infty}\frac{1}{m}\|\phi_m^{\pm}\|_{L^{\frac{10}{3}}(\RR_t\times\RR^3_{\mathbf{x}})}dm\leq
C\sum_{j=0,1}\left\|
  \mathbf{H}_0^{\frac{1}{2}\left(\frac{1}{2}-j\right)}\Phi_j\right\|_{L^2(\RR^4_{\mathbf{x},z})}.
\end{equation*}

To estimate the third term, we remark that 
$\int_1^{M}\phi_m^{\pm}(t,\mathbf{x})g^{\pm}(z,m)dm$ is solution
of the free wave equation in the Minkowski space-time
$\RR_t\times\RR^4_{\mathbf{x},z}$, thus the Strichartz estimate
assures that 
\begin{equation*}
\| \int_1^{M}\phi_m^{\pm}(t,\mathbf{x})g^{\pm}(z,m)dm \|_{L^{\frac{10}{3}}(\RR_t\times\RR^4_{\mathbf{x},z})}\leq
C \sum_{j=0,1}\left\|
  \left(-\Delta_{\mathbf{x}}-\partial_z^2\right)^{\frac{1}{2}\left(\frac{1}{2}-j\right)}\int_1^{M}\phi_m^{j,\pm}(\mathbf{x})g^{\pm}(z,m)dm\right\|_{L^2(\RR^4_{\mathbf{x},z})}.
\end{equation*}
Taking account of the expression of $g^{\pm}$, we have 
$$
 \left(-\Delta_{\mathbf{x}}-\partial_z^2\right)^{\frac{1}{2}\left(\frac{1}{2}-j\right)}\int_1^{M}\phi_m^{j,\pm}(\mathbf{x})g^{\pm}(z,m)dm= \int_1^M\left(-\Delta_{\mathbf{x}}+m^2\right)^{\frac{1}{2}\left(\frac{1}{2}-j\right)}\phi_m^{j,\pm}(\mathbf{x})g^{\pm}(z,m)dm,
$$
hence we get by  Lemma \ref{mulspec}
that for almost all $\mathbf{x}$ fixed in $\RR^3$ :
\begin{equation*}
\begin{split}
\left\|
  \left(-\Delta_{\mathbf{x}}-\partial_z^2\right)^{\frac{1}{2}\left(\frac{1}{2}-j\right)}\int_1^{M}\phi_m^{j,\pm}(\mathbf{x})g^{\pm}(z,m)dm\right\|_{L^2(\RR_{z})}
&=\left\|
  \left(-\Delta_{\mathbf{x}}+m^2\right)^{\frac{1}{2}\left(\frac{1}{2}-j\right)}\phi_m^{j,\pm}(\mathbf{x})\right\|_{L^2([1,M]_m)}\\
&\leq \left\|
  \left(-\Delta_{\mathbf{x}}+m^2\right)^{\frac{1}{2}\left(\frac{1}{2}-j\right)}\phi_m^{j,\pm}(\mathbf{x})\right\|_{L^2(\RR^+_m)}\\
&\leq \left\|
  \left(-\Delta_{\mathbf{x}}+\mathbf{h}_{\pm}\right)^{\frac{1}{2}\left(\frac{1}{2}-j\right)}\left[\Phi_j(\mathbf{x},z)\pm\Phi_j(\mathbf{x},-z)\right]\right\|_{L^2(\RR^+_z)}\\
&\leq C  \left\|
  {\mathbf{H}}_0^{\frac{1}{2}\left(\frac{1}{2}-j\right)}\Phi_j(\mathbf{x},.)\right\|_{L^2(\RR^+_z)}.
\end{split}
\end{equation*}
Finally we take the $L^2(\RR^3_{\mathbf{x}})$-norm and we get 
\begin{equation*}
\sup_{M>1}\| \int_1^{M}\phi_m^{\pm}(t,\mathbf{x})g^{\pm}(z,m)dm
\|_{L^{\frac{10}{3}}(\RR_t\times\RR^4_{\mathbf{x},z})}\leq
C\sum_{j=0,1}\left\|
  \mathbf{H}_0^{\frac{1}{2}\left(\frac{1}{2}-j\right)}\Phi_j\right\|_{L^2(\RR^4_{\mathbf{x},z})},
\end{equation*}
and the proof of (\ref{striun}) is complete.

The proof of the $L^{\infty}(L^4)$ estimate for the Kaluza-Klein tower is similar. When
$\Phi_j\in\mathfrak{K}^0\cap\mathfrak{D}(\mathbf{H}_0^{\frac{1}{2}(1-j)+\varepsilon})$,
 we deduce from 
(\ref{estf}), (\ref{estff}) and (\ref{strim}) that for any $z\in[-R,R]$ we have for any $M>1$ :
\begin{equation*}
\begin{split}
\|\Phi_M(.,.,z)\|_{L^4(\RR_t\times\RR^3_{\mathbf{x}})}&\leq
C_R\sum_{\pm}\int_0^M
\|\phi_m^{\pm}\|_{L^4(\RR_t\times\RR^3_{\mathbf{x}})}dm\\
&\leq C_R\sum_{j=0,1}\int_0^{1}\parallel
  \left(m^2-\Delta_{\mathbf{x}}\right)^{\frac{1}{2}(\frac{1}{2}-j)}\phi_m^{j,\pm}(\mathbf{x})\parallel_{L^2(\RR^3_{\mathbf{x}})}dm\\
&+\int_1^{M}\parallel
  \left(m^2-\Delta_{\mathbf{x}}\right)^{\frac{1}{2}(1-j)+\varepsilon}\phi_m^{j,\pm}(\mathbf{x})\parallel_{L^2(\RR^3_{\mathbf{x}})}m^{-\frac{1}{2}-2\varepsilon}dm\\
&\leq\frac{C_R}{\sqrt\varepsilon}\sum_{j=0,1}\left(\int_0^{\infty}\parallel
  \left(m^2-\Delta_{\mathbf{x}}\right)^{\frac{1}{2}(1-j)+\varepsilon}\phi_m^{j,\pm}(\mathbf{x})\parallel_{L^2(\RR^3_{\mathbf{x}})}^2dm\right)^{\frac{1}{2}}\\
&\leq \frac{C_R}{\sqrt\varepsilon}\sum_{j=0,1}\left\|
  \mathbf{H}_0^{\frac{1}{2}\left(1-j\right)+\varepsilon}\Phi_j\right\|_{L^2(\RR^4_{\mathbf{x},z})},
\end{split}
\end{equation*}
hence (\ref{stride}) is established. At last, estimate
(\ref{strig}) for a massless graviton with initial data
$\Phi_j(\mathbf{x},z)=\phi_j(\mathbf{x})\otimes f_0(z)$ is a direct
consequence of  (\ref{strimo}) since $\left(-\Delta_{\mathbf{x}}\right)^{\frac{1}{2}(\frac{1}{2}-j)}\phi_j\otimes
  f_0= \mathbf{H}_0^{\frac{1}{2}\left(\frac{1}{2}-j\right)}\Phi_j$.

\fin



\section{Brane-World Scattering}

In this section we prove that the finite energy solutions of
(\ref{EQB}) with data in $\mathfrak{K}^1\times  \mathfrak{K}^0$, that
are the Kaluza-Klein towers, are
asymptotic to free waves $\Phi_{\infty}^{\pm}$ with finite energy on the
five-dimensional Minkowski space-time :
\begin{equation}
\left(\partial^2_t-\Delta_{\mathbf
    x}-\partial^2_z\right)\Phi_{\infty}^{\pm}=0,\;\;(t,\mathbf{x},z)\in\RR\times\RR^3\times\RR,
  \label{freq}
\end{equation}
\begin{equation}
\Phi_{\infty}^{\pm}\in C^0\left(\RR_t;BL^1\left(\RR^4_{{\mathbf x},z}\right)\right),\;\;\partial_t\Phi_{\infty}^{\pm}\in C^0\left(\RR_t;L^2\left(\RR^4_{{\mathbf x},z}\right)\right).
  \label{freg}
\end{equation}

\begin{Theorem}
Any finite energy solution $\Phi$ with initial data
$\Phi_0\in{\mathfrak K}^1$, $\Phi_1\in{\mathfrak K}^0$, satisfies the
decay estimate (\ref{kk}) for any $a,b$, and there exist unique  free waves
$\Phi_{\infty}^{\pm}$ satisfying (\ref{freq}), (\ref{freg}) and 
\begin{equation}
\lim_{t\rightarrow\pm\infty}
\parallel\partial_t\Phi(t)- \partial_t\Phi_{\infty}^{\pm}(t)\parallel_{L^2(\RR^4_{{\mathbf x},z})}=0.
  \label{ast}
\end{equation}
Moreover, these fields satisfy :
\begin{equation}
\lim_{t\rightarrow\pm\infty}
\parallel \nabla_{t,\mathbf x}\Phi(t)- \nabla_{t,\mathbf x}\Phi_{\infty}^{\pm}(t)\parallel_{L^2(\RR^4_{{\mathbf x},z})}=0,
  \label{asx}
\end{equation}
\begin{equation}
\lim_{t\rightarrow\pm\infty}
\parallel
\partial_z\Phi(t)+\frac{3}{2}\frac{z}{\mid z\mid}\left(\frac{1}{1+\mid z \mid}\right)\Phi(t)-\partial_z\Phi_{\infty}^{\pm}(t)\parallel^2_{
  L^2\left(\RR^3_{\mathbf{x}}\times\RR_z\right)},
  \label{asz}
\end{equation}
and when $\Phi$ is a z-odd wave, we also have :
\begin{equation}
\lim_{t\rightarrow\pm\infty}
\parallel \partial_{z}\Phi(t)- \partial_{z}\Phi_{\infty}^{\pm}(t)\parallel_{L^2(\RR^4_{{\mathbf x},z})}=0.
  \label{asxz}
\end{equation}
The wave operators
\begin{equation*}
{\mathbf W^{\pm}}:\;\left(\Phi_0,\Phi_1\right)\longmapsto \left(\Phi_{\infty}^{\pm}(0,.),\partial_t\Phi_{\infty}^{\pm}(0,.)\right),
\end{equation*}
are isometries from ${\mathfrak K}^1\times{\mathfrak K}^0$ onto
$BL^1\left(\RR^4_{\mathbf{x},z}\right)\times
L^2\left(\RR^4_{\mathbf{x},z}\right)$.

  \label{tsct}
\end{Theorem}

We can introduce the Scattering Operator
\begin{equation*}
\mathbf{S}:={\mathbf W^+}\left({\mathbf W^-}\right)^{-1},
\end{equation*}
that is unitary on $BL^1\left(\RR^4_{\mathbf{x},z}\right)\times
L^2\left(\RR^4_{\mathbf{x},z}\right)$ and describes
the scattering of the Kaluza-Klein waves by the Minkowski brane. We can interpret the existence of these operators in
term of a Goursat problem on the manifolds $\mathbf{B}^{\pm}$ described by Figure 1. 
The asymptotic completeness of the wave operators $\mathbf{W}^-$ assures that the characteristic problem
with the data specified on the conformal null infinity $T=-\infty$, $y=\pm\infty$,
and the  boundary condition on the brane, is well posed, and the
existence of $\mathbf{W}^+$ implies that the solution has a trace on
the future characteristic boundary  $T=+\infty$, $y=\pm\infty$. This
approach of the Goursat problem in general relativity, was used in
\cite{maxwell} for the Maxwell system on the Schwarzschild manifold
for which the timelike infinity is singular,
and in \cite{mason} for the Maxwell and Dirac equations in a large
class of non-stationary vacuum space-times admitting a conformal
compactification that is smooth at null and timelike infinity.\\

{\it Proof of Theorem \ref{tsct}.} First we prove the uniqueness of
the asymptotic fields.  Given some finite energy solution $\Phi$ with initial data
$\Phi_0\in{\mathfrak K}^1$, $\Phi_1\in{\mathfrak K}^0$, assume there exist two free waves
$\Phi_{\infty}^1$,  $\Phi_{\infty}^2$, solutions of (\ref{freq}) and
(\ref{freg}), satisfying
$$
\lim_{t\rightarrow\infty}
\parallel \partial_t\Phi(t)- \partial_t\Phi_{\infty}^{j}(t)\parallel_{L^2(\RR^4_{{\mathbf x},z})}=0.
$$
Then $$
\lim_{t\rightarrow\infty}
\parallel \partial_t\left(\Phi_{\infty}^1- \Phi_{\infty}^{2}\right)(t)\parallel_{L^2(\RR^4_{{\mathbf x},z})}=0.
$$
The well-known result of the equipartition of the energy (see
e.g. \cite{equipart}) assures that
$$
\lim_{t\rightarrow\infty}
\parallel \partial_t\left(\Phi_{\infty}^1- \Phi_{\infty}^{2}\right)(t)\parallel^2_{L^2(\RR^4_{{\mathbf x},z})}=\frac{1}{2}\parallel \nabla_{t,\mathbf{x},z}\left(\Phi_{\infty}^1- \Phi_{\infty}^{2}\right)(0)\parallel^2_{L^2(\RR^4_{{\mathbf x},z})}.
$$
We conclude that $\Phi_{\infty}^1=\Phi_{\infty}^2$.
Conversely,  given some free finite energy solution $\Phi_{\infty}$ of (\ref{freq}) and
(\ref{freg}), assume there exists two waves
$\Phi^1$,  $\Phi^2$, solutions of (\ref{EQB}) with initial data in
${\mathfrak W}^1\times L^2$,  satisfying
$$
\lim_{t\rightarrow\infty}
\parallel \partial_t\Phi_{\infty}(t)- \partial_t\Phi^{j}(t)\parallel_{L^2(\RR^4_{{\mathbf x},z})}=0.
$$
Then $$
\lim_{t\rightarrow\infty}
\parallel \partial_t\left(\Phi^1- \Phi^{2}\right)(t)\parallel_{L^2(\RR^4_{{\mathbf x},z})}=0.
$$
Since $0$ is not eigenvalue of $\mathbf{H}_0$, an abstract result of
\cite{goldstein} assures that
$$
\lim_{T\rightarrow\infty}\frac{1}{T}\int_0^T
\parallel \partial_t\left(\Phi^1-
  \Phi^{2}\right)(t)\parallel^2_{L^2(\RR^4_{{\mathbf x},z})}dt=
\frac{1}{2}\parallel \partial_t\left(\Phi^1-
  \Phi^{2}\right)(0)\parallel^2_{L^2(\RR^4_{{\mathbf x},z})}+
\frac{1}{2}\parallel \Phi^1(0)-
  \Phi^{2}(0)\parallel^2_{{\mathfrak W}^1},
$$
and we conclude that $\Phi^1=\Phi^2$.

A similar argument shows that the wave operators are isometric since
\begin{equation}
\begin{split}
\parallel \partial_t\Phi_1\parallel^2_{L^2(\RR^4_{{\mathbf x},z})}+
\parallel \Phi_0\parallel^2_{{\mathfrak W}^1}
&=
\lim_{T\rightarrow\infty}\frac{2}{T}\int_0^T
\parallel \partial_t\Phi(t)\parallel^2_{L^2(\RR^4_{{\mathbf x},z})}dt\\
&=
\lim_{T\rightarrow\infty}\frac{2}{T}\int_0^T
\parallel \partial_t\Phi_{\infty}^+(t)\parallel^2_{L^2(\RR^4_{{\mathbf x},z})}dt\\
&=
\lim_{T\rightarrow\infty}2
\parallel \partial_t\Phi_{\infty}^+(t)\parallel^2_{L^2(\RR^4_{{\mathbf
      x},z})}\\
&=
\parallel \partial_t\Phi_{\infty}^+(0)\parallel^2_{L^2(\RR^4_{{\mathbf x},z})}+
\parallel \Phi_{\infty}^+(0)\parallel^2_{BL^1(\RR^4)}.
\end{split}
\label{iso}
\end{equation}

Thanks to this property and the conservation of the energy, it is
sufficient to construct ${\mathbf W}^{\pm}$ on a dense subspace of
${\mathfrak K}^1\times {\mathfrak K}^0$, and  $\left({\mathbf W}^{\pm}\right)^{-1}$
 on a dense subspace of $BL^1(\RR^4)\times L^2(\RR^4)$.
If $\hat{f}$ is the partial Fourier transform of $f$ with respect to
$\mathbf{x}$ defined by (\ref{fourier}), $f\in{\mathfrak W}^1$ iff
$\hat{f} \in
L^2\left(\RR^3_{\pmb\xi}\times\RR_z;\mid\pmb\xi\mid^2d\pmb\xi
  dz\right)$ and $\partial_z\hat{f}+\frac{3}{2}\frac{z}{\mid
  z\mid}\left(\frac{1}{1+\mid z \mid}\right)\hat{f}\in
L^2(\RR^4_{{\pmb\xi},z})$. We deduce from (\ref{kun}) and the Fubini
theorem that when $f\in{\mathfrak K}^1$, (respect. $f\in{\mathfrak K}^0$), we have for almost all
$\pmb\xi\in\RR^3$ :
\begin{equation*}
\hat{f}(\pmb\xi,z)\in H^1(\RR_z)\;\;\left(respect.\;\; L^2(\RR_z)\right),\;\;\int_{-\infty}^{\infty}\hat{f}(\pmb\xi,z)f_0(z)dz=0.
\end{equation*}
We introduce a space of regular data
\begin{equation*}
{\mathcal D}_{0}:=\left\{f\in H^1(\RR^4);\;\hat{f}\in
  C^{\infty}_0\left(\left(\RR^3_{\pmb\xi}\setminus\{0\}\right)\times\RR_z\right)\right\}.
\end{equation*}
By cut-off and regularization, we can easily show that
\begin{equation*}
{\mathfrak D}_0:=\left\{f\in {\mathcal D}_{0},\;\;\int_{-\infty}^{\infty}f({\mathbf x},z)f_0(z)dz=0\right\}
\end{equation*}
is dense in ${\mathfrak K}^1\times {\mathfrak K}^0$. \\

In the sequel, we take
$(\Phi_0,\Phi_1)\in {\mathfrak D}_0\times  {\mathfrak D}_0$, and
$\hat\Phi_j(\pmb\xi,z)=0$ when $\mid\pmb\xi\mid\notin[\alpha,\beta]$
for some $0<\alpha<\beta<\infty$, and we
consider the unique finite energy solution $\Phi$ with this initial
data. For any $\pmb\xi\in\RR^3$, Lemma \ref{lemah} assures there exists a unique solution
$u(t,z;\pmb\xi)\in C^0(\RR_t;H^1(\RR_z))\cap C^1(\RR_t;L^2(\RR_z))$ of the Klein-Gordon equation
\begin{equation*}
\partial_t^2u+{\mathbf h}u+\mid\pmb\xi\mid^2u=0,\;\;t,z\in\RR,
\end{equation*}
with Cauchy data
\begin{equation*}
u(0,z;\pmb\xi)=\hat\Phi_0(\pmb\xi,z),\;\;\partial_tu(0,z;\pmb\xi)=\hat\Phi_1(\pmb\xi,z),
\end{equation*}
and $u$ satisfies the constraint
\begin{equation*}
\forall t\in\RR,\;\;\int_{-\infty}^{\infty}u(t,z;\pmb\xi)f_0(z)dz=0.
\end{equation*}
Moreover the usual tools of the spectral theory (see
e.g. \cite{reed1}, Theorem VIII.25, Theorem VIII.20) imply that the map $\pmb\xi\mapsto \mathbf h+\mid\pmb\xi\mid^2$ is
$C^{\infty}$ in the strong resolvent sense, hence
\begin{equation*}
u\in C^{\infty}_0\left(\RR^3_{\pmb\xi}; C^0(\RR_t;H^1(\RR_z))\cap C^1(\RR_t;L^2(\RR_z))\right)
\end{equation*}
and we conclude that
\begin{equation*}
\Phi(t,\mathbf{x},z)=\frac{1}{(2\pi)^{\frac{3}{2}}}\int_{\alpha\leq\mid\pmb\xi\mid\leq\beta}e^{i\mathbf{x}\pmb\xi}u(t,z;\pmb\xi)d\pmb\xi.
\end{equation*}
Furthermore the map $\mathbf{x}\mapsto\Phi(t,\mathbf{x},.)\in H^1(\RR_z)$ is a
function of rapid decrease, hence
\begin{equation*}
u(t,z;\pmb\xi)=\frac{1}{(2\pi)^{\frac{3}{2}}}\int_{\RR^3}e^{-i\mathbf{x}\pmb\xi}\Phi(t,\mathbf{x},z)d\mathbf{x}.
\end{equation*}
This elementary Fourier analysis allows to reduce the study of the
asymptotic behaviour of $\Phi$ to the scattering theory for a 1+1-dimensional
Klein-Gordon equation.

Let $m$ be a strictly positive real number. We compare the solutions
of
\begin{equation}
\partial_t^2u+{\mathbf h}u+m^2u=0,
  \label{kgp}
\end{equation}
with the solutions of the free Klein-Gordon equation
\begin{equation}
\partial_t^2u-\partial_z^2u+m^2u=0,\;\;t,z\in\RR.
  \label{kgun}
\end{equation}
We denote $H^1_m(\RR)$ the usual Sobolev space $H^1(\RR)$, endowed with the norm
$$
\parallel u\parallel^2_{H^1_m}:=\int_{-\infty}^{\infty}\mid
u'(z)\mid^2+m^2\mid u(z)\mid^2dz,
$$
and we introduce the spaces
$$
K^1:=\left\{u\in
  H^1(\RR),\;\int_{-\infty}^{\infty}u(z)f_0(z)dz=0\right\},\;\;
K^0:=\left\{u\in
  L^2(\RR),\;\int_{-\infty}^{\infty}u(z)f_0(z)dz=0\right\},\;\;
$$
and $K^1_m$ denotes $K^1$ equipped with the norm
\begin{equation}
\parallel u\parallel_{K^1_m}^2:=\int_{-\infty}^{\infty}\left\vert
  u'(z)+\frac{3}{2}\frac{z}{\mid z\mid}\left(\frac{1}{1+\mid z
      \mid}\right)u(z)\right\vert^2+m^2\mid u(z)\mid^2dz.
  \label{nkun}
\end{equation}
From (\ref{coeq}), we see that $\parallel u\parallel_{K^1_m}$ and
$\parallel u\parallel_{H^1_m}$
are two equivalent norms. We now develop the scattering theory for
(\ref{kgp}). In particular, the following result provides a rigorous
justification of the numerical experiments by the physicists (see e.g. \cite{sea}).

\begin{Lemma}
For any $u_0\in K^1$,
$u_1\in K^0$, there exist unique $u^{\pm}\in C^0(\RR_t;H^1(\RR_z))\cap
C^1(\RR_t;L^2(\RR_z))$ solutions of (\ref{kgun}) such that
\begin{equation*}
\parallel u(t)-u^{\pm}(t)\parallel_{H^1_m}+\parallel\partial_t u(t)-\partial_tu^{\pm}(t)\parallel_{L^2(\RR)}\longrightarrow
0,\;\;t\rightarrow\pm\infty,
\end{equation*}
where $u\in C^0(\RR_t;H^1(\RR_z))\cap
C^1(\RR_t;L^2(\RR_z))$ is the solution of (\ref{kgp}) with the initial
data $u(0,z)=u_0(z)$, $\partial_tu(0,z)=u_1(z)$. The maps
\begin{equation*}
W^{\pm}_m:\;(u_0,u_1)\longmapsto \left(u^{\pm}(0),\partial_tu^{\pm}(0)\right)
\end{equation*}
are isometries from $K^1_m\times K^0$ onto $H^1_m(\RR)\times
L^2(\RR)$.

For any $u_0\in H^1(\RR)$,
$u_1\in L^2(\RR)$,  there exist unique $A,B\in\CC$, $u^{\pm}, \varrho^{\pm}\in C^0(\RR_t;H^1(\RR_z))\cap
C^1(\RR_t;L^2(\RR_z))$, $u^{\pm}$ solutions of (\ref{kgun}), such that
$$
u(t,z)=\left[Ae^{imt}+Be^{-imt}\right](1+\mid z\mid)^{-\frac{3}{2}}+u^{\pm}(t,z)+\varrho^{\pm}(t,z),
$$
$$
\parallel \varrho^{\pm}(t)\parallel_{H^1_m}+\parallel\partial_t\varrho^{\pm}(t)\parallel_{L^2(\RR)}\longrightarrow
0,\;\;t\rightarrow\pm\infty.
$$
  \label{lemscat}
\end{Lemma}

{\it Proof of Lemma \ref{lemscat}.} We use the scattering theory with
two Hilbert spaces by T. Kato \cite{kato}, that deals with the
comparison between two
abstract wave equations
$$
\frac{d^2}{dt^2}u+A_ju=0,\;\;j=1,2,\;\;t\in\RR.
$$
We consider some densely
defined self-adjoint operators $A_j$ on a separable Hilbert space
$(\mathfrak{h},\parallel.\parallel)$. We assume $A_j$ is strictly
positive,
\begin{equation}
\forall u\in{\mathfrak D}(A_j)\setminus\{0\},\;\;0<(A_ju,u),
  \label{pos}
\end{equation}hence
$B_j:=A_j^{\frac{1}{2}}$ is well defined and the null space of
$B_j$ is $\{0\}$. Furthermore, we suppose that
\begin{equation}
\left(A_2+1\right)^{-1}-\left(A_1+1\right)^{-1}\;\;is\;\;trace\;\;class.
  \label{trace}
\end{equation}
We refer to \cite{reed1} for the trace class ideal.
We know that with these assumptions, the principle of invariance (see
e.g. \cite{reed3}, Theorem XI.11, Corollary 2, p. 30), assures that the wave
operators
\begin{equation}
\Omega^{\pm}(A_2,A_1):=s- \lim_{t\rightarrow\pm\infty}e^{-itA_2}e^{itA_1}Q_1=s- \lim_{t\rightarrow\pm\infty}e^{-itB_2}e^{itB_1}Q_1\;\;exist\;\;and\;\;are
\;\;complete,
  \label{ws}
\end{equation}
where $Q_j$ is the projection for  $\mathfrak{h}$ on the subspace of
absolute continuity for $A_j$.

We introduce the Hilbert space
$\mathfrak{H}_j:=\left[\mathfrak{D}(B_j)\right]\times\mathfrak{h}$,
equipped with the norm
$$
\parallel(u,v)\parallel_{\mathfrak{H}_j}^2=\parallel
B_ju\parallel^2+\parallel v\parallel^2,
$$
where $\left[\mathfrak{D}(B_j)\right]$ is the closure of the domain
of $B_j$, $ \mathfrak{D}(B_j)$, for the norm $\parallel
B_ju\parallel$. We assume that
\begin{equation}
 \mathfrak{D}(B_1)= \mathfrak{D}(B_2),
  \label{dbb}
\end{equation}
and there exists $M>0$ such that for any  $u\in\mathfrak{D}(B_j)$, we
have
\begin{equation}
\parallel B_1u\parallel\leq\parallel B_2u\parallel\leq M\parallel B_1u\parallel.
  \label{bbm}
\end{equation}
As consequence, $\mathfrak{H}_1$ and  $\mathfrak{H}_2$ are the same
linear space, endowed with two equivalent norms. The finite energy
solutions of the wave equations are given by the unitary groups
$U_j(t)$ on $\mathfrak{H}_j$, specified on
$
\mathfrak{D}(B_j)\times\mathfrak{h}$
by
$$
U_j(t)
=
\left(
\begin{array}{cc}
\cos tB_j &B_j^{-1}\sin tB_j\\
-B_j\sin tB_j &\cos tB_j
\end{array}
\right).
$$
$P_j$ denoting the projection on the subspace of absolute continuity for
the selfadjoint operator that generates $U_j(t)$, we define the wave
operators
\begin{equation*}
W^{\pm}(A_2,A_1):=s-\lim_{t\rightarrow\pm\infty}U_2(-t)U_1(t)P_1.
\end{equation*}
Theorem 10.3, Theorem 10.5 and Remark 10.6 of \cite{kato} assure that
assumptions (\ref{pos}), (\ref{ws}), (\ref{dbb}) and (\ref{bbm}) imply that
$W^{\pm}(A_2,A_1)$ exist, are complete and partially isometric with
initial projection $P_1$ and final projection $P_2$. Moreover
$W^{\pm}(A_1,A_2)$ exist, are complete  and equal to
$\left[W^{\pm}(A_2,A_1)\right]^*$.
Finally, using the principle of invariance and formula (9.10)
of \cite{kato}, we obtain a nice expression of the wave operators
:
\begin{equation}
W^{\pm}(A_2,A_1)=\frac{1}{2}
\left(
\begin{array}{cc}
\Omega^+(A_2,A_1)+\Omega^-(A_2,A_1)&i\tilde{B}_2^{-1}\left[\Omega^-(A_2,A_1)-\Omega^+(A_2,A_1)\right]\\
-i\tilde{B}_2\left[\Omega^-(A_2,A_1)-\Omega^+(A_2,A_1)\right]&\Omega^+(A_2,A_1)+\Omega^-(A_2,A_1)
\end{array}
\right),
  \label{exprew}
\end{equation}
where $\tilde{B}_j$ is the unitary map from
$\left[\mathfrak{D}(B_j)\right]$ onto $\mathfrak{h}$ defined as the
unique continuous extension of $B_j$.\\

To apply these abstract results, we take $\mathfrak{h}=L^2(\RR)$ and we
consider operator $\mathbf{h}$ defined by (\ref{h!}), and we introduce
operator
\begin{equation*}
{\mathbf h}_1:={\mathbf h}+m^2,\;\;\;{\mathfrak D}({\mathbf h}_1)={\mathfrak D}({\mathbf h})
\end{equation*}
 which we want to compare with
 the free hamiltonian
\begin{equation*}
{\mathbf h}_2:=-\frac{d^2}{dz^2}+m^2,\;\;{\mathfrak D}({\mathbf h}_2)=H^2(\RR).
\end{equation*}
Because of (\ref{bbm}), we need a third operator
\begin{equation*}
{\mathbf h}_3:=-\frac{d^2}{dz^2}+m^2+\frac{15}{4}\left(\frac{1}{1+\mid
      z\mid}\right)^2,\;\;{\mathfrak D}({\mathbf h}_3)=H^2(\RR).
\end{equation*}

First  we choose $A_1={\mathbf h}_2$,
$A_2={\mathbf h}_3$.
It is obvious that ${\mathbf h}_2$ and  ${\mathbf h}_3$ are strictly positive
selfadjoint operators. Moreover  ${\mathbf h}_3$ is a short range
perturbation of  ${\mathbf h}_2$, by potential $V(z):=\frac{15}{4}\left(\frac{1}{1+\mid
      z\mid}\right)^2$ and puting
  $g(\zeta)=\left(\zeta^2+m^2\right)^{-1}$, we write
$$
\left({\mathbf h}_3+1\right)^{-1}-\left({\mathbf h}_2+1\right)^{-1}=
\left({\mathbf h}_3+1\right)^{-1}V(z)g\left(i\frac{d}{dz}\right).
$$
Since $(1+z^2)^{\frac{1}{2}}V(z)$ and $(1+\zeta^2)^{\frac{1}{2}}g(\zeta)$ are in $L^2(\RR)$, a
well known result (see \cite{reed3}, Theorem XI.21) assures that
$V(z)g\left(i\frac{d}{dz}\right)$ is trace class, so (\ref{trace}) is satisfied. Moreover ${\mathfrak
  D}({\mathbf h}_2^{\frac{1}{2}})={\mathfrak
  D}({\mathbf h}_3^{\frac{1}{2}})=H^1(\RR)$, hence ${\mathfrak H}_1= {\mathfrak
  H}_2=H^1(\RR)\times L^2(\RR)$, and we see that hypotheses (\ref{pos}), (\ref{ws}),
(\ref{dbb}) and (\ref{bbm}) are satisfied. We conclude that
\begin{equation}
W^{\pm}({\mathbf h}_2,{\mathbf h}_3) \;\;exist\;\;and\;\;are
\;\;complete.
  \label{23}
\end{equation}

Now we take  $A_1={\mathbf h}_1$,
$A_2={\mathbf h}_3$. According to Lemma \ref{lemah},  ${\mathbf h}_1$
is strictly positive.
(\ref{coeq}) assures that  ${\mathfrak
  D}({\mathbf h}_1^{\frac{1}{2}})={\mathfrak
  D}({\mathbf h}_3^{\frac{1}{2}})=H^1(\RR)$ and for $u\in H^1(\RR)$ we
have 
$$
 \min\left(1,\frac{2m^2}{11}\right)\parallel {\mathbf
   h}_2^{\frac{1}{2}}u\parallel^2
\leq
\parallel {\mathbf h}_1^{\frac{1}{2}}u\parallel^2
\leq
\parallel {\mathbf h}_3^{\frac{1}{2}}u\parallel^2
\leq
\left(1+\frac{15}{4m^2}\right)\parallel {\mathbf h}_2^{\frac{1}{2}}u\parallel^2
$$
hence  (\ref{pos}), (\ref{ws}),
(\ref{dbb}) and (\ref{bbm}) are satisfied. At last, given $f\in
L^2(\RR)$, $u=\left({\mathbf h}_3+1\right)^{-1}f-\left({\mathbf
    h}_1+1\right)^{-1}f$ is solution of
$$
-u''+\frac{15}{4}\left(\frac{1}{1+\mid
      z\mid}\right)^2u+m^2u=0,\;\;z\neq 0,
$$
thus  $\left({\mathbf h}_3+1\right)^{-1}-\left({\mathbf
    h}_1+1\right)^{-1}$ is a finite rank operator and  (\ref{trace})
is satisfied. We conclude that 
\begin{equation}
W^{\pm}({\mathbf h}_3,{\mathbf h}_1) \;\;exist\;\;and\;\;are
\;\;complete.
  \label{31}
\end{equation}
We deduce from (\ref{23}), (\ref{31}) and the Chain Rule Theorem, that
$W^{\pm}({\mathbf h}_2,{\mathbf h}_1)$ exist, are complete and partially isometric with
initial projection $P_1$ and final projection $P_2$. It is clear that
$P_2=Id$ and ${\mathfrak H}_2=H^1_m(\RR)\times L^2(\RR)$. Moreover,
Lemma 8.1 in \cite{kato} assures that $(u,v)\in P_1{\mathfrak H}_1$
iff $v\in Q_1{\mathfrak h}$ and ${\mathbf h}_1^{\frac{1}{2}}u \in
Q_1{\mathfrak h}$. Lemma (\ref{lemah}) implies that $Q_1{\mathfrak
  h}=K^0$, and since $({\mathbf h}_1^{\frac{1}{2}}u,f_0)=m(u,f_0)$, we
deduce that ${\mathbf h}_1^{\frac{1}{2}}u \in
Q_1{\mathfrak h}$ iff $u\in K^1$ and $ P_1{\mathfrak H}_1=K^1_m\times K^0$. We
conclude that $W^{\pm}_m=W^{\pm}({\mathbf h}_2,{\mathbf h}_1)$ is an isometry
from $K^1_m\times K^0$ onto $H^1_m(\RR)\times L^2(\RR)$. Finally for
general initial data $u_0\in H^1(\RR)$, $u_1\in L^2(\RR)$, we apply
the previous result to the projections on $K^1\times K^0$ and we
compute :
$$
A=\frac{k}{2}\int(1+\mid
z\mid)^{-\frac{3}{2}}\left[u_0(z)-im^{-1}u_1(z)\right]dz,\;\;
B=\frac{k}{2}\int(1+\mid
z\mid)^{-\frac{3}{2}}\left[u_0(z)+im^{-1}u_1(z)\right]dz.
$$
The uniqueness of $A,B,u^{\pm},\varrho^{\pm}$ is obvious.

\fin

We return to the asymptotic behaviour of $\Phi$. We introduce
\begin{equation*}
\Phi_{\infty}^{\pm}(t,\mathbf{x},z)=\frac{1}{(2\pi)^{\frac{3}{2}}}\int_{\alpha\leq\mid\pmb\xi\mid\leq\beta}e^{i\mathbf{x}\pmb\xi}u^{\pm}(t,z;\pmb\xi)d\pmb\xi,
\end{equation*}
where $u^{\pm}(t,z;\pmb\xi)$ is the solution of the free Klein-Gordon
equation (\ref{kgun}) with $m=\mid\pmb\xi\mid$, given by the previous Lemma and satisfying
\begin{equation*}
\parallel u(t,.;\pmb\xi)-u^{\pm}(t,.;\pmb\xi)\parallel_{H^1_{\mid\pmb\xi\mid}}+\parallel\partial_t u(t,.;\pmb\xi)-\partial_tu^{\pm}(t,.;\pmb\xi)\parallel_{L^2(\RR)}\longrightarrow
0,\;\;t\rightarrow\pm\infty.
\end{equation*}
Since $u^{\pm}\in C^0_0\left(\RR^3_{\pmb\xi};C^0(\RR_t;H^1(\RR_z))\cap
  C^1(\RR_t;L^2(\RR_z))\right)$, $\Phi_{\infty}^{\pm}$ is a free wave
satisfying (\ref{freq}) and (\ref{freg}). The Parseval equality gives
\begin{equation*}
\begin{split}
\parallel \nabla_{t,\mathbf{x},z}\Phi(t)-
\nabla_{t,\mathbf{x},z}\Phi_{\infty}^{\pm}(t)\parallel_{L^2(\RR^4_{{\mathbf
      x},z})}^2
+
\parallel\partial_t\Phi(t)-
\partial_t\Phi_{\infty}^{\pm}(t)\parallel_{L^2(\RR^4_{{\mathbf
      x},z})}^2\\
=
\int_{\alpha\leq\mid\pmb\xi\mid\leq\beta}
\parallel
u(t,.;\pmb\xi)-u^{\pm}(t,.;\pmb\xi)\parallel_{H^1_{\mid\pmb\xi\mid}}^2+\parallel\partial_t
u(t,.;\pmb\xi)-\partial_tu^{\pm}(t,.;\pmb\xi)\parallel_{L^2(\RR)}^2d\pmb\xi
\end{split}
\end{equation*}
We evaluate the integrand :
\begin{equation*}
\begin{split}
\parallel
u(t,.;\pmb\xi)-u^{\pm}(t,.;\pmb\xi)&\parallel_{H^1_{\mid\pmb\xi\mid}}^2+\parallel\partial_t
u(t,.;\pmb\xi)-\partial_tu^{\pm}(t,.;\pmb\xi)\parallel_{L^2(\RR)}^2\\
&\leq C\left(
\parallel
u(t,.;\pmb\xi)\parallel_{K^1_{\mid\pmb\xi\mid}}^2+\parallel u^{\pm}(t,.;\pmb\xi)\parallel_{H^1_{\mid\pmb\xi\mid}}^2+\parallel\partial_t
u(t,.;\pmb\xi)\parallel_{L^2(\RR)}^2+\parallel\partial_tu^{\pm}(t,.;\pmb\xi)\parallel_{L^2(\RR)}^2
\right)\\
&=C\left(
\parallel
u(0,.;\pmb\xi)\parallel_{K^1_{\mid\pmb\xi\mid}}^2+\parallel u^{\pm}(0,.;\pmb\xi)\parallel_{H^1_{\mid\pmb\xi\mid}}^2+\parallel\partial_t
u(0,.;\pmb\xi)\parallel_{L^2(\RR)}^2+\parallel\partial_tu^{\pm}(0,.;\pmb\xi)\parallel_{L^2(\RR)}^2
\right)\\
&=2C\left(\parallel\hat\Phi_0(\pmb\xi,.)\parallel^2_{K^1_{\mid\pmb\xi\mid}}+\parallel\hat\Phi_1(\pmb\xi,.)\parallel^2_{L^2(\RR_z)}\right).
\end{split}
\end{equation*}
We conclude by the dominated convergence theorem that 
$$
\parallel \nabla_{t,\mathbf{x},z}\Phi(t)-
\nabla_{t,\mathbf{x},z}\Phi_{\infty}^{\pm}(t)\parallel_{L^2(\RR^4_{{\mathbf
      x},z})}^2
+
\parallel\partial_t\Phi(t)-
\partial_t\Phi_{\infty}^{\pm}(t)\parallel_{L^2(\RR^4_{{\mathbf
      x},z})}^2\rightarrow 0,\;\;t\rightarrow\pm\infty,
$$
and the Wave
Operators $\mathbf{W}^{\pm}$ are well defined on $ {\mathfrak
  D}_0\times  {\mathfrak D}_0$. Since these operators are isometric,
they can be extended by density on
$\mathfrak{K}^1\times\mathfrak{K}^0$ and 
(\ref{ast}) and (\ref{asxz}) are met. To prove (\ref{asz}), it is
sufficient to show that when $\Phi_0,\Phi_1\in {\mathfrak
  D}_0$, then
$$
\int_{\RR^4}\left(\frac{1}{1+\mid
    z\mid}\right)^2\mid\Phi(t,\mathbf{x},z)\mid^2d\mathbf{x}dz\rightarrow 0,\;\;\mid t\mid\rightarrow\infty.
$$
To prove this result, we use the $L^{\infty}$ estimate for the
solutions of the Klein-Gordon equation (\ref{kgun}), $\mid
u(t,z)\mid\leq C(m+\frac{1}{m})\mid mt\mid ^{-\frac{1}{2}}\sum_{i=0,1}\sum_{j\leq
  2-i}\|\partial_t^i\partial_z^ju(0,.)\|_{L^1(\RR_z)}$ :
\begin{equation*}
\begin{split}
\int_{\RR^4}\left(\frac{1}{1+\mid
    z\mid}\right)^2\mid\Phi(t,\mathbf{x},z)\mid^2d\mathbf{x}dz
&=\int_{\alpha\leq
  \mid\pmb\xi\mid\leq\beta}\left(\int_{-\infty}^{\infty}\mid
  u(t,z;\pmb{\xi})\mid^2\frac{1}{(1+\mid z\mid)^2}dz\right)d\pmb\xi\\
&\leq C\mid
t\mid^{-1}\sum_{i=0,1}\sum_{j\leq 2-i}\int_{\alpha\leq
  \mid\pmb\xi\mid\leq\beta}\|
\partial_t^i\partial_z^ju(0,.;\pmb\xi)\|_{L^1}^2d\pmb\xi\\
&\leq C'\mid
t\mid^{-1}\sum_{i=0,1}\sum_{j\leq 2-i}\int_{\alpha\leq
  \mid\pmb\xi\mid\leq\beta}\|(1+\mid z\mid)
\partial_t^i\partial_z^ju(0,.;\pmb\xi)\|_{L^2}^2d\pmb\xi\\
&\leq C'\mid
t\mid^{-1}\sum_{i=0,1}\sum_{j\leq 2-i}\|(1+\mid z\mid)\partial_z^j\Phi_j\|^2_{L^2(\RR^4)}
\end{split}
\end{equation*}

Conversely we take $\Phi_{\infty}^j\in{\mathcal D}_0$ with
$\hat\Phi_{\infty}^j(\pmb\xi,z)=0$ when $\mid\pmb\xi\mid\notin[\alpha,\beta]$
for some $0<\alpha<\beta<\infty$, and we consider
$\Phi_{\infty}\in C^1(\RR_t;{\mathcal D}_0)$ solution of (\ref{freq}) and (\ref{freg}) with
$\Phi_{\infty}(0, \mathbf{x},z)=\Phi_{\infty}^0(\mathbf{x},z)$,
$\partial_t\Phi_{\infty}(0,\mathbf{x},z)=\Phi_{\infty}^1(\mathbf{x},z)$.
Then $u_{\infty}(t,z;\pmb\xi):=\hat{\Phi}_{\infty}(t,\pmb\xi,z)\in C^0_0\left(\RR^3_{\pmb\xi};C^0(\RR_t;H^1(\RR_z))\cap
  C^1(\RR_t;L^2(\RR_z))\right)$ is solution of (\ref{kgun}) with
$m=\mid\xi\mid$. Lemma \ref{lemscat} assures there exists $u^{\pm}(t,z;\pmb\xi)\in C^0_0\left(\RR^3_{\pmb\xi};C^0(\RR_t;K^1)\cap
  C^1(\RR_t;K^0)\right)$ solution of (\ref{kgp}) with $m=\mid\xi\mid$
satisfying
\begin{equation*}
\parallel u_{\infty}(t,.;\pmb\xi)-u^{\pm}(t,.;\pmb\xi)\parallel_{H^1_{\mid\pmb\xi\mid}}+\parallel\partial_t u_{\infty}(t,.;\pmb\xi)-\partial_tu^{\pm}(t,.;\pmb\xi)\parallel_{L^2(\RR)}\longrightarrow
0,\;\;t\rightarrow\pm\infty.
\end{equation*}
We put
\begin{equation*}
\Phi^{\pm}(t,\mathbf{x},z):=\frac{1}{(2\pi)^{\frac{3}{2}}}\int_{\alpha\leq\mid\pmb\xi\mid\leq\beta}e^{i\mathbf{x}\pmb\xi}u^{\pm}(t,z;\pmb\xi)d\pmb\xi.
\end{equation*}
that is solution of (\ref{EQB}) in
$C^0\left(\RR_t;\mathfrak{K}^1\right)$ and $\partial_t\Phi^{\pm}\in
C^0\left(\RR_t;\mathfrak{K}^0\right)$. As above we have
\begin{equation*}
\begin{split}
\parallel \nabla_{t,\mathbf{x},z}\Phi^{\pm}(t)-
\nabla_{t,\mathbf{x},z}\Phi_{\infty}(t)&\parallel_{L^2(\RR^4_{{\mathbf
      x},z})}^2
+
\parallel\partial_t\Phi^{\pm}(t)-
\partial_t\Phi_{\infty}(t)\parallel_{L^2(\RR^4_{{\mathbf
      x},z})}^2\\
&=
\int_{\alpha\leq\mid\pmb\xi\mid\leq\beta}
\parallel
u^{\pm}(t,.;\pmb\xi)-u_{\infty}(t,.;\pmb\xi)\parallel_{H^1_{\mid\pmb\xi\mid}}^2+\parallel\partial_t
u^{\pm}(t,.;\pmb\xi)-\partial_tu_{\infty}(t,.;\pmb\xi)\parallel_{L^2(\RR)}^2d\pmb\xi\\
&\longrightarrow0,\;\;t\rightarrow\pm\infty.
\end{split}
\end{equation*}
We conclude that we have constructed  $\left({\mathbf W}^{\pm}\right)^{-1}$
 on ${\mathcal D}_0\times{\mathcal D}_0$ that is a dense subspace of
 $BL^1(\RR^4)\times L^2(\RR^4)$. Finally (\ref{asxz}) follows from
 (\ref{asx}) since $\parallel \Phi\parallel_{BL^1(\RR^4)}=\parallel
 \Phi\parallel_{\mathfrak W^1}$ for the $z$-odd functions.

\fin


\section{Scattering Amplitude and Brane Resonances}

In this section we compute the scattering matrix and we investigate
its holomorphic continuation. We recall that the spectral representation $\mathcal{R}$ of the free wave equation
in $\RR_t\times\RR^4_{\mathbf{x},z}$ is defined by
\begin{equation*}
\mathcal{R}:\;\left(\Phi_0,\Phi_1\right)\in BL^1\left(\RR^4\right)\times
L^2\left(\RR^4\right)
\longmapsto
\frac{\mid\sigma\mid^{\frac{3}{2}}}{\sqrt{2}}\left[i\sigma\mathcal{F}_{\mathbf{x},z}\Phi_0(\sigma\pmb{\omega})+\mathcal{F}_{\mathbf{x},z}\Phi_1(\sigma\pmb{\omega})\right]\in
L^2\left(\RR_{\sigma}\times S^3_{\pmb{\omega}}\right),
\end{equation*}
where $S^3_{\pmb{\omega}}$ is the unit sphere of the euclidean space $\RR^4$,
$\sigma\pmb{\omega}=(\pmb{\xi},\zeta)\in\RR^3_{\pmb{\xi}}\times\RR_{\zeta}$,
and we have used the Fourier transform on $\RR^3_{\mathbf{x}}\times\RR_z$
defined for $\Phi\in L^1\left(\RR^3_{\mathbf{x}}\times\RR_z\right)$ by
\begin{equation*}
\mathcal{F}_{\mathbf{x},z}\Phi(\pmb{\xi},\zeta)=\frac{1}{(2\pi)^2}\int_{\RR^4}e^{-i(\mathbf{x.}\pmb{\xi}+z\zeta)}\Phi(\mathbf{x},z)d\mathbf{x}dz.
\end{equation*}
We know (see e.g. \cite{petkov}) that $\mathcal{R}$ is an isometry
from $BL^1\left(\RR^4_{\mathbf{x},z}\right)\times
L^2\left(\RR^4_{\mathbf{x},z}\right)$ onto $L^2\left(\RR_{\sigma}\times S^3_{\pmb{\omega}}\right)$
and if $\Phi$ is a finite energy solution of the wave equation
$\partial_t^2\Phi-\Delta_{\mathbf{x},z}\Phi=0$, we have
\begin{equation*}
\mathcal{R}\left(\Phi(t,.),\partial_t\Phi(t,.)\right)=e^{i\sigma t}\mathcal{R}\left(\Phi(0,.),\partial_t\Phi(0,.)\right).
\end{equation*}
It will be useful to distinguish the odd and the even part of the
scattering operator. Thus we
introduce spaces $\pmb{E}_{\pm}$ associated with
$\pmb{E}=\mathfrak{K}^1,\,\mathfrak{K}^0,\,BL^1,\,L^2$ by
\begin{equation*}
\pmb{E}_{\pm}:=\left\{\Phi\in\pmb{E},\;\;\Phi(\mathbf{x},-z)=\pm\Phi(\mathbf{x},z)\right\}.
\end{equation*}
It is obvious that $\mathfrak{K}^1_{\pm}\times\mathfrak{K}^1_{\pm}$
and $BL^1_{\pm}\times L^2_{\pm}$ are let invariant, respectively  by the
pertubed and the free dynamics. Therefore $\mathbf{S}_{\pm}$ defined
as the restriction of $\mathbf{S}$ to $BL^1_{\pm}\times L^2_{\pm}$ is an isometry
of this space. We also introduce
\begin{equation*}
L^2_{\pm}\left(\RR_{\sigma}\times S^3_{\pmb{\omega}}\right):=\left\{f\in
  L^2\left(\RR_{\sigma}\times
    S^3_{\pmb{\omega}}\right),\;\;f(\sigma,\omega_1,\omega_2,\omega_3,-\omega_4)=\pm f(\sigma,\omega_1,\omega_2,\omega_3,\omega_4)\right\}.
\end{equation*}
Then, $\mathcal{R}$ is an isometry from
$BL^1_{\pm}(\RR^4_{\mathbf{x},z})\times
L^2_{\pm}(\RR^4_{\mathbf{x},z})$ onto $L^2_{\pm}\left(\RR_{\sigma}\times S^3_{\pmb{\omega}}\right)$.
In this part we  establish first that the Scattering Operator
$\mathbf{S}=\mathbf{S}_+\oplus \mathbf{S}_-$ is
emplemented by an explicit Scattering Amplitude
 acting as a multiplication operator on
$L^2_+\left(\RR_{\sigma}\times S^3_{\pmb{\omega}}\right)\oplus L^2_-\left(\RR_{\sigma}\times S^3_{\pmb{\omega}}\right)$.

\begin{Theorem}
There exists $\mathcal{S}_+,\mathcal{S}_-\in
C^0\left(\RR^*_{\sigma}\times S^3_{\pmb{\omega}};
S^1\right)$ such that for any $f_{\pm}\in L^2_{\pm}\left(\RR_{\sigma}\times
  S^3_{\pmb{\omega}}\right)$, we have :
\begin{equation*}
{\mathcal
  R}\mathbf{S}\mathcal{R}^{-1}\left(f_+\oplus f_-\right)(\sigma,\pmb{\omega})
=\mathcal{S}_+(\sigma,\pmb{\omega})f_+(\sigma,\pmb{\omega})+\mathcal{S}_-(\sigma,\pmb{\omega})f_-(\sigma,\pmb{\omega}).
\end{equation*}
Moreover these Scattering Amplitudes are explicitely known :
  \begin{equation}
0<\sigma\Rightarrow \mathcal{S}_{+[-]}(\sigma,\pmb{\omega})
=
+[-]\frac{e^{2i\sigma\mid\omega_4\mid}}{i}\frac{H_{1[2]}^{(2)}\left(\sigma\mid\omega_4\mid\right)}{H_{1[2]}^{(1)}\left(\sigma\mid\omega_4\mid\right)},
    \label{forscat}
  \end{equation}
 \begin{equation}
\sigma<0\Rightarrow\mathcal{S}_{+[-]}(\sigma,\pmb{\omega})
=
-[+]\frac{e^{2i\sigma\mid\omega_4\mid}}{i}\frac{H_{1[2]}^{(1)}\left(-\sigma\mid\omega_4\mid\right)}{H_{1[2]}^{(2)}\left(-\sigma\mid\omega_4\mid\right)}.
    \label{forscatm}
  \end{equation}
  \label{reso}
\end{Theorem}

{\it Proof of Theorem \ref{reso}.} We compute $\mathcal{S}:={\mathcal
  R}\mathbf{S}\mathcal{R}^{-1}$. If the inverse Fourier transform on
$\RR^4_{\pmb{\xi},\zeta}$ is denoted by
$\mathcal{F}_{\pmb{\xi},\zeta}^{-1}$, we easily check that
\begin{equation*}
{\mathcal R}^{-1}=\mathcal{F}_{\pmb{\xi},\zeta}^{-1}\mathbf{U}
\end{equation*}
where $\mathbf{U}$ is the unitary map given by
\begin{equation*}
\mathbf{U}:\;L^2\left(\RR_{\sigma}\times
  S^3_{\pmb{\omega}}\right)\longrightarrow
L^2\left(\RR^4_{\pmb{\xi},\zeta},(\mid\pmb{\xi}\mid^2+\zeta^2)d\pmb{\xi}d\zeta\right)\times
L^2\left(\RR^4_{\pmb{\xi},\zeta}, d\pmb{\xi}d\zeta\right)
\end{equation*}
\begin{equation*}
\mathbf{U}f(\pmb{\xi},\zeta)=\frac{\sigma^{-\frac{3}{2}}}{i\sqrt{2}}\left(
\begin{array}{c}
\sigma^{-1}f(\sigma,\pmb{\omega})-\sigma^{-1}f(-\sigma,-\pmb{\omega})\\
if(\sigma,\pmb{\omega})+if(-\sigma,-\pmb{\omega})
\end{array}
\right),\;\;0\leq\sigma,\;\sigma\pmb{\omega}=(\pmb{\xi},\zeta).
 \end{equation*}
From the proof of Theorem \ref{tsct}, we have by using the partial
Fourier transform (\ref{fourier}) :
\begin{equation*}
\mathbf{W}^{\pm}=\mathcal{F}^{-1}_{\pmb{\xi}}W^{\pm}_{\mid\pmb{\xi}\mid}\mathcal{F}_{\mathbf{x}},
\end{equation*}
where $ W^{\pm}_{\mid\pmb{\xi}\mid}$ is given by Lemma \ref{lemscat} with
$m=\mid \pmb{\xi}\mid$, and with (\ref{exprew}) we get
\begin{equation*}
\mathbf{W}^{\pm}=\frac{1}{2}\mathcal{F}^{-1}_{\pmb{\xi}}
\left(
\begin{array}{cc}
\Omega^+(\mathbf{h}_0,\mathbf{h})+\Omega^-(\mathbf{h}_0,\mathbf{h})&i\mathbf{h}_{\mid\pmb{\xi}\mid}^{-\frac{1}{2}}\left[\Omega^-(\mathbf{h}_0,\mathbf{h})-\Omega^+(\mathbf{h}_0,\mathbf{h})\right]\\
-i\mathbf{h}_{\mid\pmb{\xi}\mid}^{\frac{1}{2}}\left[\Omega^-(\mathbf{h}_0,\mathbf{h})-\Omega^+(\mathbf{h}_0,\mathbf{h})\right]&\Omega^+(\mathbf{h}_0,\mathbf{h})+\Omega^-(\mathbf{h}_0,\mathbf{h})
\end{array}
\right)
\mathcal{F}_{\mathbf{x}}
\end{equation*}
with
\begin{equation*}
0\leq m,\;\;\mathbf{h}_m:=-\partial_z^2+m^2,\;\;\mathfrak{D}(\mathbf{h}_m)=H^2(\RR).
\end{equation*}
We deduce that the scattering operator has the form :
\begin{equation*}
\mathbf{S}=\frac{1}{2}\mathcal{F}^{-1}_{\pmb{\xi}}
\left(
\begin{array}{cc}
\mathbf{s}+\mathbf{s}^{-1}&-i\mathbf{h}_{\mid\pmb{\xi}\mid}^{-\frac{1}{2}}\left[\mathbf{s}-\mathbf{s}^{-1}\right]\\
i\mathbf{h}_{\mid\pmb{\xi}\mid}^{\frac{1}{2}}\left[\mathbf{s}-\mathbf{s}^{-1}\right]&\mathbf{s}+\mathbf{s}^{-1}
\end{array}
\right)
\mathcal{F}_{\mathbf{x}}
\end{equation*}
where $\mathbf{s}$ is the usual scattering operator associated with the
Schr\"odinger equation $i\partial_tu-\mathbf{h}u=0$ :
\begin{equation*}
\mathbf{s}:=\Omega^+(\mathbf{h}_0,\mathbf{h})\circ\left[\Omega^-(\mathbf{h}_0,\mathbf{h})\right]^{-1}
\end{equation*}
We introduce the self-adjoint
operators $\left(\mathbf{h}_{0,\pm}, {\mathfrak D}(\mathbf{h}_{0,\pm})\right)$, on $L^2(\RR^+)$ defined by :
\begin{equation*}
{\mathbf h_{0,\pm}}:=-\frac{d^2}{dz^2},
\end{equation*}
\begin{equation*}
{\mathfrak D}({\mathbf h_{0,+}}):=\left\{u_+\in
  H^2(\RR^+);\;u'_+(0)=0\right\},\;\;{\mathfrak D}({\mathbf h_{0,-}}):=\left\{u_-\in
  H^2(\RR^+);\;u_-(0)=0\right\}.
\end{equation*}
We easily see that $\left(\mathbf{h}_{0,\pm},\mathbf{h}_{0})\right)$
satisfies (\ref{trace}) and so the waves operators
$\Omega^{\pm}(\mathbf{h}_{0,\pm},\mathbf{h_\pm})$ exist and are
complete on $L^2(\RR^+_z)$.
From the intertwining relations (\ref{interp}) and
$\left(\mathbf{h}_{0,+}\oplus\mathbf{h}_{0,-}\right)\mathbf{P}=\mathbf{P}\mathbf{h}_{0}$,
we deduce that :
\begin{equation*}
\Omega^{\pm}(\mathbf{h}_0,\mathbf{h})=\mathbf{P}^{-1}\left[\Omega^{\pm}(\mathbf{h}_{0,+},\mathbf{h_+})\oplus\Omega^{\pm}(\mathbf{h}_{0,-},\mathbf{h_-})\right]\mathbf{P},
\end{equation*}
and if we introduce the isometries on $L^2(\RR^+_z)$ given by
\begin{equation*}
\mathbf{s}_{\pm}:=\Omega^+(\mathbf{h}_{0,\pm},\mathbf{h_{\pm}})\circ\left[\Omega^-(\mathbf{h}_{0,\pm},\mathbf{h_{\pm}})\right]^{-1},
\end{equation*}
we obtain the splitting of the scattering operator as the sum of  an
even part and an odd part :
\begin{equation*}
\mathbf{s}=\mathbf{P}^{-1}\left[\mathbf{s_+}\oplus\mathbf{s}_-\right]\mathbf{P}.
\end{equation*}
We now want to determine the form of
$\hat{\mathbf{s}}:=\mathcal{F}_z\mathbf{s}\mathcal{F}_{\zeta}^{-1}$ on
$L^2(\RR_{\zeta})$ where $\mathcal{F}_z$ is the Fourier transform
\begin{equation*}
f\in L^1(\RR_z),\;\;\mathcal{F}_zf(\zeta)=\frac{1}{\sqrt{2\pi}}\int_{-\infty}^{\infty}e^{-iz\zeta}f(z)dz.
\end{equation*}
We need the  Fourier sine and Fourier cosine transforms which are the
unitary maps $\mathcal{F}_{\pm}$ on  $L^2(\RR^+)$ defined by
\begin{equation*}
f\in L^1(\RR^+),\;\;\mathcal{F}_+f(m)=\sqrt{\frac{2}{\pi}}\int_{0}^{\infty}\cos(mz)f(z)dz,\;\;\mathcal{F}_-f(m)=\sqrt{\frac{2}{\pi}}\int_{0}^{\infty}\sin(mz)f(z)dz.
\end{equation*}
We now  can express the scattering matrix :
\begin{equation*}
\hat{\mathbf{s}}=\mathbf{P}^{-1}\left[\mathcal{F}_+\mathbf{s}_+\mathcal{F}_+^{-1}\oplus\mathcal{F}_-\mathbf{s}_-\mathcal{F}_-^{-1}\right]\mathbf{P}.
\end{equation*}
In a first time, we admit that
$\mathcal{F}_{\pm}\mathbf{s}_{\pm}\mathcal{F}_{\pm}^{-1}$ is a
multiplication operator $\hat{\mathbf{s}}_{\pm}(m)$ on
$L^2(\RR^+_{m})$. Given $f\in L^2(\RR_{\sigma}\times
S^3_{\pmb{\omega}})$ we write $f=f_+\oplus f_-$, $f_{\pm}\in
L^2_{\pm}(\RR_{\sigma}\times S^3_{\pmb{\omega}})$, and we compute :
$$
\mathcal{S}f_{\pm}(\sigma,\pmb{\omega})=\frac{1}{2}\left(1+\frac{\sigma}{\mid\sigma\mid}\right)\hat{\mathbf{s}}_{\pm}(\mid\sigma\omega_4\mid)f_{\pm}\left(\mid\sigma\mid,\frac{\sigma}{\mid\sigma\mid}\pmb{\omega}\right)+\frac{1}{2}\left(1-\frac{\sigma}{\mid\sigma\mid}\right)\hat{\mathbf{s}}_{\pm}^{-1}(\mid\sigma\omega_4\mid)f_{\pm}\left(-\mid\sigma\mid,-\frac{\sigma}{\mid\sigma\mid}\pmb{\omega}\right),
$$
and we deduce that
\begin{equation}
\begin{split}
\mathcal{S}f(\sigma,\pmb{\omega})
&=\left[\hat{\mathbf{s}}_+(\sigma\mid\omega_4\mid)f_+(\sigma,\pmb{\omega})+\hat{\mathbf{s}}_-(\sigma\mid\omega_4\mid)f_-(\sigma,\pmb{\omega})\right]\mathbf{1}_{]0,\infty[}(\sigma)\\
&+\left[\hat{\mathbf{s}}_+^{-1}(-\sigma\mid\omega_4\mid)f_+(\sigma,\pmb{\omega})+\hat{\mathbf{s}}_-^{-1}(-\sigma\mid\omega_4\mid)f_-(\sigma,\pmb{\omega})\right]\mathbf{1}_{]-\infty,0[}(\sigma).
\end{split}
  \label{decost}
\end{equation}
We now compute $\hat{\mathbf{s}}_{\pm}$. We know (see
e.g. \cite{pear}, Theorem 8.1), that when the free
hamiltonian is $\mathbf{h}_{0,-}$, the scattering
operator for the pertubed Schr\"odinger operator on the half line, is determined by the scattering phase shift $\delta_-(m)$  defined by
$u_-(z,m)=\sqrt{\frac{2}{\pi}}\sin\left(mz+\delta_-(m)+o(1)\right)$
    with $\lim_{z\rightarrow\infty}o(1)=0$, and the formula
\begin{equation*}
\hat{\mathbf{s}}_-(m)=e^{2i\delta_-(m)}.
\end{equation*}
 From (\ref{estmz}) we get
\begin{equation*}
\hat{\mathbf{s}}_-(m)=ie^{2im}\frac{H_2^{(2)}\left(m\right)}{H_2^{(1)}\left(m\right)}.
    \end{equation*}
The same argument assures that if
\begin{equation*}
\mathbf{s}_{+}':=\Omega^+(\mathbf{h}_{0,-},\mathbf{h_{+}})\circ\left[\Omega^-(\mathbf{h}_{0,-},\mathbf{h_{+}})\right]^{-1},
\end{equation*}
then for $f\in L^2(\RR^+_m)$ we have :
\begin{equation*}
\mathcal{F}_-\mathbf{s}_{+}'\mathcal{F}_-^{-1}f(m)=ie^{2im}\frac{H_1^{(2)}\left(m\right)}{H_1^{(1)}\left(m\right)}f(m).
\end{equation*}
Since $\cos(mz)=\sin(mz+\frac{\pi}{2})$, the scattering phase shift for the pair
$(\mathbf{h}_{0,-},\mathbf{h}_{0,+})$ is $\frac{\pi}{2}$, hence formula
(8.2.6) in \cite{pear} assures that
$\Omega^{\pm}(\mathbf{h}_{0,-},\mathbf{h}_{0,+})=\pm i\mathcal{F}_-^{-1}\circ\mathcal{F}_+$.
Now from the chain rule, we have
$\mathbf{s}_+=\left(\Omega^+(\mathbf{h}_{0,-},\mathbf{h}_{0,+})\right)^{-1}\circ\mathbf{s}'_+\circ
\Omega^-(\mathbf{h}_{0,-},\mathbf{h}_{0,+})$, then we conclude that
\begin{equation*}
\hat{\mathbf{s}}_+(m)=-ie^{2im}\frac{H_1^{(2)}\left(m\right)}{H_1^{(1)}\left(m\right)}.
    \end{equation*}
Thus (\ref{forscat}) and (\ref{forscatm}) follow
from (\ref{decost}). Moreover, by  the relation of conjugation
\begin{equation*}
H_{\nu}^{(1)}\left(\overline{z}\right)=\overline{H_{\nu}^{(2)}(z)},
\end{equation*}
and the asymptotic behaviour $\lim_{x\rightarrow
  0^+}\frac{H^{(2)}_{\nu}(x)}{H^{(1)}_{\nu}(x)}=-1$, we get
$\mathcal{S}_{\pm}\in C^0\left(\RR^*_{\sigma}\times
  S^3_{\pmb{\omega}};S^1\right)$.

\fin

We now study the analytic continuation of the Scattering Amplitudes.
We denote $\mathcal{Z}^{(j)}_j$ the
set of the zeros of $H^{(j)}_{\nu}(z)$, $\nu\in\NN$, in the Riemann surface of the logarithm
$\widetilde{\CC^*}$. Thanks to the relation of conjugation,
$z\in\mathcal{Z}_{\nu}^{(1)}\Leftrightarrow
\overline{z}\in\mathcal{Z}_{\nu}^{(2)}$. The zeros of $H_{\nu}^{(1)}$ in the
principal Riemann sheet $\{z;\;-\pi<\arg z\leq \pi\}$ are well known
(\cite{abra}, p.373, \cite{olver1}, figure 14) ; they are located in the open lower half-plane
$\{z;\;-\pi<\arg z<0\}$ and are of two types : (i) an infinite number
of zeros $z_n$ just below the negative real semiaxis with $\Re z_n<-\nu$
and $\lim_{n\rightarrow\infty}\Im z_n=-\frac{1}{2}\log 2$
; (ii) a finite set of $\nu$ zeros with  $\mid \Re z\mid<\nu$, which lie
along the lower half of the boundary of an eye-shapes domain around
the origin. By using the relation
$$
H^{(1)}_{\nu}\left(ze^{-im\pi}\right)=(-1)^{m\nu}\left((m+1)H_{\nu}^{(1)}(z)+m\overline{H_{\nu}^{(1)}(\overline{z})}\right),
$$
Olver \cite{olver1} has proved that the asymptotic repartition in the other Riemann sheets are
analogous, and the
projection of $\mathcal{Z}^{(j)}_{\nu}$ on $\CC^*$ is a lattice with
asymptotes $\mid\Im z\mid=\frac{1}{2}\log \left(\frac{\mid
    4m-1\mid+1}{\mid 4m-1\mid-1}\right)$, $ \frac{1}{2}\log \left(\frac{\mid
    4m-1\mid+3}{\mid 4m-1\mid+1}\right)$. We note that the previous
relation shows the symetry of the zeros
with respect to the imaginary axis :
$$
H_{\nu}^{(1)}(z)=0\Leftrightarrow H_{\nu}^{(1)}(e^{i\pi}\overline{z})=0.
$$
 We emphasize that on each sheet, there is
no accumulation near the real axis, but there exists a sequence $z_n\in\mathcal{Z}_{\nu}^{(j)}$
such that $\arg z_n\rightarrow\infty$, $\Im z_n\rightarrow 0$,
$n\rightarrow\infty$.  We recall the value of the firsts of them (see
e.g. \cite{cruz}).\\

\begin{center}
\begin{tabular}{|c|c|c|c|}
\hline
\multicolumn{2}{|c|}{Zeros of $H_1^{(1)}$}&
\multicolumn{2}{|c|}{Zeros of $H_2^{(1)}$}\\
\hline
 $-\pi<\arg z\leq\pi$  &  $-3\pi<\arg z\leq-\pi$ &  $-\pi<\arg z\leq\pi$  &  $-3\pi<\arg z\leq-\pi$\\
\hline
$-0,419-0,577i$ & $ 0,333+0,413i$& $0,429-1,281i$ & $-0,386+1,101i$\\
$-3,832-0,355i$ & $3,832+0,208i$ & $-1,317-0,836i$ & $ 1,146+0,652i$\\
$-7,016-0,349i$ & $7,016+0,204i$  &  $-5,138-0,372i$ & $ 5,136+0,218i$\\
$-10,173-0,348i$ & $ 10,137+0,203i$ &$-8,418-0,356i$ & $8,417+0,208i$ \\
$-13,326-0,348i$ & $13,326+0,204i$ & $-11,620-0,351i$ & $11,620+0,206i$\\
\hline
\end{tabular}
\end{center}
\vspace{0.5cm}

We show that the singularities of the Scattering Amplitudes
$\sigma\mapsto\mathcal{S}_{\pm}(\sigma)\in C^0\left(S^3_{\pmb{\omega}}\right)$, called
{\it Brane  Resonances}, form a lattice of radial half straight lines :
\begin{equation*}
\Sigma_{\nu}^{(1)}:=\left\{ z=\alpha z_*\in\widetilde{\CC^*},\;\;1\leq\alpha,\;H_{\nu}^{(1)}(z_*)=0\right\},\;\;\Sigma_{\nu}^{(2)}:=\left\{ z=\alpha z_*\in\widetilde{\CC^*},\;\;1\leq\alpha,\;H_{\nu}^{(2)}(-z_*)=0\right\},
\end{equation*}
with $\nu=1,2$. In the figure below, the first lines of resonances in the principal
sheet $-\pi<\arg z\leq\pi$, and the second sheet $-3\pi<\arg z\leq
-\pi$, are depicted.

\begin{center} 
\begin{pspicture}(0,0)(12,8) 
\psline[linewidth=0.2mm,linestyle=dotted]{->}(6,0.5)(6,7.5)
\psline[linewidth=0.2mm, linestyle=dotted]{->}(6,4)(12,4)
\psline[linewidth=0.2mm, doubleline=true](0,4)(6,4)
\rput(3,4.2){cut}
\rput(6.2,3.8){0}

\psline{*->}(5.581,3.423)(3.486, 0.538)
\psline{*->}(6.419,3.423)(8.514, 0.538)

\psline{*->}(2.168,3.645)(0.06,3.449)
\psline{*->}(9.832,3.645)(11.94,3.449)

\psline{*->}(6.429,2.719)(7.172,0.5)
\psline{*->}(5.571,2.719)(4.828,0.5)

\psline{*->}(4.683,3.164)(0.486,0.5)
\psline{*->}(7.317,3.164)(11.514,0.5)

\psline{*->}(0.862,3.628)(0,3.568)
\psline{*->}(11.138,3.628)(12,3.568)


\psline[linewidth=0.2mm,linestyle=dashed]{o->}(6.333,4.413)(9.225,8)
\psline[linewidth=0.2mm,linestyle=dashed]{o->}(5.667,4.413)(2.775,8)

\psline[linewidth=0.2mm,linestyle=dashed]{o->}(9.832,4.208)(12,4.325)
\psline[linewidth=0.2mm,linestyle=dashed]{o->}(2.168,4.208)(0,4.325)


\psline[linewidth=0.2mm,linestyle=dashed]{o->}(5.614,5.101)(4.660,8)
\psline[linewidth=0.2mm,linestyle=dashed]{o->}(6.386,5.101)(7.34,8)

\psline[linewidth=0.2mm,linestyle=dashed]{o->}(7.146,4.652)(12,7.414)
\psline[linewidth=0.2mm,linestyle=dashed]{o->}(4.854,4.652)(0,7.414)

\psline[linewidth=0.2mm,linestyle=dashed]{o->}(11.136,4.218)(12,4.254)
\psline[linewidth=0.2mm,linestyle=dashed]{o->}(0.864,4.218)(0,4.254)

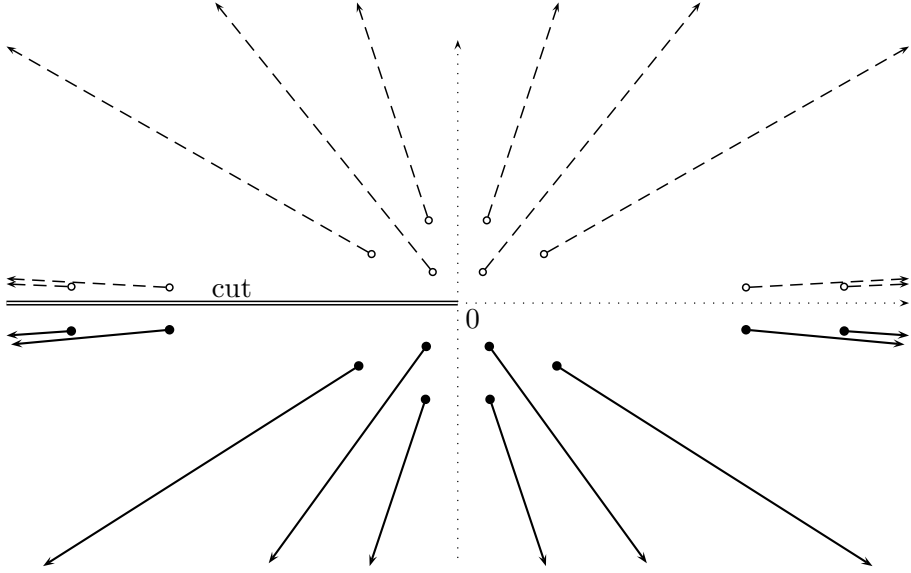
\rput(6,-0.5){{\it Figure 2.} Brane Resonances in the first
($\mathrel{\mathop\bullet}\rightarrow$) and second ($\mathrel{\mathop\circ}\dashrightarrow$)
 Riemann sheets.}

\end{pspicture}

\end{center}

\vspace{1cm}

\begin{Theorem}
The Scattering Amplitude $\mathcal{S}_{+[-]}(\sigma,\pmb{\omega})$
considered as  a $C^0\left(S^3_{\pmb{\omega}}\right)$-valued function
of $\sigma\in [0,\infty[$ (respectively $\sigma\in]-\infty,0]$) has an analytic continuation on
$\widetilde{\CC^*}\setminus\Sigma^{(1)}_{1[2]}$ (respectively $\widetilde{\CC^*}\setminus\Sigma^{(2)}_{1[2]}$). For $\sigma_*\in
\Sigma^{(j)}_{1[2]}$, there exists $C>0$ such that
$C\mid \sigma-\sigma_*\mid^{-1}\leq \|
\mathcal{S}_{+[-]}(\sigma)\|_{L^{\infty}(S^3_{\pmb{\omega}})}$ as $\sigma\rightarrow
\sigma_*$, $\sigma\in\widetilde{\CC^*}\setminus\Sigma^{(j)}_{1[2]}$.
  \label{theores}
\end{Theorem}

{\it Proof of Theorem \ref{theores}.} First we fix $\pmb{\omega}\in
S^3$ with $\omega_4\neq 0$. We know that $H_{\nu}^{(j)}(z)$ is
holomorphic on $\widetilde{\CC^*}$ and its $z$-zeros are non real and
simple (\cite{olver} p. 244). Moreover $H_{\nu}^{(1)}(z)$ and $H_{\nu}^{(2)}(z)$ have no common
zero, because otherwise such a complex would be a non real zero of
$J_{\nu}$ that has only real zeros (\cite{olver} p. 245). We conclude
that $\sigma\in[0,\infty[\mapsto\mathcal{S}_{+[-]}(\sigma,\pmb{\omega})$ has a
meromorphic continuation on $\widetilde{\CC^*}$, is holomorphic
 in  $\mathcal{D}_{+[-]}^{(1)}:=\widetilde{\CC^*}\setminus\Sigma_{1[2]}^{(1)}$, and  its poles are
simple. Now for $\sigma_0$ fixed in
$\widetilde{\CC^*}\setminus\Sigma_{1[2]}^{(1)}$, $\sigma_0\mid\omega_4\mid$
belongs to $\mathcal{D}_{+[-]}^{(1)}$ for $\omega_4\neq
0$. Moreover the Neumann expansions of the Bessel  functions of first
and second kind (\cite{olver} p.243), imply that
$H^{(2)}_{\nu}(\sigma_0\mid\omega_4\mid)/H^{(1)}_{\nu}(\sigma_0\mid\omega_4\mid)\rightarrow
-1$ as $\omega_4\rightarrow 0$. We deduce that
$\mathcal{S}_{\pm}(\sigma_0)\in
C^0\left(S^3_{\pmb{\omega}}\right)$. To prove that the Scattering
Amplitudes are $C^0\left(S^3_{\pmb{\omega}}\right)$-valued holomorphic
functions on $\mathcal{D}_{+[-]}^{(1)}$, we use the Dunford theorem
(\cite{yoshida}, p. 128). We take a Radon measure $\mu$ on
$S^3_{\pmb{\omega}}$. Then for any piecewise-$C^1$ loop $\gamma$ in
$\mathcal{D}_{\pm}^{(1)}$, the Fubini theorem and the Cauchy theorem give
$$
\oint_{\gamma}\left(\int_{S^3}\mathcal{S}_{\pm}(\sigma,\pmb{\omega})d\mu(\pmb{\omega})\right)d\sigma=\int_{S^3}\left(\oint_{\gamma}\mathcal{S}_{\pm}(\sigma,\pmb{\omega})d\sigma\right)d\mu(\pmb{\omega})=0,
$$
thus the Morera theorem assures that
$\mu\left(\mathcal{S}_{\pm}(\sigma,.)\right)$ is holomorphic and we
get the result by
the Dunford theorem. To prove that the resonances are actually
singularities, we pick $\sigma_*=\alpha z_*\in\Sigma_{1[2]}^{(1)}$ and we
choose $\pmb{\omega}_*\in S^3$ with $\omega_4=\alpha^{-1}$. Since
$z_*$ is a simple zero of $H^{(1)}_{1[2]}(z)$ and not a zero for
$H^{(2)}_{1[2]}(z)$, there exists $\varepsilon,\delta,C>0$ such that :
$$
\mid\sigma-\sigma_*\mid\leq\delta\Rightarrow
\left\vert H^{(2)}_{1[2]}\left(\sigma\mid
    \omega_4\mid\right)\right\vert\geq \varepsilon,\;\;\left\vert H^{(1)}_{1[2]}\left(\sigma\mid
    \omega_4\mid\right)\right\vert\leq  C\mid\sigma-\sigma_*\mid.
$$
We conclude that $\mid
\mathcal{S}_{+[-]}(\sigma,\pmb{\omega}_*)\mid\geq \frac{\alpha}{C}\mid
\sigma-\sigma_*\mid^{-1}$ when
$\mid\sigma-\sigma_*\mid\leq\delta$. The proof of the holomorphic
continuation of
$\sigma\in]-\infty,0]\mapsto\mathcal{S}_{+[-]}(\sigma,\pmb{\omega})$ on
$\widetilde{\CC^*}\setminus\Sigma_{1[2]}^{(2)}$
is
similar.
\fin

We end this part by some remarks on the {\it brane quasimodes}. In the
physical litterature, this
term means a solution of the master equation with an exponential
dumping in time as $t\rightarrow\infty$ and this solution is associated
with some resonance. We can easily construct such solutions. For any
$\zeta_{\nu}^{(j)}\in \mathcal{Z}_{\nu}^{(j)}$ the functions
$$u_1^{(j)}(z):=\sqrt{1+\mid z\mid}H_2^{(j)}\left(\zeta_1^{(j)}(1+\mid z\mid)\right),\;\;u_2^{(j)}(z):=\frac{z}{\mid z\mid}\sqrt{1+\mid z\mid}H_2^{(j)}\left(\zeta_2^{(j)}(1+\mid z\mid)\right)$$
are solution of $\mathbf{h}u_{\nu}^{(j)}=(\zeta_{\nu}^{(j)})^2u_{\nu}^{(j)}$. If we
choose any solution $\phi(\zeta_{\nu}^{(j)};t,\mathbf{x})$ of the Klein-Gordon
equation with complex mass $k\zeta_{\nu}^{(j)}$,
\begin{equation}
\partial_t^2\phi-\Delta_{\mathbf{x}}\phi+(\zeta_{\nu}^{(j)})^2\phi=0,
  \label{kgc}
\end{equation}
then
$ \Phi(t,x,z)=\phi(\zeta_{\nu}^{(j)};t,\mathbf{x})u_{\nu}^{(j)}(z)$ is solution of the master
equation (\ref{EQB}). In particular, for any singularity $\sigma_{\nu}^{(j)}\in\Sigma_{\nu}^{(j)}$ of the
Scattering Matrix, we can associate a quasimode
$\Phi(\sigma_{\nu}^{(j)};t,\mathbf{x},z)$ such that
$$
\Phi(\sigma_{\nu}^{(j)};t,\mathbf{x},z)\sim C_{\pm}
e^{i\sigma_{\nu}^{(j)}(\mathbf{x}.\pmb{\omega'}+\mid z\omega_4\mid-t)},\;\;z\rightarrow\pm\infty
$$
where $(\pmb{\omega'},\omega_4)\in S^3$. Given $\sigma_{\nu}^{(1)}=\alpha 
\zeta_{\nu}^{(1)}\in\Sigma_{\nu}^{(1)}$ or  $\sigma_{\nu}^{(2)}=-\alpha 
\zeta_{\nu}^{(2)}\in\Sigma_{\nu}^{(2)}$, $1\leq\alpha$, we choose $\pmb{\omega}\in S^3$ with
$\mid\omega_4\mid=\alpha^{-1}$, and we put
$$
\Phi(\sigma_{\nu}^{(j)};t,\mathbf{x},z):=
e^{i\sigma_{\nu}^{(j)}(\mathbf{x}.\pmb{\omega'}-t)}u_{\nu}^{(j)}(z).
$$

An elementary Fourier analysis shows that the
finite energy solutions of (\ref{kgc}) are superpositions of  plane waves solutions,
$
e^{i(\lambda_{\nu}t+\mathbf{x}.\pmb{\xi})}
$
with $\pmb{\xi}\in\RR^3$, and $\lambda_{\nu}$ satisfies the relation
of dispersion 
\begin{equation}
\lambda_{\nu}^2=\mid\pmb{\xi}\mid^2+\left(\zeta_{\nu}^{(j)}\right)^2.
  \label{reldi}
\end{equation}
The behaviour in time of these plane waves is exponential since
$\Im \lambda_{\nu}\neq 0$. In the same spirit, a second kind of quasimodes that play an important role in physics of
the branes, are the solutions that propagate on the brane as an
harmonic plane wave $e^{i\mathbf{x}.\pmb{\xi}}$ :
$$
\Phi(\lambda_{\nu};t,\mathbf{x},z):=e^{i(\lambda_{\nu} t+\mathbf{x}.\pmb{\xi})}u_{\nu}^{(j)}(z).
$$
These last modes can be associated with the singularities of the
analytic continuation of the truncated resolvents (see Corollary \ref{comres}) of
which the common domain of analyticity $\cap_{R_0>0}\mathcal{O}_{R_0}$
is depicted on 
Figure 3 (the curves are defined by the relation of dispersion
(\ref{reldi})). We remark that the
usual definitions of the resonances, on the one hand as singularities
of the scattering matrix, and on the other hand as singularities of the
truncated resolvent, are not equivalent in the case of the brane
scattering. The common singularities are the zeros of the Hankel
functions, that are the origins of the half-hyperbolas and the half
straight lines.\\

\begin{center}
\psset{xunit=1.1cm, yunit=1.1cm}
%
\end{center}
\vspace{1cm}

We let open the hard problem consisting to
give a rigorous meaning to the following possible asymptotic representation of
the solutions of (\ref{EQB}) :
\begin{equation}
\Phi(t,\mathbf{x},z)\sim\sum_{\zeta_{\nu}^{(j)}\in\mathcal{Z}_{\nu}^{(j)},-\pi<\arg\zeta_{\nu}^{(j)}<0}\phi(\zeta_{\nu}^{(j)};t,\mathbf{x})u_{\nu}^{(j)}(z),\;\;t\rightarrow\infty.
  \label{fff}
\end{equation}
 A
similar Lax-Phillips formula has been established for another
important cosmological model, the De Sitter-Schwarzschild manifold
\cite{bony}. It  allows to perform a numerical scheme of
computation of the resonances and leads to a direct method to observe
a black-hole from the detection of the gravitational waves
\cite{reson}. In the case of the brane scattering  that we have
investigated in this paper, expansion (\ref{fff}) is expected by the physicists  (see
e.g. \cite{sea}). It would be interpreted as a
``dark radiation'' on the brane, constituted of metastable  massive gravitons with
mass equal to $\Re\zeta_{\nu}^{(j)}$ and mean lifetime $\sim \mid
\Im\zeta_{\nu}^{(j)}\mid^{-1}$ (\cite{sea}, formula (12)). 
While the space dimension is even and the brane is  a non-decaying
perturbation, such a resonances expansion is not totally  hopeless since there is no
accumulation of singularities of the scattering matrix near the real
axis on the first sheet, although it is the case for the singularities of
the resolvent.
We refer to the deep works by Burq and Zworski \cite{burq} and Tang
and Zworski \cite{tang} for similar results in the euclidean framework.



\end{document}